\documentclass[aps,prb,onecolumn,nofootinbib,citeautoscript,10pt]{revtex4-2}

\usepackage{amsmath,amssymb}
\usepackage{comment}

\usepackage[tight]{subfigure}

\usepackage[dvipsnames]{xcolor}
\usepackage[papersize={8.5in,11in}]{geometry}
\usepackage[colorlinks=true]{hyperref}
\hypersetup{
    bookmarks=true,
    unicode=false,
    pdftoolbar=true,
    pdfmenubar=true,
    pdffitwindow=false,
    pdfstartview={FitH},
    pdfkeywords={plasmons} {vortex nodal-ring semimetal} {random phase approximation},
    pdfnewwindow=true,
    colorlinks=true,
    linkcolor=magenta,
    citecolor=blue,
    filecolor=magenta,
    urlcolor=blue
}

\usepackage{dcolumn}
\usepackage{color}
\usepackage{amssymb,amsmath}
\usepackage{tabularx,graphicx,makecell}
\usepackage{epstopdf}
\usepackage{latexsym}
\usepackage{colortbl}
\usepackage{psfrag}
\usepackage{bbm,bm,array,physics}
\usepackage{dsfont}
\usepackage{float, mathrsfs}
\newcommand{\bs}[1]{\boldsymbol{#1}}

\def \nn{\nonumber \\}
\def\*#1{\boldsymbol{#1}}

\begin{document}

\title{Plasmon modes in tilted three-dimensional nodal-ring semimetals. II. Vortex nodal ring}

\author{Ipsita Mandal}
\email{ipsita.mandal@snu.edu.in}

\affiliation{Department of Physics, Shiv Nadar Institution of Eminence (SNIoE), Gautam Buddha Nagar, Uttar Pradesh 201314, India}

%%%%%%%%%%%%%%%%%%%%%%%%%%%%%%%%%%%%%%%%%%%%%%%%%%%%%%%%
\begin{abstract}
We investigate collective charge excitations in three-dimensional semimetals with dipole-like Berry-curvature (BC) profiles, focusing on vortex nodal-ring semimetals (VNRs) and Hopf semimetals (HSMs). The VNR hosts a continuous ring of dipole-like BC sources generated by its winding pseudospin texture, whereas the HSM hosts an ideal BC dipole. This allows us to examine how distinct realisations of dipole-like BC structure affect collective charge dynamics. While plasmon modes in $\mathcal{PT}$-symmetric nodal-ring semimetals (PTNRs) and their gapped cousins were studied in a companion work, \href{https://arxiv.org/abs/2609.35373}{arXiv:2609.35373}, here we analyse the density response, Drude weights, dielectric function, and plasmon modes of VNRs and HSMs at low doping within the random-phase approximation. For VNRs, we consider untilted, in-plane tilted, and axially tilted systems. Although VNR and PTNR have identical energy dispersions, their distinct eigenstates introduce a toroidal-angle dependence in the band-overlap factor, yielding additional interband contributions for in-plane wavevectors and weak corrections to the screened plasmon frequencies. The toroidal Fermi surface also produces intrinsically anisotropic bare plasmons. In contrast, HSMs have isotropic bare plasmons, with anisotropy arising only through their interband response. Our results show that collective charge dynamics can distinguish different realisations of dipole-like BC structure and reveal pseudospin information beyond the single-particle dispersion.
\end{abstract}

\maketitle

\tableofcontents

%========================= BODY OF THE ARTICLE =========================
\section{Introduction}

Topological semimetals are gapless phases whose low-energy excitations are governed by symmetry-protected band touchings in momentum space. The best-known examples for three-dimensional (3d) systems are the Weyl semimetals (WSMs), in which two nondegenerate bands touch at isolated points of the Brillouin zone (BZ)~\cite{burkov11_Weyl, yan17_topological, bernevig}. Each Weyl node is a monopole source or sink of the Berry curvature (BC)~\cite{xiao_review, sundaram99_wavepacket}, and its charge is the Chern number evaluated over any closed surface that encloses the node. This picture accounts for the intrinsic anomalous Hall effect~\cite{haldane04_berry, goswami13_axionic, burkov14_anomolous} and for the unconventional magnetotransport of WSMs~\cite{zhang16_linear, amit_magneto}. In a doped three-dimensional (3d) WSM, the Fermi surface (FS) of a node is a sphere, which a tilt of the cone deforms into an ellipsoid.

Band touchings can also extend along lines. Nodal-ring semimetals (NRSs) are gapless phases in which symmetry protects a touching along a closed one-dimensional (1d) curve in the BZ~\cite{balents-nodal, fu_nlsm, yang_review_nlsm}. Candidate materials include SrAs$_3$~\cite{arpes-nlsm}, Ca$_3$P$_2$~\cite{expt1_nlsm}, CaAgP and CaAgAs~\cite{expt2_nlsm}, Co$_3$Sn$_2$S$_2$~\cite{enke}, Fe$_2$MnX compounds~\cite{claudia_nlsm}, and the alkaline-earth metals Ca, Sr, and Yb~\cite{alkaline_nlsm}. Nodal rings have also been engineered in photonic~\cite{biao_nodal} and acoustic~\cite{nlr-acoustic} crystals. First-principles calculations place a nearly circular ring with weak dispersion close to the Fermi level of CuTeO$_3$~\cite{cuteo_nlsm}. A finite chemical potential turns the ring into a toroidal FS, whereas the intrinsic limit leaves only the ring itself. The responses explored so far include anisotropic magnetoconductivity~\cite{ips-nlsm-ph, phe_nlsm, gnr-hall-ah, ips-gnr-strain}, Hall-type responses~\cite{ips-magnus, gnr-hall-ah}, and Friedel oscillations~\cite{rahmipoor-2d-nlsm, rhim-kim-friedel}. Transport and scattering in massive nodal rings and related systems have been examined as well~\cite{yang1, ips-sanskar-2dgnr, ips-sandip-fano}.

Berry geometry is not restricted to monopoles. The three-band Hopf semimetals (HSMs) host threefold nodes with a vanishing Chern number, and their topology is encoded in the Hopf invariant~\cite{graf-hopf, hopf2, hopf3, ips-fa, ips-fa-lett}. This invariant is the linking number of the preimages of the Bloch-state map. The low-energy spectrum consists of a Dirac cone and a flat band, and the FS of the doped system is a sphere. The BC around the node is a point-like dipole in place of a monopole.

The vortex nodal-ring semimetal (VNR) of Ref.~\cite{vortex-nrsm} provides a second route to a dipolar BC. In the conventional two-band NRS, the pseudospin turns within a fixed plane as one circles the ring. This produces a Berry phase of $\pi$ and a vanishing BC away from the ring~\cite{schnyder_nodal, rhim-kim-friedel, ips-dipole-vnr, lpge-ptnr}. In the VNR, the in-plane pseudospin component points radially. The pseudospin then forms a toroidal, smoke-ring-like field around the nodal ring, which acts as a vortex line in momentum space~\cite{vortex-nrsm, chang_nlsm_review}. On a $k_z$ slice away from the plane of the ring, the pseudospin map carries a skyrmion number of $+1$ on one side of the plane and $-1$ on the other~\cite{vortex-nrsm}. The two contributions cancel, and the VNR has no net monopole charge. The BC is nevertheless smooth and finite throughout the BZ~\cite{ips-dipole-vnr}. Every point of the ring acts as a dipole-like BC source, and the VNR thus realises a continuous ring of such sources, in contrast to the single point-like BC-dipole at a node of the HSM~\cite{graf-hopf, ips-dipole-vnr}. The magnetoelectric conductivity and the electrochemical response of this structure have been explored in Refs.~\cite{ips-dipole-vnr, ips-firdous-vnr}.

Plasmons, the collective oscillations of the charge density, offer a complementary probe of these systems. In the doped regime, their frequency is controlled by the Drude weight, which is fixed by the Fermi surface (FS) and the band velocities. Their damping is controlled by the interband continuum, in analogy with the plasmons of graphene~\cite{hwang-dassarma-graphene}. Among nodal-point semimetals, the plasmons of the untilted and type-I tilted Weyl semimetal (WSM) were derived in the companion paper~\cite{ips-nr-plasmon}. They are isotropic without tilt, and the interband screening contains a logarithmic dependence on the momentum cut-off of the linearised description. The plasmons of the HSM were found to be anisotropic although the band structure is isotropic, and the anisotropy was attributed to the dipolar Berry curvature (BC)~\cite{hopf-plasmon}.

Plasmons of 3d Luttinger semimetals have been analysed in the isotropic limit~\cite{ips-plasmon-luttinger-aop} and in anisotropic settings~\cite{ips-plasmon-luttinger-epjb}. Nodal rings add a distinctive feature, because the toroidal FS makes the plasmon dispersion strongly anisotropic. Plasmons have been studied in two-dimensional (2d) and 3d nodal-ring systems~\cite{rahmipoor-2d-nlsm, rhim-kim-friedel}. At the intrinsic point, the polarisability differs qualitatively from that of nodal-point semimetals. In 2d, a finite tilt modifies the collective modes~\cite{rahmipoor-2d-nlsm}. These works primarily involve pseudospin textures whose winding is confined to a fixed plane. The question we address is whether the richer momentum-space texture and the dipolar BC of the VNR also leave an imprint on the collective charge excitations.

While the anisotropic plasmons of the HSM have been analysed previously~\cite{hopf-plasmon}, our treatment extends that work in several directions. We derive the full RPA dielectric function of the untilted HSM, including both the intraband Drude term and the interband contribution, which allows us to obtain the screened plasmon frequencies in addition to the bare ones. Using the same local mapping onto two-dimensional Dirac cones that we employ for the VNR, we show that the anisotropy of the screened plasmons originates from the quantum metric of the interband transitions between the flat band and the Dirac cone. We give a closed-form expression for the anisotropy ratio. We also place the HSM in direct comparison with the VNR and the WSM. This comparison shows that the bare plasmons of the HSM are isotropic like those of the WSM, whereas the screened plasmons are anisotropic because of the interband background. In the VNR, by contrast, the anisotropy is already present in the bare frequencies owing to the toroidal Fermi surface. These results are collected in our comparison tables.

As a reference, we use the 3d $\mathcal{PT}$-symmetric nodal-ring (PTNR) semimetal and its gapped counterpart (GNR), whose plasmons were studied for in-plane and axial band tilts in Ref.~\cite{ips-nr-plasmon}. The VNR differs from the PTNR in its Hamiltonian, its pseudospin texture, and the momentum-space distribution of its BC. After linearisation about the nodal ring, however, the two systems have identical energy dispersions. Quantities fixed by the spectrum and the phase-space measure, such as the FS and the density of states (DOS), therefore coincide, which makes the comparison particularly clean. Quantities sensitive to the eigenstates can still distinguish the two systems. The overlap factor of the VNR depends on the toroidal angle, and its quantum metric contains an additional term associated with the pseudospin winding. These features can modify the interband part of the density response. The GNR, in turn, acquires a finite BC from its mass term, but with in-plane components only~\cite{ips-nlsm-ph, ips-gnr-strain}.

We compute the noninteracting density response of the VNR at zero temperature, with the chemical potential either at the nodal ring or small compared with $v_0\,k_0$. The plasmon dispersion follows from the dielectric function within the random-phase approximation (RPA), as in Ref.~\cite{rahmipoor-2d-nlsm}. The configurations are the untilted system, an in-plane tilt along $k_x$, and an axial tilt along $k_z$. The shared spectrum fixes the DOS, the Drude weights, and the intraband plasmon frequencies to their PTNR values. The differences come from the eigenstates. A wavevector along $k_z$ leaves the toroidal angle unchanged, and the complete response then coincides with that of the PTNR. An in-plane wavevector changes the toroidal angle. The vortex texture then modifies the overlap factor, which adds interband contributions and shifts the background dielectric screening. We determine these corrections and assess their effect on the plasmons. To place the results in a broader context, we also derive the plasmons of the untilted HSM and compare them with those of the VNR and the WSM.

The remainder of the paper is organised as follows. Sec.~\ref{secmodel} presents the VNR Hamiltonian and toroidal coordinates, the pseudospin texture with its overlap factor, quantum metric, and BC, the density-response formalism, and the FS geometry. The untilted VNR is treated in Sec.~\ref{secuntilted}, and the in-plane and axial tilts follow in Secs.~\ref{secinplane} and~\ref{secaxial}, respectively. Sec.~\ref{sechopf} treats the plasmons of the untilted HSM and compares them with those of the untilted VNR and the WSM. Sec.~\ref{seccompare} compares the results across various systems, including the PTNR and GNR of the companion paper, Ref.~\cite{ips-nr-plasmon}. Sec.~\ref{secsum} contains the summary and outlook.
%%%%%%%%%%%%%%%% 
Appendices~\ref{app:untilted}, \ref{app:xtilt}, and~\ref{app:ztilt} collect the intermediate steps of the calculations in the main text. We set $\hbar=c=k_B=e=1$ and keep the symbol $e$ in analytical expressions as a bookkeeping device.

%%%%%%%%%%%%%%%%%%%%%%%%%%%
\section{Model}
\label{secmodel}

The minimal model of a VNR, comprising two bands and a circular nodal loop lying in the $k_x k_y$-plane at $k_z = 0$, is captured by~\cite{vortex-nrsm}
\begin{align}
\mathcal{H}_0 (\bs{k}) = \bs{d}_0 (\bs{k}) \cdot \bs{\sigma}\,,
\quad
\bs{d}_0 (\bs{k}) = \left\lbrace \frac{-\,k_x\,k_z}{m_z},\, \frac{-\,k_y\,k_z}{m_z},\,
\frac{k_\perp^2 - k_0^2 - k_z^2}{2\,m_\perp} \right\rbrace,
\quad k_\perp = \sqrt{k_x^2 + k_y^2}\,,
\end{align}
where $\bs{\sigma} = \{\sigma_x, \sigma_y, \sigma_z\}$ denotes the vector of Pauli matrices. The parameters $k_0$, $m_z$, and $m_\perp$ are material-dependent. In the $k_z = 0$ plane, the two bands cross at $k_\perp = k_0$, which defines a nodal ring of radius $k_0$. The Fermi velocities perpendicular and parallel to the nodal plane are $v_z \equiv k_0/m_z$ and $v_\perp \equiv k_0/m_\perp$, respectively. For simplicity, we set $v_z = v_\perp = v_0$ in what follows. In contrast to the NRS model with a mass term, no further matrix structure is available to open a gap, because all three Pauli matrices already appear in $\bs{d}_0$. The VNR therefore has only a gapless realisation.

For low-energy excitations near the FS, the dominant transport signatures are captured by a linearised Hamiltonian $\mathcal{H}$ that deviates from the nodal line only at linear order in momentum~\cite{linearize-nlsm, ips-nlsm-ph}. This is achieved by transforming to cylindrical coordinates,
\begin{align}
\label{eqtrs}
k_x = k_\perp \cos\phi\,, \quad
k_y = k_\perp \sin\phi\,, \quad k_\perp = k_0 + \kappa_\perp\,,
\end{align}
with $\kappa_\perp \in [-k_0, \infty)$. Low-energy processes require $|\kappa_\perp| \ll k_0$, since $k_0$ is an ultraviolet momentum scale. Linearising in $\kappa_\perp$ and $k_z$, we obtain
\begin{align}
\label{eqham}
\mathcal{H}_\ell (\bs{k}) = \bs{d} (\bs{k}) \cdot \bs{\sigma}\,,
\quad
\bs{d} (\bs{k}) = v_0 \left\lbrace -\,k_z \cos\phi,\, -\,k_z \sin\phi,\, \kappa_\perp \right\rbrace.
\end{align}
The eigenvalues and band-velocities of $\mathcal{H}_\ell$ are
\begin{align}
\label{eqev}
\varepsilon_s (\bs{k}) &= (-1)^s\, \epsilon_k\,, \quad
\epsilon_k = v_0\,\sqrt{\kappa_\perp^2 + k_z^2}\,, \quad 
s \in \lbrace 1, 2 \rbrace \,, \quad
%%%%%%%%%%%%%
\bs{v}^{(0,s)} (\bs{k})
\equiv \nabla_{\bs{k}} \varepsilon_s (\bs{k})
= \frac{(-1)^s\, v_0^2}{\epsilon_k} \left\lbrace \kappa_\perp \cos\phi,\,
\kappa_\perp \sin\phi,\, k_z \right\rbrace .
\end{align}
%%%%%%
These coincide with the spectrum and velocities of the linearised PTNR at vanishing gap~\cite{ips-nr-plasmon}. This coincidence holds only at the linearised level. The two Hamiltonians are different, because the vector $\bs{d}_0$ of the VNR has three nonvanishing components, whereas that of the PTNR lies in a fixed plane. Beyond the linearised regime, the spectra differ as well.

The toroidal coordinates are obtained by supplementing Eq.~\eqref{eqtrs} with
\begin{align}
\label{eqtor}
\kappa_\perp = \kappa \cos\gamma\,, \quad k_z = \kappa \sin\gamma\,, \quad
\kappa = \sqrt{\kappa_\perp^2 + k_z^2}\,.
\end{align}
The Jacobian is $J = \kappa\,(k_0 + \kappa \cos\gamma)$, and $\epsilon_k = v_0\,\kappa$. Here $k_0$ is the major radius of the torus, and $\kappa$ is the minor radius, which measures the cross-sectional extent. The toroidal angle $\phi$ and the poloidal angle $\gamma$ each range over $[0, 2\pi)$. They describe rotations around the torus axis and around the ring, respectively. In these coordinates,
\begin{align}
\label{eqd-tor}
\bs{d} = v_0\,\kappa\, \bs{n}\,, \quad
\bs{n} (\gamma, \phi) = \left\lbrace -\sin\gamma \cos\phi,\, -\sin\gamma \sin\phi,\, \cos\gamma \right\rbrace,
\end{align}
where $\bs{n}$ is the unit vector along which the pseudospin of the upper band $s = 2$ points. The pseudospin of the band $s = 1$ points along $-\bs{n}$. The vortex referred to above is the factor $(\cos\phi, \sin\phi)$ in the in-plane components of $\bs{n}$. For the PTNR, the same two angles would appear as $\bs{n} = \{\cos\gamma, \sin\gamma, 0\}$ (up to a fixed choice of axes), which has no dependence on $\phi$.

We add a tilt to the linearised Hamiltonian, which gives
\begin{align}
\label{eq:hamvnr}
\mathcal{H} (\bs{k}) = \bs{d} (\bs{k}) \cdot \bs{\sigma} + v_0\, \bs{\eta} \cdot \bs{k}\; \mathbb{I}_{2 \times 2}\,,
\end{align}
where $\bs{\eta} = \eta_x\, \bs{\hat x} + \eta_y\, \bs{\hat y} + \eta_z\, \bs{\hat z}$ is the dimensionless tilt vector. The tilt term is proportional to the identity in band space. It shifts the eigenvalues to $\varepsilon_s (\bs{k}) = (-1)^s\, \epsilon_k + v_0\, \bs{\eta} \cdot \bs{k}$ and the velocities to $\bs{V}^{(s)} = \bs{v}^{(0,s)} + v_0\, \bs{\eta}$. It leaves the eigenstates unchanged.

For $\bs{\eta} = 0$, the Hamiltonian of Eq.~\eqref{eqham} is invariant under two operations. The first is a rotation of $\bs{k}$ about the $k_z$-axis, accompanied by the same rotation of the pseudospin about $\sigma_z$. The second is the reflection $k_z \to -k_z$, accompanied by the conjugation $\mathcal{H} \to \sigma_z\, \mathcal{H}\, \sigma_z$. In contrast to the PTNR, the rotation is therefore a combined operation, because the vortex texture is not invariant under a rotation of $\bs{k}$ alone. An in-plane tilt $\bs{\eta} = \eta_x\, \bs{\hat x}$ breaks the rotation and preserves the reflection. An axial tilt $\bs{\eta} = \eta_z\, \bs{\hat z}$ does the reverse.

%%%%%%%%%%%%%%%%%%%%%%%%%%%
\subsection{Pseudospin texture: overlap factor, quantum metric, and Berry curvature}
\label{secvortex}

The eigenstates of $\mathcal{H}$ are the spin-$1/2$ coherent states along $\pm\, \bs{n}$. The squared overlap between the state at $\bs{k}$ in band $s$ and the state at $\bs{k}'$ in band $s'$ is
\begin{align}
\label{eq:formfactor}
\mathcal{F}_{ss'} (\bs{k}, \bs{k}') = \frac{1 + (-1)^{s + s'}\, \bs{n} \cdot \bs{n}'}{2}\,,
\quad
\bs{n} \cdot \bs{n}' = \cos\gamma \cos\gamma' + \sin\gamma \sin\gamma' \cos (\phi - \phi')\,.
\end{align}
The interband factor is $\mathcal{F} \equiv \mathcal{F}_{12} = \mathcal{F}_{21}$. Its dependence on $\phi - \phi'$ distinguishes the VNR from the PTNR, where $\mathcal{F} = [1 - \cos (\gamma - \gamma')]/2$. For $\bs{q} = q\, \bs{\hat z}$, the momentum transfer leaves $\phi$ unchanged, and $\mathcal{F}$ takes the same form as the PTNR expression. For an in-plane $\bs{q}$, the toroidal angle shifts by $\phi' - \phi \simeq \bs{q} \cdot \bs{\hat\phi} / k_\perp$, where $\bs{\hat\phi} = \{-\sin\phi, \cos\phi, 0\}$. The shift is of order $q/k_0$.

For $q \to 0$, $\mathcal{F} \simeq q_i\, q_j\, g_{ij} (\bs{k})$, where the quantum metric is $g_{ij} = \partial_i \bs{n} \cdot \partial_j \bs{n} / 4$. Using Eq.~\eqref{eqd-tor},
\begin{align}
\label{eq:qmetric}
g_{ij} = \frac{\hat\gamma_i\, \hat\gamma_j}{4\,\kappa^2}
+ \frac{\sin^2\gamma}{4\,k_\perp^2}\, \hat\phi_i\, \hat\phi_j\,,
\quad
\bs{\hat\gamma} = \left\lbrace -\sin\gamma \cos\phi,\, -\sin\gamma \sin\phi,\, \cos\gamma \right\rbrace,
\end{align}
with $k_\perp = k_0 + \kappa \cos\gamma$. The unit vector $\bs{\hat\gamma}$ points along $\bs{\nabla}_{\bs{k}} \gamma$. Its components coincide with those of $\bs{n}$ in Eq.~\eqref{eqd-tor}, but the two vectors are different objects, since $\bs{n}$ lives in pseudospin space and $\bs{\hat\gamma}$ in momentum space. The first term has the same form as the quantum metric of the PTNR. The second term has no PTNR counterpart and arises from the winding in $\phi$. It is smaller than the first by a factor of order $(\kappa / k_0)^2$ and vanishes in the $k_z = 0$ plane.

The BC of the band $s$ is $\bs{\Omega}^{(s)} = \pm\, \tfrac{1}{2}\, \sin\gamma\, \nabla_{\bs{k}} \gamma \times \nabla_{\bs{k}} \phi$, where the overall sign depends on the band and on convention. It is directed along the radial unit vector $\bs{\hat\kappa} = \{\cos\gamma \cos\phi,\, \cos\gamma \sin\phi,\, \sin\gamma\}$ of the poloidal cross-section, with magnitude
\begin{align}
\label{eq:berry}
|\bs{\Omega}| = \frac{|\sin\gamma|}{2\,\kappa\, k_\perp}\,.
\end{align}
The BC is nonzero everywhere except on the $k_z = 0$ plane, where $\sin\gamma = 0$. This contrasts with the PTNR, where the BC vanishes away from the ring. The flux through any torus $\kappa = \text{const}$ is proportional to $\int_0^{2\pi} \sin\gamma\, d\gamma$ and vanishes. The VNR therefore has no net monopole charge, and the BC acts as dipole-like sources distrbuted along the nodal ring~\cite{ips-dipole-vnr, ips-firdous-vnr}.

%%%%%%%%%%%%%%%%%%%%%%%%%%%
\subsection{Density response, Drude weight, and plasmon dispersion}
\label{secresponse}

The noninteracting density-density response function at temperature $T$ and chemical potential $\mu$ is given by the retarded bare bubble diagram, evaluated with the eigenstates of Eq.~\eqref{eq:hamvnr}~\cite{giuliani-vignale}. It is the 3d version of the expression used in Ref.~\cite{rahmipoor-2d-nlsm} for a 2d tilted nodal-line semimetal. Explicitly,
\begin{align}
\label{eq:polgeneral}
\Pi (\bs{q}, \omega) = g \int \frac{d^3 \bs{k}}{(2\,\pi)^3}
\sum_{s, s'} \mathcal{F}_{ss'} (\bs{k}, \bs{k}')\,
\frac{f_0 (\varepsilon_s (\bs{k})) - f_0 (\varepsilon_{s'} (\bs{k}'))}
{\omega + \varepsilon_s (\bs{k}) - \varepsilon_{s'} (\bs{k}') + i\, 0^+}\,,
\quad \bs{k}' = \bs{k} + \bs{q}\,,
\end{align}
where $g$ is the degeneracy factor and $f_0$ is the Fermi-Dirac distribution. The model is spinless, and $g = 1$. We keep $g$ to retain generality. The complex response function splits as $\Pi = \Re \Pi + i\, \Im \Pi$. The real part is the physical polarisability that enters the RPA dielectric function, and the imaginary part governs the particle-hole damping. All calculations are performed at $T = 0$. The sum over $s, s'$ then splits into the intraband part $\Pi_{\rm intra}$ with $s = s'$ and the interband part $\Pi_{\rm inter}$ with $s \neq s'$. Hence, $\Pi = \Pi_{\rm intra} + \Pi_{\rm inter}$.

We consider low but nonzero doping, $0 < \mu \ll v_0\,k_0$. The band-dependent DOS is
\begin{align}
\label{eqdos}
\rho_s (\mu) = g \int \frac{d^3 \bs{k}}{(2\,\pi)^3}\, \delta (\varepsilon_s - \mu)\,.
\end{align}
It enters the Drude-weight tensor,
\begin{align}
\label{eqdrude}
D_{ij} = \pi\, g\, e^2 \sum_s \int \frac{d^3 \bs{k}}{(2\,\pi)^3}\,
\delta (\varepsilon_s - \mu)\, V_i^{(s)}\, V_j^{(s)}\,,
\end{align}
which fixes the Drude peak of the optical conductivity, $\Re \sigma_{ij} (\omega) = D_{ij}\, \delta (\omega)$. Both quantities depend only on $\varepsilon_s$ and $\bs{V}^{(s)}$. Since these coincide with their PTNR counterparts, $\rho_s$ and $D_{ij}$ are the same as in the PTNR for every tilt. Taking the limit $q \to 0$ first and thereafter $\omega \to 0$, the intraband part gives~\cite{giuliani-vignale}
\begin{align}
\label{eqintra}
\Pi_{\rm intra} (\bs{q}, \omega) \simeq q_i\, q_j\, D_{ij} / (\pi\, e^2\, \omega^2)\,.
\end{align}
The opposite order of limits gives the static result $\Pi (\bs{q} \to 0, 0) = -\, \rho (\mu)$, with $\rho = \sum_s \rho_s$. Within the RPA, the dielectric function is
\begin{align}
\label{eq:epsrpa}
\varepsilon^{\rm RPA} (\bs{q}, \omega) = 1 - v (q)\, \Pi (\bs{q}, \omega)\,,
\quad v (q) = \frac{4\,\pi\, e^2}{q^2}\,,
\end{align}
where $v (q)$ is the Fourier transform of the 3d Coulomb interaction $e^2/r$. Undamped collective modes lie outside the particle-hole continuum, where $\Im \Pi = 0$. They are the zeros of $\Re \varepsilon^{\rm RPA}$. At long wavelengths, $\Re \Pi$ is dominated by the intraband part of Eq.~\eqref{eqintra}. Inserting it into $\Re \varepsilon^{\rm RPA} = 0$ gives the plasmon frequency along the principal direction $i$,
\begin{align}
\label{eqplasmon}
\omega_{p, i}^2 = 4\, D_{ii}\,,
\end{align}
with no sum over $i$. The interband part is real and negative below the interband threshold. It acts as a background dielectric constant $\epsilon_{b, i}$ and lowers the bare frequencies to $\tilde\omega_{p, i}^2 = \omega_{p, i}^2 / \epsilon_{b, i}$.

The tensor $D_{ij}$ is diagonal in the Cartesian frame set by the tilt direction. The argument depends only on $\varepsilon_s$, and it gives the same result as for the PTNR. For $\bs{\eta} = 0$ and for the axial tilt, the combined rotation about $k_z$ gives $D_{xx} = D_{yy}$. For the in-plane tilt, $\varepsilon_s$ depends on $\phi$ only through $\cos\phi$. It is therefore invariant under $\phi \to -\phi$ and, independently, under $\gamma \to -\gamma$. These two reflections flip the signs of $V_y^{(s)}$ and $V_z^{(s)}$, respectively. The off-diagonal entries then vanish, while $D_{xx} \neq D_{yy}$.

At leading order, the vortex acts on the interband part. The intraband overlap $\mathcal{F}_{ss} \simeq 1 - q_i\,q_j\,g_{ij}$ also differs from its PTNR counterpart, but this affects $\Pi_{\rm intra}$ only at higher order in $q$. Outside any electron- or hole-pocket, the band $s = 1$ is completely filled and the band $s = 2$ is completely empty. At the nodal ring, this holds throughout for $\bs{\eta} = 0$ and for the axial tilt. For the in-plane tilt, pockets exist even at $\mu = 0$, as shown in Sec.~\ref{secfermisurface}. The momenta $\bs{k}$ and $\bs{k}'$ must then both lie outside the pockets, and we obtain
\begin{align}
\label{eq:polinter}
\Pi_{\rm inter} (\bs{q}, \omega) = g \int \frac{d^3 \bs{k}}{(2\,\pi)^3}\, \mathcal{F} (\bs{k}, \bs{k}')
\left (\frac{1}{\omega - \epsilon_k - \epsilon_{k'} - v_0\, \bs{\eta} \cdot \bs{q} 
+ i\, 0^+}
- \frac{1}{\omega + \epsilon_k + \epsilon_{k'} - v_0\, \bs{\eta} \cdot \bs{q} + i\, 0^+} \right ),
\end{align}
where $\mathcal{F}$ is given by Eq.~\eqref{eq:formfactor} and $\epsilon_{k'}$ is $\epsilon_k$ evaluated at $\bs{k}'$. The tilt enters only through the shift $v_0\, \bs{\eta} \cdot \bs{q}$ of the transition energies. In the static limit, the two shifts cancel to leading order in $q$, and
\begin{align}
\label{eq:polinterstatic}
\Pi_{\rm inter} (\bs{q}, 0) \simeq -\, g \int \frac{d^3 \bs{k}}{(2\,\pi)^3}\,
\frac{q_i\, q_j\, g_{ij} (\bs{k})}{\epsilon_k}
 \left[ f_0 (\varepsilon_1) - f_0 (\varepsilon_2) \right],
\end{align}
with $g_{ij}$ from Eq.~\eqref{eq:qmetric}. The occupation difference vanishes inside any electron- or hole-pocket, which is then excluded from the integral. The replacement of $\mathcal{F}$ by $q_i\,q_j\,g_{ij}$ requires $q \ll \kappa$ for all contributing states, which means $q \ll \kappa_F$ in the doped case. At $\mu = 0$, the integrand scales as $\kappa^{-3}$ near the ring. The radial integral with measure $\kappa\,d\kappa$ then diverges at small $\kappa$. The expansion fails in this limit, and the response is non-analytic in $q$, scaling linearly instead of quadratically, as for a 2d Dirac cone. The tilt corrections to Eq.~\eqref{eq:polinterstatic} are of relative order $(v_0\,\bs{\eta} \cdot \bs{q} / \epsilon_k)^2$.

%%%%%%%%%%%%%%%%%%%%%%%%%%%
\subsection{Validity of the low-energy description}
\label{secvalidity}

We work at low doping, $\mu \ll v_0\,k_0$, which is equivalent to $\kappa_F \ll k_0$. The corrections to Eq.~\eqref{eqham} that come from the terms dropped in $\bs{d}_0$ are of relative order $\kappa / k_0$. They include the quadratic terms in $\kappa_\perp$ and $k_z$ in the third component of $\bs{d}_0$, and the factor $k_\perp / k_0$ in the first two components. The same ratio controls the mapping of the response onto decoupled copies of a 2d Dirac cone, one for each $\phi$, with the ring supplying the overall multiplicity $k_0$ through the Jacobian. This mapping requires $q \ll k_0$. For the VNR, the dependence of $\bs{n}$ on $\phi$ gives the second term of the quantum metric, which is smaller than the first by a factor of order $(\kappa / k_0)^2$, as seen from Eq.~\eqref{eq:qmetric}. After integration over momentum, it enters the real part of the in-plane response at relative order $\kappa_F\,\Lambda / k_0^2$, where $\Lambda$ is the momentum cut-off of the linearised description. For $\Lambda \sim k_0$, this is of order $\kappa_F / k_0$, as shown in Sec.~\ref{secvortexcorr}. This is the same order as the corrections from the terms dropped in $\bs{d}_0$. The contribution of the $\phi$-dependent part of the overlap factor is therefore meaningful only together with the leading terms evaluated with the same accuracy. The leading-order results of the following sections hold within these limits. We do not compute Friedel oscillations, since they require the response at $q = 2\,\kappa_F$, which can be comparable to $k_0$ for realistic parameters~\cite{ips-gnr-strain, ips-dipole-vnr}. The numerical estimates in this paper use $v_0 = 0.0004$ and $k_0 = 160$ eV in natural units, taken from Ref.~\cite{phe_nlsm}. We set $g = 1$, although we keep $g$ in the analytical expressions for the sake of generality. We choose the chemical potential $\sim  0.1\,v_0\,k_0$, which gives $\kappa_F/k_0 \sim 0.1$ without tilt, obeying the constraint $\kappa_F \ll k_0$. The PTNR results shown for comparison are evaluated at the same values.

%%%%%%%%%%%%%%%%%%%%%%%%%%%%%%%%%%%%%%%%%%%%%%%%%
\subsection{Fermi surface under chemical potential and tilt}
\label{secfermisurface}

The FS follows from $\varepsilon_s (\bs{k}) = \mu$. Since the bands are those of the PTNR at $\Delta = 0$, the geometry of the FS is also the same~\cite{ahn-tilted-nl, ips-dipole-vnr}. Without tilt, it is a torus with major radius $k_0$ and minor radius $\kappa_F = \mu / v_0$. The condition $\mu \ll v_0\,k_0$ ensures that this is a ring torus. For $\mu \geq v_0\,k_0$, it would become a horn torus and then a spindle torus, which self-intersects. Such values lie outside the validity of the linearised Hamiltonian.

\begin{figure*}[t!]
\centering
\includegraphics[width=0.9\textwidth]{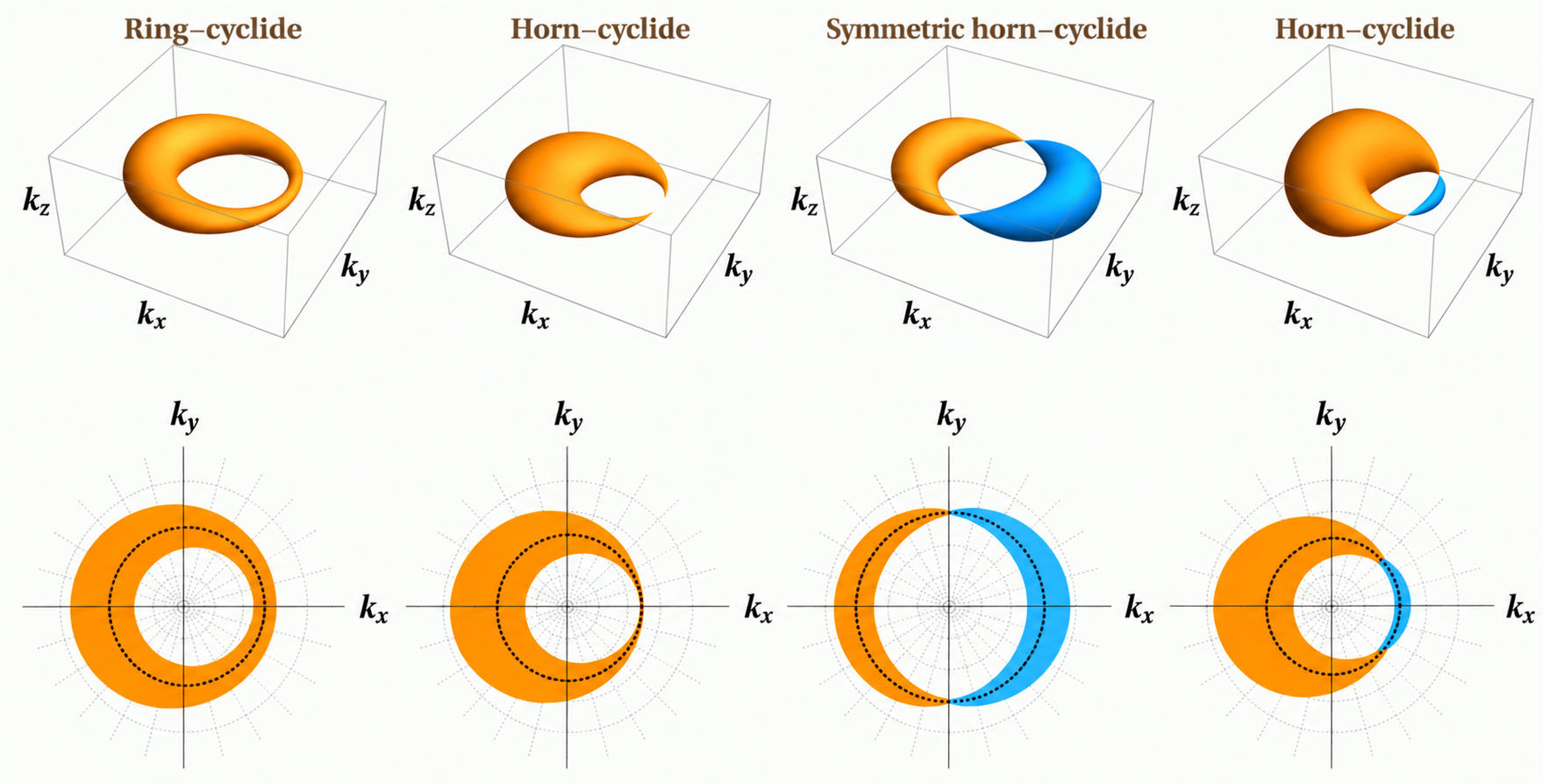}
\caption{\label{figfsvnr} Evolution of the FS of the VNR under an in-plane tilt with respect to the $k_x$-axis~\cite{ahn-tilted-nl, ips-dipole-vnr}. The upper row shows the 3d view and the lower row shows the projection onto the $k_x k_y$-plane. The panels show, from left to right, the ring-cyclide regime, the horn-cyclide regime at $\mu = \tilde{\eta}$, where the hole-pocket has shrunk to a point, the symmetric horn-cyclide regime ($\mu = 0$), and the horn-cyclide regime with $0 < \mu < \tilde{\eta}$. The orange and blue colours depict the $s=2$ (electron) and $s=1$ (hole) pockets, respectively. The dotted black circle marks the original nodal ring at $\mu = \eta_x = 0$. The FS is identical to that of the PTNR.}
\end{figure*}

An in-plane tilt, $\bs{\eta} = \eta_x\, \bs{\hat x}$, shifts the energy by $v_0\, \eta_x\, k_x$. On the nodal ring, this equals $\tilde{\eta} \cos\phi$, where $\tilde{\eta} = v_0\, \eta_x\, k_0$. The ring is no longer at constant energy. For the upper band, the FS is located at
\begin{align}
\label{eq:kappaF-x}
\kappa_F (\gamma, \phi) = \frac{\mu - \tilde{\eta} \cos\phi}{v_0\, (1 + \eta_x \cos\gamma \cos\phi)}\,,
\end{align}
where the numerator must be positive. At $\mu = 0$, the FS consists of an electron-pocket and a hole-pocket, which touch at the two points $\phi = \pm \pi / 2$. The resulting surface is a symmetric horn-cyclide. For $0 < \mu < \tilde{\eta}$, the two pockets coexist on a horn-cyclide, and the contact points, located at $\cos\phi = \mu / \tilde{\eta}$, move from $\phi = \pm\, \pi / 2$ towards the $k_x$-axis as $\mu$ increases. The hole-pocket shrinks to a point at $\mu = \tilde{\eta}$ and disappears for larger $\mu$. For $\mu > \tilde{\eta}$, the FS is a single ring-cyclide that encloses the nodal ring. Fig.~\ref{figfsvnr} shows these regimes. The classification depends only on $|\tilde{\eta}|$, because $\eta_x \to -\eta_x$ is equivalent to $\phi \to \phi + \pi$.

%%%%%%%%%%%%%%%%%%%%%%%%%%%%%%%%%%%%%%%%%%%%%%%%%%%%%%%%%%
%%%%%%%%%%%%%%%%%%%%

An axial tilt, $\bs{\eta} = \eta_z\, \bs{\hat z}$, shifts the energy by $v_0\, \eta_z\, k_z$. This vanishes on the nodal ring, which therefore stays at zero energy. The FS is at $\kappa_F (\gamma) = \mu / [v_0\, (1 + \eta_z \sin\gamma)]$. For $|\eta_z| < 1$, no pocket of the opposite carrier type forms, and the FS is a single closed surface of revolution about the $k_z$-axis. At fixed $\phi$, its cross-section is an ellipse with eccentricity $|\eta_z|$, and the nodal ring sits at one of its foci. For $|\eta_z| > 1$, the factor $1 + \eta_z \sin\gamma$ vanishes at some $\gamma$. The FS then opens into two unbounded sheets. This type-II regime is outside the scope of the present work, and we restrict to $|\eta_x| < 1$ and $|\eta_z| < 1$ throughout.

%%%%%%%%%%%%%%%%%%%%%%%%%%%
\section{Untilted VNR}
\label{secuntilted}

In this section, we set $\bs{\eta} = 0$. The chemical potential either lies at the nodal ring, $\mu = 0$, or is small and positive, $0 < \mu \ll v_0\,k_0$. The FS is then a ring torus of minor radius $\kappa_F = \mu/v_0$. All results are even in $\mu$, because the electron and hole sectors are related by $s \to 3 - s$. We therefore restrict to $\mu \geq 0$. The residual symmetries of Sec.~\ref{secmodel} keep the response isotropic in the $k_x k_y$-plane. Only the directions $\bs{\hat z}$ and $\bs{\hat x}$ are needed. For $\bs{q} = q\,\bs{\hat z}$, the overlap factor of Eq.~\eqref{eq:formfactor} coincides with that of the PTNR. The energies and the measure are the same as well. The response along $\bs{\hat z}$ is therefore identical to the PTNR one~\cite{ips-nr-plasmon}. For an in-plane $\bs{q}$, the overlap factor also depends on $\phi$, and we treat this dependence in Sec.~\ref{secvortexcorr}. The results that follow hold to leading order in $q/k_0$.

%%%%%%%%%%%%%%%%%%%%%%%%%%%
\subsection{Reduction to decoupled Dirac cones}
\label{secreduction}

At fixed $\phi$, the Hamiltonian of Eq.~\eqref{eqham} depends only on the local coordinates $u = k_\perp - k_0$ and $w = k_z$. It reads $\mathcal{H} = v_0\,(u\,\sigma_z - w\,\sigma_\phi)$, with $\sigma_\phi = \cos\phi\,\sigma_x + \sin\phi\,\sigma_y$. This is a massless 2d Dirac cone of velocity $v_0$. The pseudospin axes are rotated with $\phi$, but the energies and the overlap factor between states with the same $\phi$ are those of the standard cone. The Jacobian weight $k_0 + \kappa\cos\gamma \to k_0$ is independent of $\phi$ for $\kappa \ll k_0$. The ring therefore supplies the multiplicity $\int_0^{2\pi} k_0\,d\phi/(2\,\pi) = k_0$. A wavevector $\bs{q} = q\,\bs{\hat z}$ lies in the local plane at every $\phi$. A wavevector $\bs{q} = q\,\bs{\hat x}$ has the local component $q\cos\phi$ along $u$. Its component along the ring changes the energies at higher order in $q/k_0$. It also shifts $\phi$ and thereby probes the pseudospin winding. Neglecting this shift for the moment, the copies decouple, and
\begin{align}
\label{eq:ringmapvnr}
\Pi (q\,\bs{\hat z}, \omega) = k_0\, \Pi_{\rm 2D} (q, \omega)\,, \quad
\Pi (q\,\bs{\hat x}, \omega) \simeq k_0 \int_0^{2\pi} \frac{d\phi}{2\,\pi}\, \Pi_{\rm 2D} \big( q\,|\cos\phi|, \omega \big)\,,
\end{align}
where $\Pi_{\rm 2D} (q, \omega)$ is the noninteracting polarisability of a single massless Dirac cone~\cite{hwang-dassarma-graphene}. The first relation is exact in the linearised model, up to the replacement of the Jacobian by $k_0$. The second holds up to the contribution of the $\phi$-dependence of the overlap factor, which we treat in Sec.~\ref{secvortexcorr}. The steps are given in Appendix~\ref{app:vnrdirac}.

%%%%%%%%%%%%%%%%%%%%%%%%%%%
\subsection{Intrinsic limit}
\label{secintrinsic}

At $\mu = 0$, the FS collapses to the nodal ring. Only interband transitions contribute, and $\Pi = \Pi_{\rm inter}$ of Eq.~\eqref{eq:polinter}, with $\epsilon_k = v_0\,\kappa$. This is $k_0$ times the polarisability of a single undoped massless Dirac cone. The evaluation of this polarisability and of the quantities derived from it below is given in Appendix~\ref{app:vnrintrinsic}. For $\bs{q} = q\,\bs{\hat z}$ and below the interband threshold,
\begin{align}
\label{eq:polmu0vnr}
\Pi (q\,\bs{\hat z}, \omega) = -\, \frac{g\,k_0\,q^2}{16\,\sqrt{v_0^2\,q^2 - \omega^2}}\,,
\quad 0 \leq \omega < v_0\,q\,, \quad q \ll k_0\,.
\end{align}
For $\omega > v_0\,q$, the polarisability becomes purely imaginary, $\Pi (q\,\bs{\hat z}, \omega) = -\,i\,g\,k_0\,q^2 / \big( 16\,\sqrt{\omega^2 - v_0^2\,q^2} \big)$, which signals damping by interband particle-hole excitations. Deep inside the continuum,
\begin{align}
\label{eq:imvnr}
\Im \Pi (q\,\bs{\hat z}, \omega) \simeq -\, \frac{g\,k_0\,q^2}{16\,\omega}\,,
\quad v_0\,q \ll \omega \ll v_0\,k_0\,.
\end{align}
For $\bs{q} = q\,\bs{\hat x}$, the local wavevector is $q\cos\phi$. Since $\Im \Pi_{\rm 2D} \propto q^2$ in this window, the angular average $\langle \cos^2\phi \rangle = 1/2$ halves the result, and the intrinsic damping is anisotropic by a factor of two. This holds up to corrections of relative order $q/k_0$, which arise from the tangential component of $\bs{q}$ and from the second term of the overlap factor. The static limit of Eq.~\eqref{eq:polmu0vnr} is
\begin{align}
\label{eq:pistaticvnr}
\Pi (q\,\bs{\hat z}, 0) = -\, \frac{g\,k_0\,q}{16\,v_0}\,,
\quad \Pi (q\,\bs{\hat x}, 0) = -\, \frac{g\,k_0\,q}{8\,\pi\,v_0}\,.
\end{align}
Both are linear in $q$, which reflects the two-dimensional Dirac dispersion transverse to the nodal ring~\cite{rhim-kim-friedel}. The in-plane value is smaller by the factor $2/\pi$, which is the angular average of $|\cos\phi|$.

Inserting Eq.~\eqref{eq:polmu0vnr} into Eq.~\eqref{eq:epsrpa} gives
\begin{align}
\label{eq:epsintrvnr}
\varepsilon^{\rm RPA} (q\,\bs{\hat z}, \omega) = 1 + \frac{\pi\,g\,e^2\,k_0}{4\,\sqrt{v_0^2\,q^2 - \omega^2}}\,,
\quad 0 \leq \omega < v_0\,q\,,
\end{align}
which exceeds unity throughout the window below the continuum. There is no zero, and hence no undamped plasmon. For $\bs{q} = q\,\bs{\hat x}$, the local threshold $v_0\,q\,|\cos\phi|$ vanishes at $\phi = \pi/2$, and the continuum extends down to $\omega = 0$. The decoupled-cone description fails in the narrow window $|\cos\phi| \lesssim q/k_0$ around this point, where the tangential component of $\bs{q}$ matters. This window has a relative weight of order $q/k_0$ and does not change the conclusion. The intrinsic VNR therefore supports no undamped RPA plasmon, in the same way as undoped graphene and the intrinsic PTNR. There is no FS and, hence, no Drude weight. The positive intraband contribution to $\Pi$ at finite frequency, which is needed to make $\Re \varepsilon^{\rm RPA}$ vanish, is therefore absent.

%%%%%%%%%%%%%%%%%%%%%%%%%%%
\subsection{Doped regime: density-of-states, Drude weights, and plasmons}
\label{secdopedvnr}

For $0 < \mu \ll v_0\,k_0$, the weight $\kappa_F\,(k_0 + \kappa_F\cos\gamma) \to \kappa_F\,k_0$ gives the DOS
\begin{align}
\label{eq:dosvnr}
\rho (\mu) = \frac{g}{(2\,\pi)^3\,v_0} \int_0^{2\pi} \! d\phi \int_0^{2\pi} \! d\gamma\; \kappa_F\,k_0
= \frac{g\,k_0\,\mu}{2\,\pi\,v_0^2}\,,
\end{align}
which is $k_0$ times the DOS of a single doped Dirac cone. The term proportional to $\kappa_F\cos\gamma$ in the weight integrates to zero over $\gamma$. Hence Eq.~\eqref{eq:dosvnr} is exact for $\mu < v_0\,k_0$ within the linearised model. The Drude weight of Eq.~\eqref{eqdrude} for the band $s = 2$, with the velocity of Eq.~\eqref{eqev}, has the nonzero components
\begin{align}
\label{eq:dvnr}
D_{zz} = \frac{g\,e^2\,k_0\,\mu}{4}\,, \quad
D_{xx} = D_{yy} = \frac{g\,e^2\,k_0\,\mu}{8}\,.
\end{align}
The $\kappa_F\cos\gamma$ term in the weight drops out here as well, since $\int_0^{2\pi} \sin^2\gamma\,\cos\gamma\,d\gamma = \int_0^{2\pi} \cos^3\gamma\,d\gamma = 0$. The ratio $D_{zz}/D_{xx} = 2$ comes from the relative weights of $\langle \sin^2\gamma \rangle$ and $\langle \cos^2\gamma \cos^2\phi \rangle$ over the FS. The intraband polarisabilities follow from Eq.~\eqref{eqintra} as $\Pi_{\rm intra} (q\,\bs{\hat z}, \omega) = g\,k_0\,\mu\,q^2/(4\,\pi\,\omega^2)$ and $\Pi_{\rm intra} (q\,\bs{\hat x}, \omega) = g\,k_0\,\mu\,q^2/(8\,\pi\,\omega^2)$. The plasmon frequencies of Eq.~\eqref{eqplasmon} are
\begin{align}
\label{eq:wpvnr}
\omega_{p,z}^2 = g\,e^2\,k_0\,\mu\,, \quad
\omega_{p,x}^2 = \omega_{p,y}^2 = \frac{g\,e^2\,k_0\,\mu}{2}\,,
\end{align}
with the anisotropy ratio $\omega_{p,z}/\omega_{p,x} = \sqrt{2}$. This is a signature of the toroidal FS. Both frequencies are finite for $q \to 0$, as appropriate for a 3d system, and they scale as $\sqrt{\mu}$ since $D_{ij} \propto \mu$ throughout the low-doping window. Since $\rho_s$ and $D_{ij}$ are fixed by $\varepsilon_s$ and $\bs{V}^{(s)}$ alone, they are the same as in the PTNR, exactly within the linearised model. These are the bare frequencies, which apply as such only for $\lambda \ll 1$, with $\lambda = g\,e^2\,k_0/(4\,\mu)$ defined in Sec.~\ref{secscreenvnr}. For the parameters of Sec.~\ref{secvalidity}, $\lambda \gg 1$, and $\omega_{p,z}$ exceeds both $2\,\mu$ and $v_0\,k_0$. The interband screening of Sec.~\ref{secscreenvnr} must then be included. The steps of this subsection are given in Appendix~\ref{app:vnrdrude}.

%%%%%%%%%%%%%%%%%%%%%%%%%%%
\subsection{Interband screening and the screened plasmons}
\label{secscreenvnr}

For $\mu > 0$, both bands contribute, and the interband part must be kept alongside $\Pi_{\rm intra}$. Through Eq.~\eqref{eq:ringmapvnr}, only the states with $\epsilon_k > \mu$ are unblocked. To leading order in $q$,
\begin{align}
\label{eq:pinterdopedvnr}
\Pi_{\rm inter} (q\,\bs{\hat z}, \omega) = -\, \frac{g\,k_0\,q^2}{16\,\pi\,\omega}
\left[ \ln \left| \frac{\omega + 2\,\mu}{\omega - 2\,\mu} \right| + i\,\pi\,\Theta (\omega - 2\,\mu) \right],
\end{align}
and $\Pi_{\rm inter} (q\,\bs{\hat x}, \omega)$ is half of this. The interband onset moves from $\omega = v_0\,q$ to $\omega_{\rm th} = 2\,\mu$. For $\omega < 2\,\mu$, the polarisability is real and negative, with the static value $-\,g\,k_0\,q^2/(16\,\pi\,\mu)$. Together with the Drude term of Eq.~\eqref{eq:wpvnr}, it gives, for $v_0\,q \ll \omega \ll v_0\,k_0$,
\begin{align}
\label{eq:epsdopedvnr}
\varepsilon^{\rm RPA} (q\,\bs{\hat z}, \omega) = 1 - \frac{g\,e^2\,k_0\,\mu}{\omega^2}
+ \frac{g\,e^2\,k_0}{4\,\omega} \left[ \ln \left| \frac{\omega + 2\,\mu}{\omega - 2\,\mu} \right| + i\,\pi\,\Theta (\omega - 2\,\mu) \right].
\end{align}
The $\bs{\hat x}$ channel follows by halving the prefactor $g\,e^2\,k_0$ in all three terms. Below the onset, the interband part acts as the background dielectric constant $\epsilon_{b,z} (\omega) = 1 + g\,e^2\,k_0\,\operatorname{arctanh} (\omega/2\,\mu)/(2\,\omega)$. Its static values are
\begin{align}
\label{eq:ebvnr}
\epsilon_{b,z} = 1 + \frac{g\,e^2\,k_0}{4\,\mu}\,, \quad
\epsilon_{b,x} = \epsilon_{b,y} = 1 + \frac{g\,e^2\,k_0}{8\,\mu}\,.
\end{align}
We define $\lambda = g\,e^2\,k_0/(4\,\mu)$ and $\tilde{x} = \omega/(2\,\mu)$. The $\bs{\hat z}$ channel then reads $\varepsilon^{\rm RPA} = 1 + \lambda\,[\operatorname{arctanh} \tilde{x}/\tilde{x} - 1/\tilde{x}^2]$. It tends to $-\infty$ for $\tilde{x} \to 0$ and to $+\infty$ for $\tilde{x} \to 1^-$. Hence it has a zero below the onset for every coupling. The same holds along $\bs{\hat x}$ with $\lambda \to \lambda/2$. For $\lambda \ll 1$, the zero lies at the bare frequency of Eq.~\eqref{eq:wpvnr}. In the static approximation for $\epsilon_b$, the screened frequencies are
\begin{align}
\label{eq:wpscreenedvnr}
\tilde\omega_{p,z}^2 = \frac{4\,\mu^2\,\lambda}{1 + \lambda}\,, \quad
\tilde\omega_{p,x}^2 = \frac{2\,\mu^2\,\lambda}{1 + \lambda/2}\,.
\end{align}
They always lie below $\omega_{\rm th}$. The anisotropy ratio becomes $\tilde\omega_{p,z}/\tilde\omega_{p,x} = \sqrt{2\,(1 + \lambda/2)/(1 + \lambda)}$, which is below $\sqrt{2}$ and tends to unity for $\lambda \gg 1$. In that limit, the static-approximation values approach the onset $2\,\mu$ from below. This approximation then ceases to be accurate, since $\epsilon_b (\omega)$ diverges logarithmically at $\omega_{\rm th}$. The exact zero of the dielectric function instead tends, for both channels, to $\omega = 2\,\mu\,x_*$, where $x_*$ is the root of $\operatorname{arctanh} x = 1/x$. This root is unique and satisfies $x_* < 1$, because the left-hand side increases from zero to infinity on $(0, 1)$, whereas the right-hand side decreases from infinity to unity. The zero therefore stays below the onset, and the ratio of the exact $\bs{\hat z}$ and $\bs{\hat x}$ zeros tends to unity. The steps are given in Appendix~\ref{app:vnrscreen}.

%%%%%%%%%%%%%%%%%%%%%%%%%%%
\subsection{Toroidal-angle dependence of the overlap factor for in-plane wavevectors}
\label{secvortexcorr}

The results above follow from the first term of the overlap factor of Eq.~\eqref{eq:formfactor}, which depends only on $\gamma$ and $\gamma'$. The second term arises from the winding of the pseudospin in $\phi$. Explicitly, the overlap factor separates as
\begin{align}
\label{eq:fsplit}
\mathcal{F} = \frac{1 - \cos (\gamma - \gamma')}{2} + \frac{\sin\gamma \sin\gamma'}{2}\, \left[ 1 - \cos (\phi - \phi') \right].
\end{align}
The first term has the same form as the PTNR factor. The second term has no PTNR counterpart. It vanishes for $\bs{q} = q\,\bs{\hat z}$, where $\phi' = \phi$. For $\bs{q} = q\,\bs{\hat x}$, the shift is $\phi' - \phi \simeq -\, q \sin\phi/k_0$, and the second term becomes $\sin^2\gamma\; q^2 \sin^2\phi/(4\,k_0^2)$. Inserting it into Eq.~\eqref{eq:polinter} and performing the angular integrals, we obtain the contribution of the second term
\begin{align}
\label{eq:dpivnr}
\delta\Pi (q\,\bs{\hat x}, \omega) \simeq \frac{g\,v_0\,q^2}{8\,\pi\,k_0}
\int_{\kappa_F}^{\Lambda} \frac{\kappa^2\, d\kappa}{(\omega + i\,0^+)^2 - 4\,v_0^2\,\kappa^2}\,,
\end{align}
where $\Lambda \lesssim k_0$ is the momentum scale at which the linearised Hamiltonian ceases to hold. The imaginary part is controlled by the resonant radius $\kappa_0 = \omega/(2\,v_0)$ and is free of this scale,
\begin{align}
\label{eq:imdpivnr}
\Im \delta\Pi (q\,\bs{\hat x}, \omega) \simeq -\, \frac{g\,q^2\,\omega}{128\,v_0^2\,k_0}\, \Theta (\omega - 2\,\mu)\,.
\end{align}
Compared with the leading term, $-\,g\,k_0\,q^2/(32\,\omega)$ in the same channel, it is smaller by $(\omega/2\,v_0\,k_0)^2$. This is the local suppression $(\kappa/k_0)^2$ of the second term of Eq.~\eqref{eq:qmetric}, evaluated at $\kappa = \kappa_0$.

The real part is different. For $\omega \ll v_0\,\Lambda$, the integral is dominated by large $\kappa$ and gives the frequency-independent shift $\Re \delta\Pi \simeq -\, g\,q^2\,(\Lambda - \kappa_F)/(32\,\pi\,v_0\,k_0)$. The same value follows from Eq.~\eqref{eq:polinterstatic}, where the first term of Eq.~\eqref{eq:qmetric} reproduces the static $\Pi_{\rm inter}$ of Eq.~\eqref{eq:pinterdopedvnr} and the second term gives this shift. It raises the in-plane background dielectric constant by
\begin{align}
\label{eq:debvnr}
\delta\epsilon_{b,x} = \delta\epsilon_{b,y} \simeq \frac{g\,e^2\,(\Lambda - \kappa_F)}{8\,v_0\,k_0}\,,
\end{align}
and leaves $\epsilon_{b,z}$ unchanged. Relative to $\epsilon_{b,x} - 1$ of Eq.~\eqref{eq:ebvnr}, it is of order $\kappa_F\,\Lambda/k_0^2$. The ratio is exactly $(\Lambda - \kappa_F)\,\kappa_F/k_0^2$ with $\kappa_F = \mu/v_0$. It is thus small at low doping but linear, rather than quadratic, in $\kappa_F/k_0$. For $\Lambda = k_0$, it equals $(1 - \kappa_F/k_0)\,\kappa_F/k_0$. Because it is set by $\Lambda$, it is of the same order as the corrections of relative size $\kappa/k_0$ that were dropped from the leading terms. We therefore regard Eqs.~\eqref{eq:wpvnr} and~\eqref{eq:wpscreenedvnr} as the controlled predictions of the model. Relative to the PTNR, the bare plasmon frequencies $\omega_{p,i}$ are the same, the in-plane absorption differs at relative order $(\omega/2\,v_0\,k_0)^2$, and the in-plane screening can only be stronger. The steps of this subsection are given in Appendix~\ref{app:vnrvortex}.

\section{VNR tilted with respect to the \texorpdfstring{$k_x$}{kx}-axis}
\label{secinplane}

We now include the in-plane tilt by setting $\bs{\eta} = \eta_x\,\bs{\hat x}$. The tilt term becomes $v_0\,\eta_x\,k_x = v_0\,\eta_x\,(k_0 + \kappa\cos\gamma)\cos\phi$, and the eigenvalues are $\varepsilon_s (\bs{k}) = (-1)^s\,v_0\,\kappa + v_0\,\eta_x\,(k_0 + \kappa\cos\gamma)\cos\phi$. Setting $\varepsilon_s = \mu$ gives the Fermi momentum
\begin{align}
\label{eq:cyclidevnr}
\kappa_s (\phi, \gamma) = \frac{\mu - \tilde{\eta}\cos\phi}{v_0\,\zeta_s (\phi, \gamma)}\,,
\quad \zeta_s (\phi, \gamma) = (-1)^s + \eta_x \cos\gamma \cos\phi\,,
\quad \tilde{\eta} = v_0\,\eta_x\,k_0\,,
\end{align}
and the delta function becomes $\delta (\varepsilon_s - \mu) = \delta (\kappa - \kappa_s)/(v_0\,|\zeta_s|)$. The eigenvalues, the band velocities, and the integration measure coincide with those of the PTNR at vanishing gap. The FS, the DOS, the Drude weights, and the intraband plasmon frequencies therefore coincide with those of the PTNR~\cite{ips-nr-plasmon} for every $\eta_x$, as stated in Sec.~\ref{secresponse}. We quote them below and concentrate on the interband response, where the pseudospin texture matters. The FS is a ring-cyclide for $|\tilde{\eta}| < \mu$ and a horn-cyclide for $\mu < |\tilde{\eta}|$ (Fig.~\ref{figfsvnr}). We assume the type-I condition $|\eta_x| < 1$ throughout. The classification depends only on $|\tilde{\eta}|$, because $\eta_x \to -\eta_x$ is equivalent to $\phi \to \phi + \pi$. For $\eta_x < 0$, the conditions below are read with $\tilde{\eta} \to |\tilde{\eta}|$. The part of $\Im \Pi (q\,\bs{\hat x}, \omega)$ that is odd in $\eta_x$ changes sign together with $q \to -q$. For numerical estimates, we use the parameters of Sec.~\ref{secvalidity}. We take $\mu = 0.1\,v_0\,k_0$ and $0.2\,v_0\,k_0$, which obey the low-doping condition. The tilt $\tilde{\eta} = 0.03$ eV exceeds both values of $\mu$, which places the FS in the horn-cyclide regime, and the corresponding $\eta_x = \tilde{\eta}/(v_0\,k_0)$ is not small. The expansions of this section need $\omega \ll v_0\,k_0$ and, in the ring-cyclide regime, $|\eta_x| < \mu/(v_0\,k_0) \ll 1$. We therefore use the tilted parameters only for indicative estimates.

For $\bs{q} = q\,\bs{\hat z}$ and $\bs{q} = q\,\bs{\hat y}$, the momentum transfer has no $x$-component. Then $k_x' = k_x$, and the tilt contribution to the interband transition energy cancels exactly. For $\bs{q} = q\,\bs{\hat x}$, the tilt shifts the transition energy by $v_0\,\eta_x\,q$, and the frequency $\omega$ is replaced by $\Omega = \omega - v_0\,\eta_x\,q$ in the resonance condition $v_0\,(\kappa + \kappa') = \Omega$. The overlap factor of Eq.~\eqref{eq:formfactor} does not involve the tilt, because the tilt term is proportional to the identity. The Fermi momentum of Eq.~\eqref{eq:cyclidevnr}, the delta function, and the transition energies are derived in Appendix~\ref{app:vnrxfs}.

%%%%%%%%%%%%%%%%%%%%%%%%%%%
\subsection{Intrinsic point}
\label{secxintr}

At $\mu = 0$, the FS is the nodal ring only for $\tilde{\eta} \to 0$. At finite $\tilde{\eta}$, it is the symmetric horn-cyclide of Sec.~\ref{secxhorn}. Here we consider the interband response at frequencies well above the pocket scale. At fixed $\phi$ and in the local plane $(u, w) = (\kappa\cos\gamma, k_z)$, the two bands are a tilted Dirac cone with the tilt along $u$ and a constant offset $\tilde{\eta}\cos\phi$. The offset induces small local pockets. These do not overlap the region of phase space with $\kappa + \kappa' \sim \omega/v_0$ that the resonance selects, provided $\omega \gg v_0\,q$ and $\omega \gg |\tilde{\eta}|$. For the parameters of Sec.~\ref{secvalidity}, this window is essentially empty, since $|\tilde{\eta}|$ is comparable to $v_0\,k_0$. The statements below are therefore parametric ones for small $\eta_x$. In this window, the local mapping of Eq.~\eqref{eq:ringmapvnr} gives
\begin{align}
\label{eq:imintrxvnr}
\Im \Pi (q\,\bs{\hat z}, \omega) \simeq -\, \frac{g\,k_0\,q^2}{16\,\omega}\,, \quad
\Im \Pi (q\,\bs{\hat x}, \omega) \simeq -\, \frac{g\,k_0\,q^2}{32\,\omega} \left( 1 + \frac{v_0\,\eta_x\,q}{\omega} \right),
\quad
\Im \Pi (q\,\bs{\hat y}, \omega) \simeq -\, \frac{g\,k_0\,q^2}{32\,\omega}\,.
\end{align}
The $\bs{\hat z}$ and $\bs{\hat y}$ channels do not depend on $\eta_x$. The $\bs{\hat x}$ channel acquires a correction that is odd in $\eta_x$ and of first order in $v_0\,\eta_x\,q/\omega$. The frequency shift $v_0\,\eta_x\,q$ is the same for every $\phi$. The local component $q\cos\phi$ alone would give only $v_0\,\eta_x\,q\cos^2\phi$. The remainder $v_0\,\eta_x\,q\sin^2\phi$ comes from the displacement of $\phi$ along the ring by $\phi' - \phi \simeq -\,q\sin\phi/k_0$, which changes the offset $\tilde{\eta}\cos\phi$. This term is of leading order in $\eta_x$, because $\tilde{\eta} = v_0\,\eta_x\,k_0$ compensates the factor $1/k_0$. For $\bs{q} = q\,\bs{\hat z}$ and $\omega \gg v_0\,q$, the imaginary part in Eq.~\eqref{eq:imintrxvnr} is nonzero, and $\varepsilon^{\rm RPA}$ is complex in this window. For $\omega < v_0\,q$, the interband part is real, and the pockets add the correction
\begin{align}
\label{eq:dpipocketmainvnr}
\delta\Pi (q\,\bs{\hat z}, \omega) = -\,\frac{4\,g\,k_0\,q^2\,|\tilde{\eta}|^3}{9\,\pi^2 \left( v_0^2\,q^2 - \omega^2 \right)^2}
\end{align}
to Eq.~\eqref{eq:polmu0vnr}. It is negative and cubic in $|\tilde{\eta}|$, with the relative size $64\,|\tilde{\eta}|^3/[9\,\pi^2\,(v_0^2\,q^2 - \omega^2)^{3/2}]$. The pockets thus add a positive term to Eq.~\eqref{eq:epsintrvnr}. Hence $\varepsilon^{\rm RPA} > 1$ persists and has no zero. The derivation is given in Appendix~\ref{app:vnrxintr}. For $\bs{q} = q\,\bs{\hat x}$, the shifted frequency $\Omega$ enters, and the local threshold $v_0\,q\,|\cos\phi|$ vanishes at $\phi = \pi/2$. The interband continuum then extends down to the lower edge of the window $\omega \gg |\tilde{\eta}|$, and $\varepsilon^{\rm RPA}$ is complex throughout it. For $\omega \lesssim |\tilde{\eta}|$, the pockets must be treated explicitly, as in Sec.~\ref{secxhorn}. In the window above, the interband response thus gives no undamped plasmon in any channel.

The vortex adds to the in-plane channels the term $\delta\Pi$ of Sec.~\ref{secvortexcorr}. Its imaginary part at the intrinsic point is $\Im \delta\Pi (q\,\bs{\hat x}, \omega) \simeq \Im \delta\Pi (q\,\bs{\hat y}, \omega) \simeq -\, g\,q^2\,\omega/(128\,v_0^2\,k_0)$, which is smaller than the leading terms of Eq.~\eqref{eq:imintrxvnr} by $(\omega/2\,v_0\,k_0)^2$. The tilt enters it only through $\Omega$. The resulting change is of relative order $v_0\,\eta_x\,q/\omega$ on a term that is already suppressed, and we drop it.

%%%%%%%%%%%%%%%%%%%%%%%%%%%
\subsection{Ring-cyclide regime}
\label{secxring}

For $0 < |\tilde{\eta}| < \mu \ll v_0\,k_0$, the FS is a single closed surface that encloses the nodal ring. The condition $|\tilde{\eta}| < \mu$ is equivalent to $|\eta_x| < \mu/(v_0\,k_0)$. It is far tighter than the type-I condition and forces the tilt to be small at low doping. The DOS follows from Eq.~\eqref{eq:cyclidevnr} and the exact Jacobian $\kappa\,(k_0 + \kappa\cos\gamma)$. To $\mathcal{O} (\eta_x^2)$, including the cross-term between the offset $\tilde{\eta}\cos\phi$ and the term $\kappa_F^2 \cos\gamma$, which is of the same order in $\eta_x$ since $\tilde{\eta}/k_0 = v_0\,\eta_x$, we obtain
\begin{align}
\label{eq:dosxvnr}
\rho (\mu) = \frac{g\,k_0\,\mu}{2\,\pi\,v_0^2} \left( 1 + \frac{9\,\eta_x^2}{4} \right).
\end{align}
The compressibility sum rule fixes $\Pi (\bs{q} \to 0, 0) = -\,\rho (\mu)$ for every direction of $\bs{q}$. The band velocity of $s = 2$ has the components $V_x^{(2)} = v_0\,(\cos\gamma\cos\phi + \eta_x)$, $V_y^{(2)} = v_0\,\cos\gamma\sin\phi$, and $V_z^{(2)} = v_0\,\sin\gamma$. The reflections $\phi \to -\phi$ and $\gamma \to -\gamma$ flip $V_y^{(2)}$ and $V_z^{(2)}$, respectively, and keep $D_{ij}$ diagonal. To the same order, the Drude weights of Eq.~\eqref{eqdrude} are
\begin{align}
\label{eq:dxvnr}
D_{zz} \simeq \frac{g\,e^2\,k_0\,\mu}{4} \left( 1 + \frac{9\,\eta_x^2}{8} \right), \quad
D_{xx} \simeq \frac{g\,e^2\,k_0\,\mu}{8} \left( 1 + \frac{17\,\eta_x^2}{16} \right), \quad
D_{yy} \simeq \frac{g\,e^2\,k_0\,\mu}{8} \left( 1 + \frac{27\,\eta_x^2}{16} \right).
\end{align}
The correction to $D_{yy}$ comes entirely from the deformation of the FS, since $V_y^{(2)}$ does not involve $\eta_x$. The plasmon frequencies of Eq.~\eqref{eqplasmon} are
\begin{align}
\label{eq:wpxvnr}
\omega_{p,z}^2 = g\,e^2\,k_0\,\mu \left( 1 + \frac{9\,\eta_x^2}{8} \right), \quad
\omega_{p,x}^2 = \frac{g\,e^2\,k_0\,\mu}{2} \left( 1 + \frac{17\,\eta_x^2}{16} \right), \quad
\omega_{p,y}^2 = \frac{g\,e^2\,k_0\,\mu}{2} \left( 1 + \frac{27\,\eta_x^2}{16} \right),
\end{align}
and their ratios are
\begin{align}
\frac{\omega_{p,z}}{\omega_{p,x}} \simeq \sqrt{2} \left( 1 + \frac{\eta_x^2}{32} \right), \quad
\frac{\omega_{p,x}}{\omega_{p,y}} \simeq 1 - \frac{5\,\eta_x^2}{16}\,, \quad
\frac{\omega_{p,z}}{\omega_{p,y}} \simeq \sqrt{2} \left( 1 - \frac{9\,\eta_x^2}{32} \right).
\end{align}
At $\eta_x = 0$, $\omega_{p,x} = \omega_{p,y}$, as required by the combined rotation of Sec.~\ref{secmodel}. The in-plane tilt splits these two frequencies at order $\eta_x^2$ and enhances all three relative to their untilted values. All results of Eqs.~\eqref{eq:dosxvnr}--\eqref{eq:wpxvnr} coincide with those of the PTNR, since they follow from $\varepsilon_s$, $\bs{V}^{(s)}$, and the integration measure, which are the same in the two linearised models. Since $|\eta_x| < \kappa_F/k_0$ in this regime, the $\eta_x^2$ terms are of order $(\kappa_F/k_0)^2$. They are exact within the linearised model, but they are smaller than the generic corrections of relative order $\kappa_F/k_0$ from the terms dropped in $\bs{d}_0$. They describe the tilt dependence within the model rather than a quantitative prediction. The steps behind Eqs.~\eqref{eq:dosxvnr}--\eqref{eq:wpxvnr} and the ratios below them are given in Appendix~\ref{app:vnrxring}.

The interband response follows from the local mapping. At fixed $\phi$, the local cone has the chemical potential $\mu_{\rm eff} (\phi) = \mu - \tilde{\eta}\cos\phi$, and Pauli-blocking removes the states with $\epsilon_k < |\mu_{\rm eff}|$. We drop the tilt term $v_0\,\eta_x\,\cos\phi\,u$ inside the local cone, which modifies the Pauli-blocking boundary at relative order $\eta_x$. For $v_0\,q \ll \omega \ll v_0\,k_0$, Eq.~\eqref{eq:pinterdopedvnr} generalises to
\begin{align}
\label{eq:pinterxvnr}
\Pi_{\rm inter} (q\,\bs{\hat e}_i, \omega) \simeq -\, \frac{g\,k_0\,q^2}{16\,\pi\,\omega} \int_0^{2\pi} \frac{d\phi}{2\,\pi}\, \check w_i (\phi)
\left[ \ln \left| \frac{\omega + 2\,|\mu_{\rm eff}|}{\omega - 2\,|\mu_{\rm eff}|} \right| + i\,\pi\,\Theta \left( \omega - 2\,|\mu_{\rm eff}| \right) \right],
\end{align}
with the weights $\check w_z = 1$, $\check w_x = \cos^2\phi$, and $\check w_y = \sin^2\phi$. The interband onset is $\omega_{\rm th} = 2\,(\mu - |\tilde{\eta}|)/(1 + |\eta_x|)$. It is set by the point $\phi = \gamma = 0$ of the FS for $\eta_x > 0$, and by $\phi = \pi$, $\gamma = 0$ for $\eta_x < 0$. This is where $\kappa_F$ is smallest, and the onset vanishes as $|\tilde{\eta}| \to \mu$. Eq.~\eqref{eq:pinterxvnr} itself has the onset $2\,(\mu - |\tilde{\eta}|)$, which differs from this value by the factor $1 + |\eta_x|$ and is accurate only to leading order in $\eta_x$. For $\omega \ll 2\,(\mu - |\tilde{\eta}|)$, the logarithm reduces to $\omega/|\mu_{\rm eff}|$. With $\tilde w \equiv \sqrt{\mu^2 - \tilde{\eta}^2}$, the static background dielectric constants are
\begin{align}
\label{eq:ebxvnr}
\epsilon_{b,z} = 1 + \frac{g\,e^2\,k_0}{4\,\tilde w}\,, \quad
\epsilon_{b,x} = 1 + \frac{g\,e^2\,k_0\,\mu}{4\,\tilde w\,(\mu + \tilde w)}\,, \quad
\epsilon_{b,y} = 1 + \frac{g\,e^2\,k_0}{4\,(\mu + \tilde w)}\,.
\end{align}
They reduce to Eq.~\eqref{eq:ebvnr} for $\tilde{\eta} \to 0$. The real part of the dielectric function is $\Re \varepsilon^{\rm RPA} \simeq \epsilon_{b,i} - \omega_{p,i}^2/\omega^2$ for $\omega_{p,i}$ well below the onset, which defines the screened frequencies $\tilde\omega_{p,i}^2 = \omega_{p,i}^2/\epsilon_{b,i}$. Below the onset, $\epsilon_{b,i} (\omega)$ increases with $\omega$ and the Drude term $-\,\omega_{p,i}^2/\omega^2$ increases as well. The real part is therefore monotonic, its zero is unique, and the static screened frequency is an upper bound of this zero. Unlike in the untilted case, $\Re \varepsilon^{\rm RPA}$ stays finite at the onset, because the logarithmic singularity of Eq.~\eqref{eq:pinterxvnr} is integrable over the ring. The zero exists below the onset if the static screened frequency lies below it. Since $\epsilon_{b,i}$ grows as $|\tilde{\eta}| \to \mu$, the screening becomes stronger as the ring-cyclide approaches the horn-cyclide. The steps behind Eqs.~\eqref{eq:pinterxvnr} and~\eqref{eq:ebxvnr}, and the statements on the screened frequencies, are given in Appendix~\ref{app:vnrxringscreen}.

%%%%%%%%%%%%%%%%%%%%%%%%%%%
\subsection{Horn-cyclide regime}
\label{secxhorn}

At $\mu = 0$, the FS is the symmetric horn-cyclide. It consists of an electron-pocket and a hole-pocket of equal size, which touch at $\phi = \pm\pi/2$. It carries a finite DOS and finite Drude weights, in contrast to the untilted VNR at $\mu = 0$. To leading order in $\eta_x$,
\begin{align}
\label{eq:horn0vnr}
\rho = \frac{g\,k_0\,\tilde{\eta}}{\pi^2\,v_0^2}\,, \quad
D_{zz} = \frac{g\,e^2\,k_0\,\tilde{\eta}}{2\,\pi}\,, \quad
D_{xx} = \frac{g\,e^2\,k_0\,\tilde{\eta}}{3\,\pi}\,, \quad
D_{yy} = \frac{g\,e^2\,k_0\,\tilde{\eta}}{6\,\pi}\,.
\end{align}
The $\eta_x^2$ corrections multiply these by $1 + 2\,\eta_x^2$ for $\rho$, $1 + \eta_x^2$ for $D_{zz}$, $1 + 3\,\eta_x^2/5$ for $D_{xx}$, and $1 + 9\,\eta_x^2/5$ for $D_{yy}$. The plasmon frequencies follow from Eq.~\eqref{eqplasmon} as $\omega_{p,z}^2 = 2\,g\,e^2\,k_0\,\tilde{\eta}/\pi$, $\omega_{p,x}^2 = 4\,g\,e^2\,k_0\,\tilde{\eta}/(3\,\pi)$, and $\omega_{p,y}^2 = 2\,g\,e^2\,k_0\,\tilde{\eta}/(3\,\pi)$.

For $0 < \mu < |\tilde{\eta}|$, the FS splits into an electron-pocket of the band $s = 2$ and a hole-pocket of the band $s = 1$. The local doping $\mu_{\rm eff} (\phi)$ is positive on the electron arc, $\phi \in (\phi_b, 2\,\pi - \phi_b)$, and negative on the hole arc, $|\phi| < \phi_b$, with $\cos\phi_b = \mu/\tilde{\eta}$. The Drude weight is quadratic in the velocity, and the hole arc therefore adds to the electron arc. Dropping the $\eta_x^2$ terms for the moment, $\kappa_F = |\mu_{\rm eff}|/v_0$, and with $\bar{w} = \sqrt{\tilde{\eta}^2 - \mu^2}$ we obtain
\begin{align}
\label{eq:horncycvnr}
\rho (\mu) &= \frac{g\,k_0}{\pi^2\,v_0^2} \left[ \mu \arcsin \!\left( \frac{\mu}{\tilde{\eta}}\right) + \bar{w} \right], \quad
D_{zz} = \frac{g\,e^2\,k_0}{2\,\pi} 
\left [ \mu \arcsin \!\left( \frac{\mu}{\tilde{\eta}}\right) + \bar{w} \right], \nn
D_{xx} &= \frac{g\,e^2\,k_0}{4\,\pi} 
\left[ \mu \arcsin \!\left( \frac{\mu}{\tilde{\eta}}\right) + \frac{\bar{w}\,(4\,\tilde{\eta}^2 - \mu^2)}{3\,\tilde{\eta}^2} \right], \quad
D_{yy} = \frac{g\,e^2\,k_0}{4\,\pi} 
\left[ \mu \arcsin \!\left( \frac{\mu}{\tilde{\eta}}\right) + \frac{\bar{w}\,(2\,\tilde{\eta}^2 + \mu^2)}{3\,\tilde{\eta}^2} \right].
\end{align}
The anisotropy ratio $D_{zz}/D_{xx}$ increases from $3/2$ at $\mu = 0$ to $2$ at $\mu = |\tilde{\eta}|$, and $D_{yy}/D_{xx}$ increases from $1/2$ to $1$. At $\mu = |\tilde{\eta}|$, the hole-pocket shrinks to a point, and Eq.~\eqref{eq:horncycvnr} joins continuously onto the ring-cyclide values. With $ \varsigma  = \bar{w}/\tilde{\eta}$ and $\Xi = [\bar{w} + \mu \arcsin (\mu/\tilde{\eta})]/\tilde{\eta}$, the $\eta_x^2$ corrections are
\begin{align}
\label{eq:horncorrvnr}
&\rho (\mu) = \frac{g\,k_0\,\tilde{\eta}}{\pi^2\,v_0^2} \left[ \left( 1 + \frac{9\,\eta_x^2}{4} \right) \Xi - \frac{\eta_x^2\, \varsigma ^3}{4} \right], \quad
D_{zz} = \frac{g\,e^2\,k_0\,\tilde{\eta}}{2\,\pi} \left[ \left( 1 + \frac{9\,\eta_x^2}{8} \right) \Xi - \frac{\eta_x^2\, \varsigma ^3}{8} \right], \nn
&D_{xx} = \frac{g\,e^2\,k_0\,\tilde{\eta}}{4\,\pi} \left[ \Xi + \frac{ \varsigma ^3}{3} + \eta_x^2 \left( \frac{17\,\Xi}{16} - \frac{3\, \varsigma ^3}{16} - \frac{3\, \varsigma ^5}{40} \right) \right], \quad
D_{yy} = \frac{g\,e^2\,k_0\,\tilde{\eta}}{4\,\pi} \left[ \Xi - \frac{ \varsigma ^3}{3} + \eta_x^2 \left( \frac{27\,\Xi}{16} - \frac{9\, \varsigma ^3}{16} + \frac{3\, \varsigma ^5}{40} \right) \right].
\end{align}
At $\eta_x = 0$, these reduce to Eq.~\eqref{eq:horncycvnr}. At $\mu = |\tilde{\eta}|$, where $ \varsigma  = 0$ and $\Xi = \pi/2$, they reproduce the ring-cyclide corrections of Eqs.~\eqref{eq:dosxvnr} and~\eqref{eq:dxvnr}. The relations $\Pi_{\rm intra} (q\,\bs{\hat e}_i, \omega) \simeq q^2\,D_{ii}/(\pi\,e^2\,\omega^2)$ and $\omega_{p,i}^2 = 4\,D_{ii}$ remain valid. The steps behind Eqs.~\eqref{eq:horn0vnr}, \eqref{eq:horncycvnr}, and~\eqref{eq:horncorrvnr} are given in Appendix~\ref{app:vnrxhorn}.

The interband response in this regime follows from the same local mapping. For $v_0\,q \ll \omega$, Pauli-blocking removes the transitions with $\omega < 2\,|\mu_{\rm eff} (\phi)|$. Let
\begin{align}
\label{eq:fnvnr}
F_n (\omega) = \int_0^{2\pi} \frac{d\phi}{2\,\pi}\, \cos^n\phi\; \Theta \left( \omega - 2\,|\mu - \tilde{\eta}\cos\phi| \right).
\end{align}
Then
\begin{align}
\label{eq:imhornvnr}
& \Im \Pi (q\,\bs{\hat z}, \omega) \simeq -\, \frac{g\,k_0\,q^2}{16\,\omega}\, F_0\,, \quad
\Im \Pi (q\,\bs{\hat x}, \omega) \simeq -\, \frac{g\,k_0\,q^2}{16\,\omega}\, F_2\,, \quad
\Im \Pi (q\,\bs{\hat y}, \omega) \simeq -\, \frac{g\,k_0\,q^2}{16\,\omega}\, (F_0 - F_2)\,,
%%%%%%%%%%%%%%%%%%%%%%%%%%%
\nn & F_0 = \frac{\arcsin b_\omega - \arcsin a_\omega}{\pi}\,, \quad
F_2 = \frac{\arcsin b_\omega - \arcsin a_\omega 
+ a_\omega \, \sqrt{1 - a_\omega^2} - b_\omega \, \sqrt{1 - b_\omega^2}}{2\,\pi}\,,
\end{align}
where $a_\omega = \max [-1, (\mu - \omega/2)/\tilde{\eta}]$ and $b_\omega = \min [1, (\mu + \omega/2)/\tilde{\eta}]$. For $\omega > 2\,(|\tilde{\eta}| + \mu)$, $F_0 = 1$ and $F_2 = 1/2$, which recovers Eq.~\eqref{eq:imintrxvnr}. The horn-cyclide has no interband onset, $\omega_{\rm th} = 0$, because the local Dirac point crosses the chemical potential at $\phi_b$. At small frequencies, $F_0 \simeq \omega/(\pi\,\bar{w})$, and $\Im \Pi (q\,\bs{\hat z}, \omega)$ tends to the constant $-\,g\,k_0\,q^2/(16\,\pi\,\bar{w})$. The plasmons in this regime are therefore always damped by interband processes.

\begin{figure*}[t!]
\centering
\includegraphics[width=1.0\textwidth]{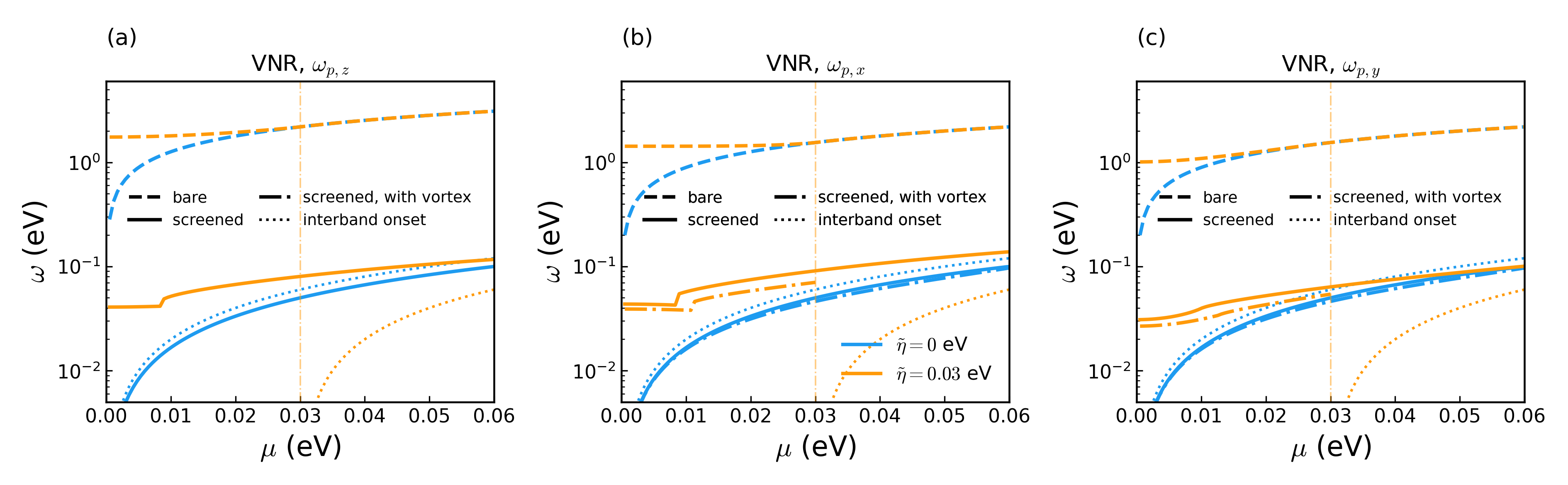}
\caption{\label{fig:plasmaxvnr} Bare and screened plasma frequencies of the VNR under an in-plane tilt as functions of $\mu$. Panels (a), (b), and (c) show $\omega_{p,z}$, $\omega_{p,x}$, and $\omega_{p,y}$, respectively. The dashed and solid lines are the bare frequencies and the screened frequencies from the zero of $\Re \varepsilon^{\rm RPA}$ of the terms that do not involve $\Lambda$. The dotted lines mark the interband onset, which is absent in the horn-cyclide regime. The blue and orange colours correspond to $\tilde{\eta} = 0$ and $0.03$ eV, and the thin vertical line marks $\mu = \tilde{\eta}$. The dash-dotted lines in (b) and (c) are the zeros of $\Re \varepsilon^{\rm RPA}$ including the second term of the overlap factor with the cut-off $\Lambda = k_0$. They are shown only where $\kappa_F < \Lambda$. The parameters are those of Sec.~\ref{secvalidity}.}
\end{figure*}

\begin{figure*}[t!]
\centering
\includegraphics[width=1.0\textwidth]{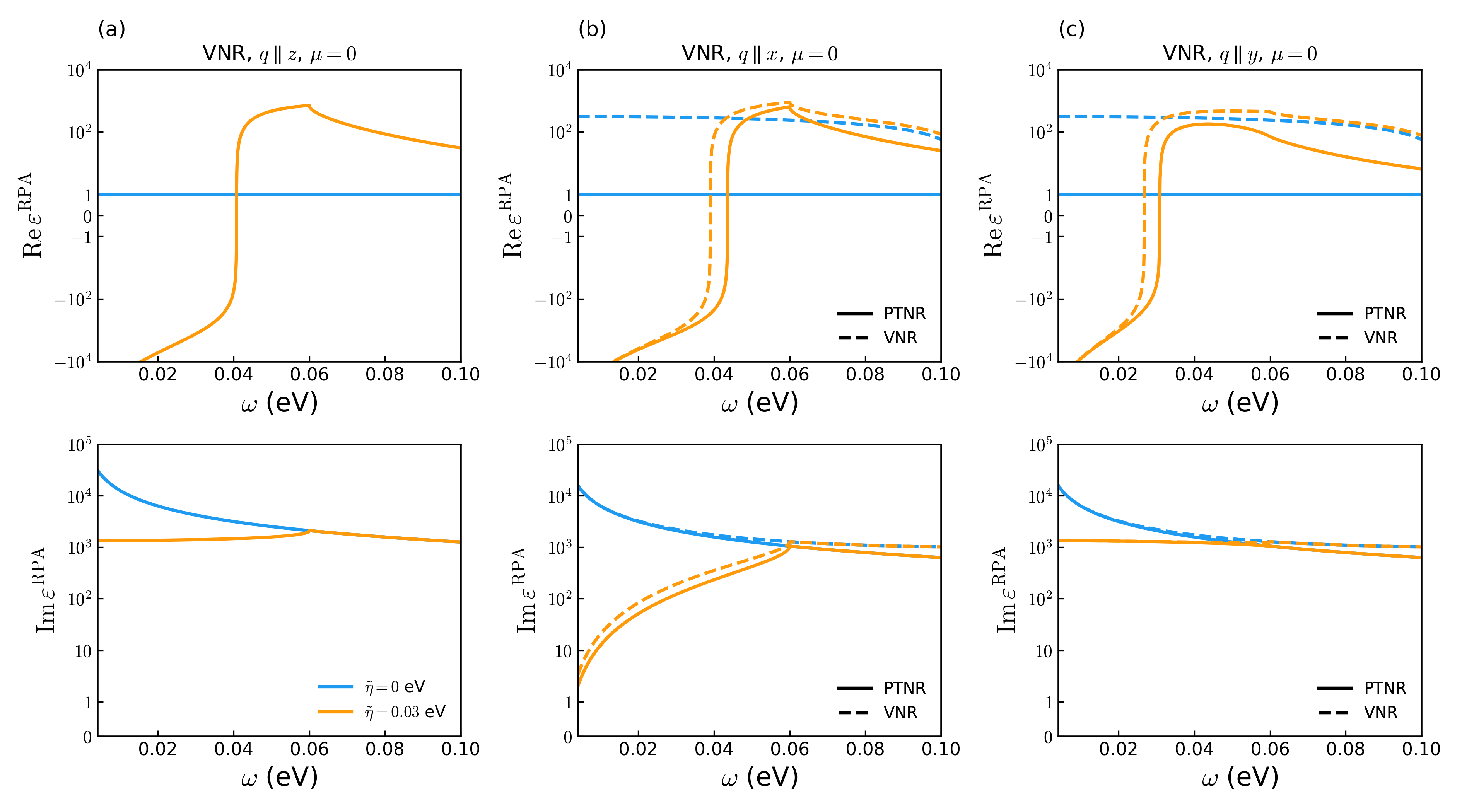}
\caption{\label{fig:epsxvnr} RPA dielectric function of the VNR under an in-plane tilt at $\mu = 0$. Panels (a), (b), and (c) show the $\bs{\hat z}$, $\bs{\hat x}$, and $\bs{\hat y}$ channels. The upper and lower rows show $\Re \varepsilon^{\rm RPA}$ and $\Im \varepsilon^{\rm RPA}$ on symmetric logarithmic scales. The blue and orange colours correspond to $\tilde{\eta} = 0$ and $0.03$ eV. The solid lines are the terms that do not involve $\Lambda$, and the $\bs{\hat z}$ channel is the full VNR result. The dashed lines in (b) and (c) include the second term of the overlap factor with the cut-off $\Lambda = k_0$, and they stop where $\omega$ approaches $2\,v_0\,\Lambda$. The parameters are those of Sec.~\ref{secvalidity}.}
\end{figure*}

For $v_0\,q \ll \omega$, Eqs.~\eqref{eqintra}, \eqref{eq:epsrpa}, and \eqref{eq:imhornvnr} give
\begin{align}
\label{eq:epshornvnr}
& \varepsilon^{\rm RPA} (q\,\bs{\hat z}, \omega) \simeq 1 - \frac{4\,D_{zz}}{\omega^2} + i\,\frac{\pi\,g\,e^2\,k_0}{4\,\omega}\, F_0\,, \quad 
%%%%%%%%%%
\varepsilon^{\rm RPA} (q\,\bs{\hat x}, \omega) 
\simeq 1 - \frac{4\,D_{xx}}{\omega^2} + i\,\frac{\pi\,g\,e^2\,k_0}{4\,\omega}\, F_2\,,
\nn & \varepsilon^{\rm RPA} (q\,\bs{\hat y}, \omega) \simeq 1 - \frac{4\,D_{yy}}{\omega^2} + i\,\frac{\pi\,g\,e^2\,k_0}{4\,\omega}\, \left( F_0 - F_2 \right).
\end{align}
Equation~\eqref{eq:epshornvnr} contains the Drude term, which is the real part of the intraband response, and the imaginary part of the interband response. It does not contain the real part of the interband response, and we describe this part now. It adds the positive term $\epsilon_{b,i} (\omega) - 1$ to $\Re \varepsilon^{\rm RPA}$. The unity in Eq.~\eqref{eq:epshornvnr} is thereby replaced by the background dielectric function $\epsilon_{b,i} (\omega)$. This background screens the Drude term and lowers the plasmon frequency below its bare value $\omega_{p,i}$. At $\mu = 0$, the static interband part is finite along $\bs{\hat x}$, because the weight $\cos^2\phi$ vanishes at the touching points. It gives $\epsilon_{b,x} = 1 + g\,e^2\,k_0/(2\,\pi\,\tilde{\eta})$. Along $\bs{\hat z}$ and $\bs{\hat y}$, the weights $1$ and $\sin^2\phi$ do not vanish at the touching points, and the real part is logarithmically enhanced. For $\omega \ll |\tilde{\eta}|$, the interband contribution to the dielectric function in the $\bs{\hat z}$ channel is
\begin{align}
\varepsilon^{\rm RPA}_{\rm inter} (q\,\bs{\hat z}, \omega) \simeq \frac{g\,e^2\,k_0}{2\,\pi\,\bar{w}} \left[ \ln \frac{\tilde{\eta}}{\omega} + c + i\,\frac{\pi}{2} \right],
\end{align}
with a real constant $c$ of order unity that depends only on $\mu/|\tilde{\eta}|$. To leading logarithmic accuracy, the $\bs{\hat x}$ and $\bs{\hat y}$ channels carry the extra factors $\mu^2/\tilde{\eta}^2$ and $\bar{w}^2/\tilde{\eta}^2$, which are the weights at $\phi_b$. The corrections are of relative order $1/\ln (\tilde{\eta}/\omega)$. The zero of $\Re \varepsilon^{\rm RPA}$ is given by the self-consistent solution of $\tilde\omega_{p,i}^2 = \omega_{p,i}^2/\epsilon_{b,i} (\tilde\omega_{p,i})$. Since $F_0$, $F_2$, and $F_0 - F_2$ are positive for all $\omega > 0$, the imaginary part never vanishes, and the mode is always damped. Neglecting the logarithmic derivative of $\epsilon_{b,i}$, the damping rate is $\Gamma_i = \left( \pi\,g\,e^2\,k_0/(8\,\epsilon_{b,i}) \right) F_n (\tilde\omega_{p,i})$, with $\epsilon_{b,i}$ taken at $\tilde\omega_{p,i}$. Here $F_n$ stands for $F_0$ for $i = z$, for $F_2$ for $i = x$, and for $F_0 - F_2$ for $i = y$. At $\mu = 0$ and for $g\,e^2\,k_0 \ll |\tilde{\eta}|$, the $\bs{\hat z}$ mode lies well below $2\,|\tilde{\eta}|$, where $\arcsin b_\omega \simeq \omega/(2\,|\tilde{\eta}|)$. With $\epsilon_{b,z} \simeq 1$ up to logarithmic factors, its damping rate reduces to $\Gamma_z = g\,e^2\,k_0\,\omega_{p,z}/(8\,|\tilde{\eta}|)$. The ratio $\Gamma_z/\omega_{p,z} = g\,e^2\,k_0/(8\,|\tilde{\eta}|)$ is then small. The representative parameters above lie in the opposite limit, $g\,e^2\,k_0 \gg |\tilde{\eta}|$. The real part of the interband response then dominates, and the plasmons are strongly screened. The steps behind Eqs.~\eqref{eq:fnvnr} to~\eqref{eq:epshornvnr}, and behind the statements on the background dielectric function and the damping, are given in Appendix~\ref{app:vnrxhorninter}. Figs.~\ref{fig:drudexvnr}, \ref{fig:plasmaxvnr}, and~\ref{fig:epsxvnr} show the Drude weights, the plasma frequencies, and the dielectric function at $\mu = 0$, respectively.

%%%%%%%%%%%%%%%%%%%%%%%%%%%
\begin{figure*}[t!]
\centering
\includegraphics[width=0.7 \textwidth]{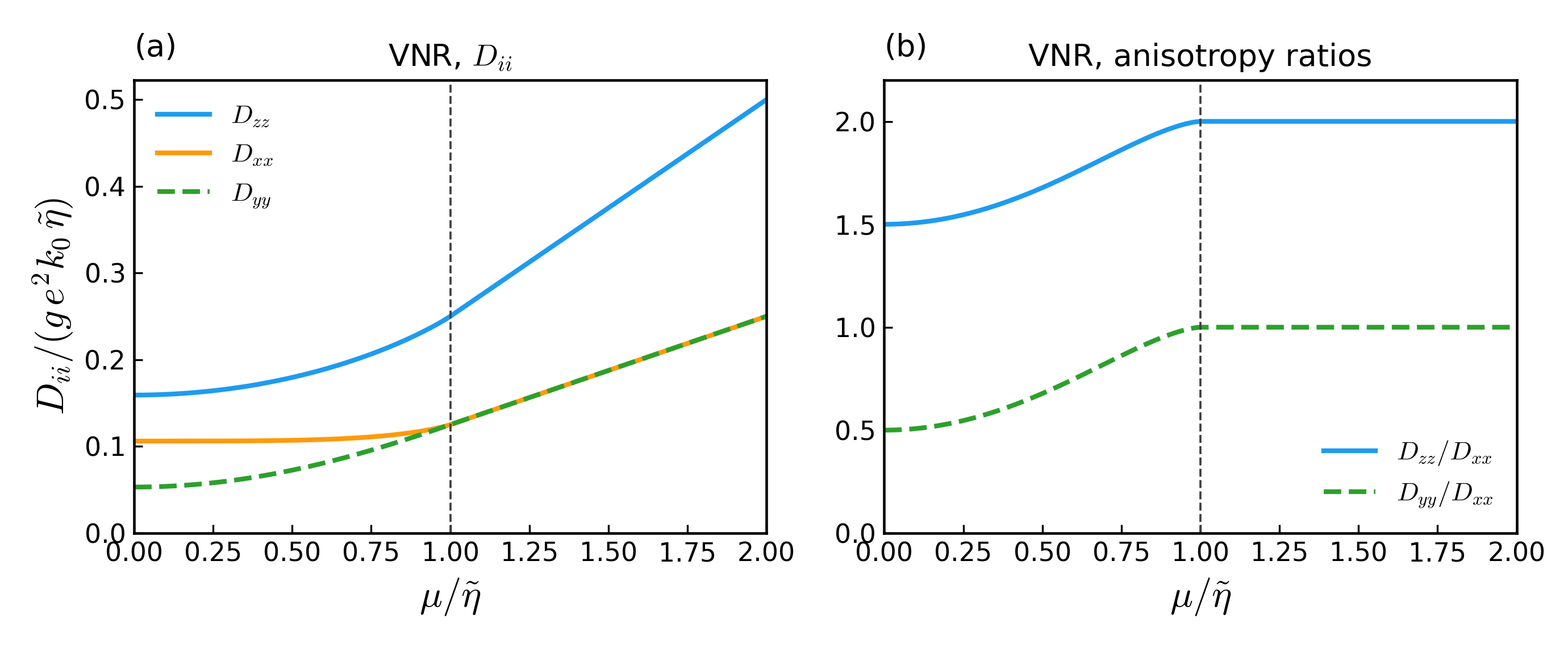}
\caption{\label{fig:drudexvnr} Drude weights of the VNR under an in-plane tilt, at leading order in $\eta_x$ and at fixed $\tilde{\eta}$. Panel (a) shows $D_{zz}$, $D_{xx}$, and $D_{yy}$ in units of $g\,e^2\,k_0\,\tilde{\eta}$ as functions of $\mu/\tilde{\eta}$. Panel (b) shows the ratios $D_{zz}/D_{xx}$ and $D_{yy}/D_{xx}$. The dashed vertical lines mark $\mu = \tilde{\eta}$, which separates the horn-cyclide and ring-cyclide regimes. All curves coincide with those of the PTNR.}
\end{figure*}
%%%%%%%%%%%%%%%%%%%%%%%%%%%

%%%%%%%%%%%%%%%%%%%%%%%%%%%
\subsection{Toroidal-angle dependence of the overlap factor at finite tilt}
\label{secxvortex}

The dependence of the overlap factor on $\phi$ enters only through the second term of Eq.~\eqref{eq:fsplit}. For $\bs{q} = q\,\bs{\hat z}$, the toroidal angle is unchanged, and this term vanishes. The $\bs{\hat z}$ channel of the VNR is therefore identical to that of the PTNR in every regime of this section. For $\bs{q} = q\,\bs{\hat x}$ and $\bs{q} = q\,\bs{\hat y}$, the shifts are $\phi' - \phi \simeq -\,q\sin\phi/k_0$ and $\phi' - \phi \simeq q\cos\phi/k_0$, respectively. The second term of Eq.~\eqref{eq:fsplit} then carries the weights $\sin^2\phi$ and $\cos^2\phi$. They are interchanged relative to the weights $\check w_x$ and $\check w_y$ of Eq.~\eqref{eq:pinterxvnr}. The tilt term is proportional to the identity. It therefore leaves the overlap factor unchanged and enters only through the occupations and through the shift $v_0\,\eta_x\,q$ of the transition energies in the $\bs{\hat x}$ channel.

The imaginary part is controlled by the resonant radius $\kappa_0 = \omega/(2\,v_0)$. Pauli-blocking acts locally at each $\phi$, and Eq.~\eqref{eq:imdpivnr} generalises to
\begin{align}
\label{eq:imdpixvnr}
\Im \delta\Pi (q\,\bs{\hat x}, \omega) \simeq -\, \frac{g\,q^2\,\omega}{64\,v_0^2\,k_0}\, (F_0 - F_2)\,, \quad
\Im \delta\Pi (q\,\bs{\hat y}, \omega) \simeq -\, \frac{g\,q^2\,\omega}{64\,v_0^2\,k_0}\, F_2\,,
\end{align}
with $F_n$ of Eq.~\eqref{eq:fnvnr}. In both regimes, the terms are suppressed by $(\omega/2\,v_0\,k_0)^2$ relative to the leading terms of Eqs.~\eqref{eq:imintrxvnr} and~\eqref{eq:imhornvnr}. For $\omega > 2\,(|\tilde{\eta}| + \mu)$, $F_0 - F_2 = F_2 = 1/2$, and both channels reduce to the untilted value of Eq.~\eqref{eq:imdpivnr} at $\mu = 0$. In the ring-cyclide regime, the terms vanish below $\omega_{\rm th}$. In the horn-cyclide regime, they are nonzero at all frequencies, with the same lack of an onset as the leading terms.

%%%%%%%%%%%%%%%%%%%%%%%%%%%
%%%%%%%%%%%%%%%%%%%%%%%%%%%

The real part is a frequency-independent shift that is dominated by $\kappa \sim \Lambda$. The tilt does not enter the energy denominators of the static limit, as in Eq.~\eqref{eq:polinterstatic}, and it modifies the occupations only for $\kappa < \kappa_F (\phi, \gamma)$. The tilt dependence therefore enters at relative order $\kappa_F/\Lambda$ only, and the in-plane background dielectric constants receive
\begin{align}
\label{eq:debxvnr}
\delta\epsilon_{b,x} \simeq \delta\epsilon_{b,y} \simeq \frac{g\,e^2\,\Lambda}{8\,v_0\,k_0}\,, \quad \delta\epsilon_{b,z} = 0\,,
\end{align}
to leading order in $\kappa_F/\Lambda$. This is the value of Eq.~\eqref{eq:debvnr}, without the tilt-induced splitting between $x$ and $y$, which appears first at relative order $\kappa_F/\Lambda$. In the ring-cyclide regime, the ratio of $\delta\epsilon_{b,x}$ to $\epsilon_{b,x} - 1$ of Eq.~\eqref{eq:ebxvnr} is $\Lambda\,\tilde w\,(\mu + \tilde w)/(2\,v_0\,k_0^2\,\mu)$, with $\tilde w = \sqrt{\mu^2 - \tilde{\eta}^2}$. Since $\tilde w \leq \mu$, this ratio is at most $\kappa_F\,\Lambda/k_0^2$ with $\kappa_F = \mu/v_0$, and it tends to this value for $|\tilde{\eta}| \ll \mu$. The shift remains of the same order as the corrections of relative size $\kappa/k_0$ that were dropped from the leading terms. We therefore regard the results of this section that do not depend on $\Lambda$ as the controlled predictions of the model. The in-plane tilt leaves the conclusion of Sec.~\ref{secvortexcorr} intact. Relative to the PTNR, the bare plasmon frequencies $\omega_{p,i}$ are the same, the in-plane absorption differs at relative order $(\omega/2\,v_0\,k_0)^2$, and the in-plane screening can only be stronger. The steps behind Eqs.~\eqref{eq:imdpixvnr} and~\eqref{eq:debxvnr} are given in Appendix~\ref{app:vnrxvortex}.

%%%%%%%%%%%%%%%
\section{VNR tilted with respect to the \texorpdfstring{$k_z$}{kz}-axis}
\label{secaxial}

We now set $\bs{\eta} = \eta_z\,\bs{\hat z}$. The tilt term becomes $v_0\,\eta_z\,k_z = v_0\,\eta_z\,\kappa\sin\gamma$, which vanishes on the nodal ring. The eigenvalues are $\varepsilon_s (\bs{k}) = v_0\,\kappa \left[ (-1)^s + \eta_z \sin\gamma \right]$, and the band velocities are $\bs{V}^{(s)} = \bs{v}^{(0,s)} + v_0\,\eta_z\,\bs{\hat z}$. For $|\eta_z| < 1$, we have $\varepsilon_1 < 0 < \varepsilon_2$ for all $\gamma$, and no pocket of the opposite carrier type forms, as described in Sec.~\ref{secfermisurface}. The eigenvalues, the band velocities, and the integration measure again coincide with those of the PTNR at vanishing gap. The FS, the DOS, the Drude weights, and the intraband plasmon frequencies therefore coincide with those of the PTNR~\cite{ips-nr-plasmon} for every $\eta_z$, as stated in Sec.~\ref{secresponse}. We quote them below and concentrate on the interband response, where the pseudospin texture matters.

%%%%%%%%%%%%%%%%%%%%%%%%%%%
%%%%%%%%%%%%%%%%%%%%%%%%%%%

The axial tilt preserves the combined rotation about the $k_z$-axis of Sec.~\ref{secmodel}, and it breaks the reflection $k_z \to -k_z$. The directions $\bs{\hat x}$ and $\bs{\hat y}$ therefore remain equivalent, and this holds for the second term of the overlap factor as well. This is in contrast with the in-plane tilt of Sec.~\ref{secinplane}, where the two directions differ. The tilt enters the interband transition energy only through the shift $v_0\,\bs{\eta} \cdot \bs{q} = v_0\,\eta_z\,q_z$. For $\bs{q} = q\,\bs{\hat x}$ and $\bs{q} = q\,\bs{\hat y}$ this shift vanishes. For $\bs{q} = q\,\bs{\hat z}$ it replaces $\omega$ by $\Omega = \omega - v_0\,\eta_z\,q$ in the resonance condition $v_0\,(\kappa + \kappa') = \Omega$. For $\bs{q} = q\,\bs{\hat z}$, the toroidal angle does not shift, and the second term of Eq.~\eqref{eq:fsplit} vanishes. The $\bs{\hat z}$ channel of the VNR is therefore identical to that of the PTNR in every regime of this section, and the second term of the overlap factor matters only in the in-plane channels, as treated in Sec.~\ref{secazvortex}. The overlap factor of Eq.~\eqref{eq:formfactor} does not involve the tilt, because the tilt term is proportional to the identity. The eigenvalues, the band velocities, the measure, and the shift of the transition energy are derived in Appendix~\ref{app:ztilt}.

For numerical estimates, we use the parameters of Sec.~\ref{secvalidity}, with $\mu = 0.1\,v_0\,k_0$ and $0.2\,v_0\,k_0$, and the tilt $\eta_z = 0.75$ used for the PTNR in Ref.~\cite{ips-nr-plasmon}. The FS now extends up to $\kappa_F^{\max} = \mu / \left[ v_0\,(1 - |\eta_z|) \right]$, which is reached at $\sin\gamma = -\,\mathrm{sgn}\, \eta_z$. The expansion in $\kappa/k_0$ requires $\kappa_F^{\max} \ll k_0$, that is $\mu \ll v_0\,k_0\,(1 - |\eta_z|)$. This is stricter than the low-doping condition, and it becomes more restrictive as $|\eta_z| \to 1$. For the chosen values of $\mu$, $\kappa_F/k_0$ is small without tilt, whereas $\kappa_F^{\max}/k_0$ is larger by the factor $1/(1 - |\eta_z|)$ with tilt. The tilted case at large $|\eta_z|$, particularly at the larger $\mu$, therefore lies at the edge of the validity of the expansions. We use it only for indicative estimates, as in Sec.~\ref{secinplane}.

%%%%%%%%%%%%%%%%%%%%%%%%%%%
\subsection{Intrinsic point}
\label{secazintr}

At $\mu = 0$, the FS remains the nodal ring for every $\eta_z$, because the offset $v_0\,\eta_z\,k_z$ vanishes at $k_z = 0$. At fixed $\phi$ and in the local plane $(u, w) = (\kappa\cos\gamma, k_z)$, the tilt is along $w$ and produces no energy offset. Unlike the in-plane tilt, the axial tilt therefore creates no pockets and gives no Drude weight. For $\bs{q} = q\,\bs{\hat x}$ and $\bs{q} = q\,\bs{\hat y}$, the transition energy is unshifted, and the result is that of the untilted ring. For $\bs{q} = q\,\bs{\hat z}$, the phase space is that of the untilted ring at the shifted frequency $\Omega$. For $\omega \gg v_0\,q$, we obtain
\begin{align}
\label{eq:imintrzvnr}
\Im \Pi (q\,\bs{\hat z}, \omega) = -\, \frac{g\,k_0\,q^2}{16\,\sqrt{\left( \omega - v_0\,\eta_z\,q \right)^2 - v_0^2\,q^2}}
\simeq -\, \frac{g\,k_0\,q^2}{16\,\omega} \left( 1 + \frac{v_0\,\eta_z\,q}{\omega} \right), 
%%%%%%%%%%%%%%%%%%
\quad \Im \Pi (q\,\bs{\hat x}, \omega) 
= \Im \Pi (q\,\bs{\hat y}, \omega) \simeq -\, \frac{g\,k_0\,q^2}{32\,\omega}\,.
\end{align}
%%%%%%%%%%%%%%%%%%%%%
The $\bs{\hat x}$ and $\bs{\hat y}$ channels do not depend on $\eta_z$. The $\bs{\hat z}$ channel acquires a correction that is odd in $\eta_z$ and of first order in $v_0\,\eta_z\,q/\omega$. This correction is small for $\omega \gg v_0\,q$. The axial tilt also moves the onset of Landau damping along $\bs{\hat z}$ from $\omega = v_0\,q$ to $\omega = v_0\,q\,(1 + \eta_z)$. Below this onset, the same frequency shift applies to Eq.~\eqref{eq:polmu0vnr}, and Eq.~\eqref{eq:epsintrvnr} generalises to
\begin{align}
\label{eq:epsintrzvnr}
\varepsilon^{\rm RPA} (q\,\bs{\hat z}, \omega) = 1 + \frac{\pi\,g\,e^2\,k_0}{4\,\sqrt{v_0^2\,q^2 - \left( \omega - v_0\,\eta_z\,q \right)^2}}\,,
\quad 0 \leq \omega < v_0\,q\,(1 + \eta_z)\,.
\end{align}
For $|\eta_z| < 1$, the argument of the square root is positive throughout this window, and the second term is positive as well. Hence $\varepsilon^{\rm RPA} > 1$ has no zero, and the axially-tilted VNR at $\mu = 0$ supports no undamped plasmon, as in the untilted case. For $\bs{q} = q\,\bs{\hat x}$, the local threshold $v_0\,q\,|\cos\phi|$ vanishes at $\phi = \pi/2$. The continuum then reaches down to $\omega = 0$, and $\varepsilon^{\rm RPA}$ is complex at all frequencies.

The vortex adds to the in-plane channels the term $\delta\Pi$ of Sec.~\ref{secazvortex}. At the intrinsic point, its imaginary part is the untilted value of Eq.~\eqref{eq:imdpivnr} without the step function, $\Im \delta\Pi (q\,\bs{\hat x}, \omega) = \Im \delta\Pi (q\,\bs{\hat y}, \omega) \simeq -\, g\,q^2\,\omega/(128\,v_0^2\,k_0)$. It is independent of $\eta_z$, because the tilt does not enter the energy denominators of the in-plane channels. The steps behind Eqs.~\eqref{eq:imintrzvnr} and~\eqref{eq:epsintrzvnr}, and behind the vortex term, are given in Appendix~\ref{app:vnrzintr}.

%%%%%%%%%%%%%%%%%%%%%%%%%%%
\begin{figure*}[t!]
\centering
\includegraphics[width=0.7 \textwidth]{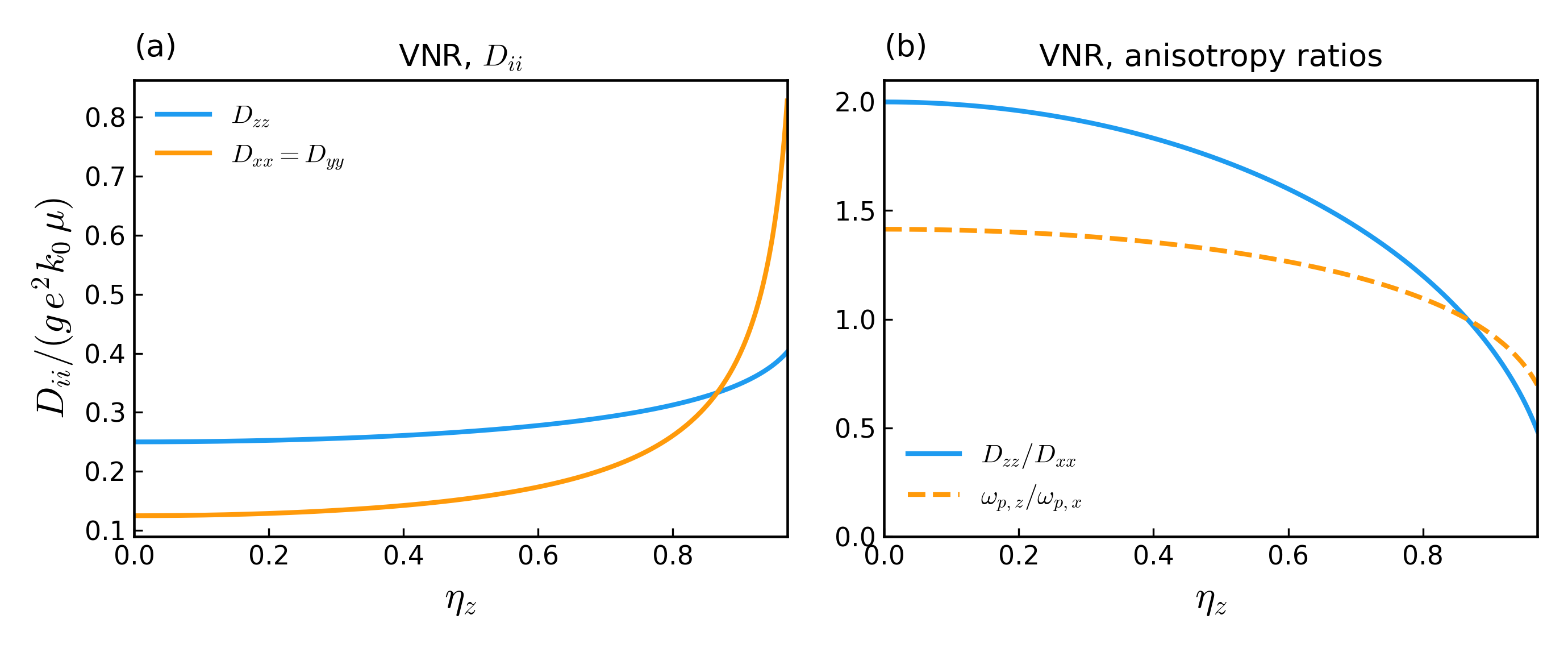}
\caption{\label{fig:drudezvnr} Drude weights of the VNR under an axial tilt. Panel (a) shows $D_{zz}$ and $D_{xx} = D_{yy}$ in units of $g\,e^2\,k_0\,\mu$ as functions of $\eta_z$. Panel (b) shows the anisotropy ratios $D_{zz}/D_{xx}$ and $\omega_{p,z}/\omega_{p,x}$. All curves coincide with those of the PTNR.}
\end{figure*}
%%%%%%%%%%%%%%%%%%%%%%%%%%%

%%%%%%%%%%%%%%%%%%%%%%%%%%%
\subsection{Doped regime: density-of-states, Drude weights, and plasmons}
\label{secazdoped}

For low doping, $0 < \mu \ll v_0\,k_0\,(1 - |\eta_z|)$, only the band $s = 2$ crosses the Fermi level. The Fermi momentum and the DOS, which follows from the Jacobian $\kappa\,k_0$, are
\begin{align}
\label{eq:kappaFzvnr}
\kappa_F (\gamma) = \frac{\mu}{v_0\,(1 + \eta_z \sin\gamma)}\,, \quad
\rho (\mu) = \frac{g\,k_0\,\mu}{2\,\pi\,v_0^2\,(1 - \eta_z^2)^{3/2}}\,.
\end{align}
The DOS is even in $\eta_z$ and increases with $|\eta_z|$. It diverges as $|\eta_z| \to 1$, which is the boundary of the type-II regime. The band velocity of $s = 2$ has the components $V_x^{(2)} = v_0\,\cos\gamma\cos\phi$, $V_y^{(2)} = v_0\,\cos\gamma\sin\phi$, and $V_z^{(2)} = v_0\,(\sin\gamma + \eta_z)$. The off-diagonal entries of $D_{ij}$ vanish after the integration over $\phi$, and $D_{xx} = D_{yy}$, as required by the combined rotation. The Drude weights of Eq.~\eqref{eqdrude} are
\begin{align}
\label{eq:dzvnr}
D_{zz} = \frac{g\,e^2\,k_0\,\mu}{2\,\eta_z^2} \left( 1 - \sqrt{1 - \eta_z^2} \right), \quad
D_{xx} = D_{yy} = \frac{g\,e^2\,k_0\,\mu}{4\,\eta_z^2} \left( \frac{1}{\sqrt{1 - \eta_z^2}} - 1 \right).
\end{align}
For $\eta_z \to 0$, they reduce to the untilted values of Eq.~\eqref{eq:dvnr}. The anisotropy ratio is $D_{zz}/D_{xx} = 2\,\sqrt{1 - \eta_z^2}$. The intraband polarisabilities follow from Eq.~\eqref{eqintra} as $\Pi_{\rm intra} (q\,\bs{\hat e}_i, \omega) \simeq q^2\,D_{ii}/(\pi\,e^2\,\omega^2)$, and the plasmon frequencies of Eq.~\eqref{eqplasmon} are
\begin{align}
\label{eq:wpzvnr}
& \omega_{p,z}^2 = \frac{2\,g\,e^2\,k_0\,\mu}{\eta_z^2} 
\left( 1 - \sqrt{1 - \eta_z^2} \right), \quad
\omega_{p,x}^2 = \omega_{p,y}^2 = \frac{g\,e^2\,k_0\,\mu}{\eta_z^2} 
\left( \frac{1}{\sqrt{1 - \eta_z^2}} - 1 \right),\nn &
%%%%%%%%%%%%%%%%%%%%%%%%%%%%
\frac{\omega_{p,z}}{\omega_{p,x}} = \sqrt{2\,\sqrt{1 - \eta_z^2}} 
\simeq \sqrt{2} \left( 1 - \frac{\eta_z^2}{4} \right).
\end{align}
To order $\eta_z^2$, the squared frequencies are $\omega_{p,z}^2 \simeq g\,e^2\,k_0\,\mu\,(1 + \eta_z^2/4)$ and $\omega_{p,x}^2 \simeq \tfrac{1}{2}\,g\,e^2\,k_0\,\mu\,(1 + 3\,\eta_z^2/4)$. The axial tilt enhances both frequencies, and it lowers their anisotropy ratio from its untilted value $\sqrt{2}$ at quadratic order in $\eta_z$. As for the in-plane tilt, these results coincide with those of the PTNR, since they follow from $\varepsilon_s$, $\bs{V}^{(s)}$, and the integration measure, which are the same in the two linearised models. The steps behind Eqs.~\eqref{eq:kappaFzvnr}--\eqref{eq:wpzvnr} are given in Appendix~\ref{app:vnrzdrude}. Fig.~\ref{fig:drudezvnr} shows the Drude weights and their anisotropy as functions of $\eta_z$.

%%%%%%%%%%%%%%%%%%%%%%%%%%%
\subsection{Interband screening and the screened plasmons}
\label{secazscreen}

%%%%%%%%%%%%%%%%%%%%%%%%%%%
\begin{figure*}[t!]
\centering
\includegraphics[width=0.7 \textwidth]{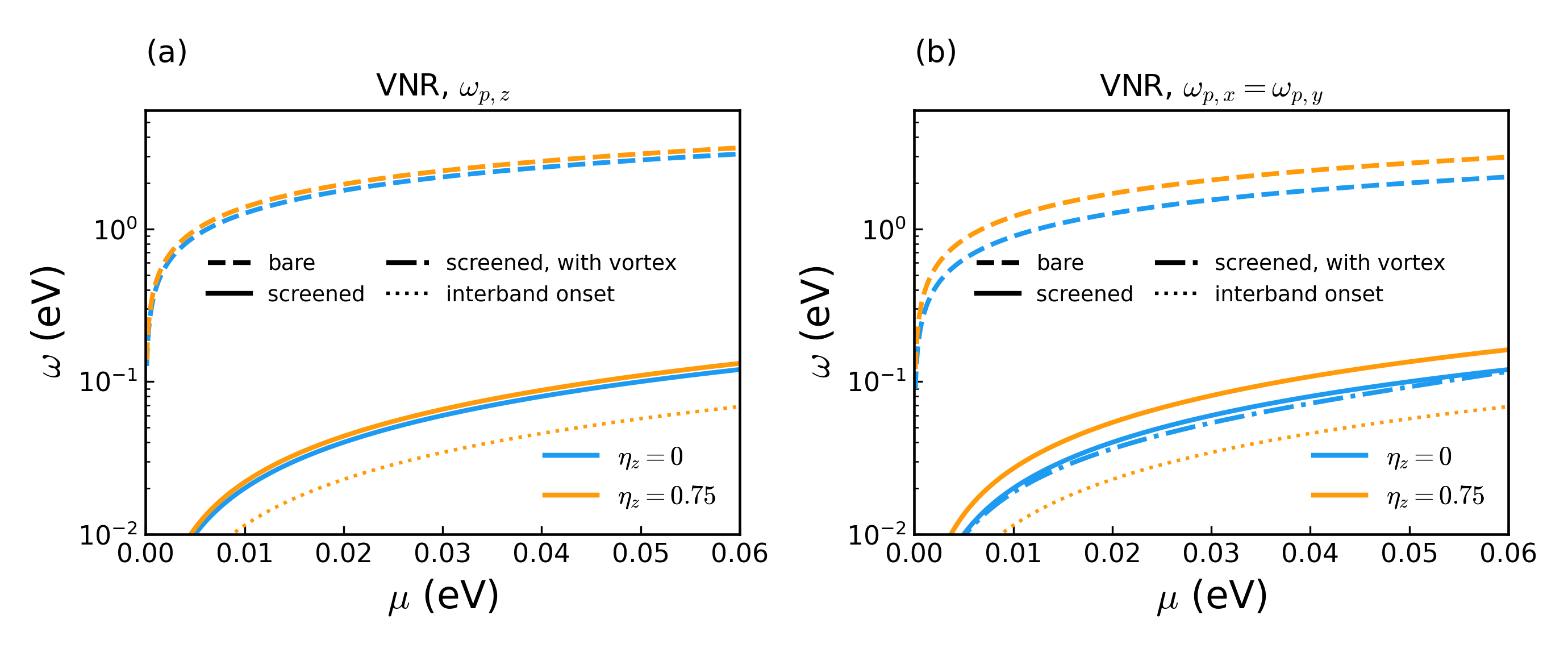}
\caption{\label{fig:plasmazvnr} Bare and screened plasma frequencies of the VNR under an axial tilt as functions of $\mu$. Panels (a) and (b) show $\omega_{p,z}$ and $\omega_{p,x} = \omega_{p,y}$, respectively. The dashed and solid lines are the bare and the screened frequencies, the latter with the static background constants that do not involve $\Lambda$. The dotted lines mark the interband onset. The blue and orange colours correspond to $\eta_z = 0$ and $0.75$. The dash-dotted line in (b) includes the shift of the in-plane background constant from the second term of the overlap factor for $\eta_z = 0$, with the cut-off $\Lambda = k_0$. The parameters are those of Sec.~\ref{secvalidity}.}
\end{figure*}
%%%%%%%%%%%%%%%%%%%%%%%%%%%

The interband response follows from the local mapping of Eq.~\eqref{eq:ringmapvnr}. For $\bs{q} = q\,\bs{\hat z}$, the mapping is exact in the linearised model, and the second term of the overlap factor does not contribute. The unblocked states are those with $\kappa > \kappa_F (\gamma)$, and the tilt does not alter the transition energy $2\,v_0\,\kappa$ at $q \to 0$, as seen in Eq.~\eqref{eq:polinterstatic}. The interband onset is therefore set by the smallest value of $2\,v_0\,\kappa_F (\gamma)$, namely $\omega_{\rm th} = 2\,\mu/(1 + |\eta_z|)$. It is lowered linearly in $|\eta_z|$ relative to the untilted value $2\,\mu$. For $\omega < \omega_{\rm th}$, we define $A_\pm = 2\,\mu \pm \omega$, $r_\pm = \sqrt{A_\pm^2 - \eta_z^2\,\omega^2}$, and $\varrho_\pm = |\eta_z|\,\omega/(A_\pm + r_\pm)$. The radial integral over $\kappa > \kappa_F (\gamma)$ is elementary, and the angular integrals follow from the Fourier series of $\ln (A + B \sin\gamma)$. We obtain
\begin{align}
\label{eq:pinterzvnr}
\Pi_{\rm inter} (q\,\bs{\hat z}, \omega) &= -\, \frac{g\,k_0\,q^2}{16\,\pi\,\omega} 
\left[ \ln \! \left( \frac{A_+ + r_+}{A_- + r_-} \right) + \frac{\varrho_+^2 - \varrho_-^2}{2} \right], \nn
\Pi_{\rm inter} (q\,\bs{\hat x}, \omega) &= \Pi_{\rm inter} (q\,\bs{\hat y}, \omega) 
= -\, \frac{g\,k_0\,q^2}{32\,\pi\,\omega} \left[ \ln \! \left( \frac{A_+ + r_+}{A_- + r_-} \right) - \frac{\varrho_+^2 - \varrho_-^2}{2} \right],
\end{align}
where the in-plane channels exclude the vortex term $\delta\Pi$. For $\eta_z \to 0$, these reduce to Eq.~\eqref{eq:pinterdopedvnr}. At $\omega \to 0$, they give $-\,g\,k_0\,q^2/(16\,\pi\,\mu)$ and half of this, independent of $\eta_z$. The static background dielectric constants are therefore the untilted ones of Eq.~\eqref{eq:ebvnr},
\begin{align}
\label{eq:ebzvnr}
\epsilon_{b,z} = 1 + \frac{g\,e^2\,k_0}{4\,\mu}\,, \quad
\epsilon_{b,x} = \epsilon_{b,y} = 1 + \frac{g\,e^2\,k_0}{8\,\mu}\,.
\end{align}
The reason is that $\kappa_F (\gamma) \propto 1/(1 + \eta_z \sin\gamma)$, and the term odd in $\sin\gamma$ drops out of the $\gamma$-average of $1/\kappa_F$. For $\omega$ above the onset, the imaginary part follows from the delta function $\delta (\omega - 2\,v_0\,\kappa)$ restricted to the unblocked arc of the circle $\kappa = \omega/(2\,v_0)$. With $\tau = \left( 2\,\mu/\omega - 1 \right)/|\eta_z|$, it reads
\begin{align}
\label{eq:imzvnr}
\Im \varepsilon^{\rm RPA} (q\,\bs{\hat z}, \omega) = \frac{\pi\,g\,e^2\,k_0}{4\,\omega}\, \mathcal{G}_+ (\tau)\,, \quad
\Im \varepsilon^{\rm RPA} (q\,\bs{\hat x}, \omega) = \frac{\pi\,g\,e^2\,k_0}{8\,\omega}\, \mathcal{G}_- (\tau)\,,
\end{align}
with $\mathcal{G}_\pm (\tau) = 1/2 - \left[ \arcsin \tau \pm \tau\,\sqrt{1 - \tau^2} \right]/\pi$ for $|\tau| < 1$, $\mathcal{G}_\pm = 0$ for $\tau \geq 1$, and $\mathcal{G}_\pm = 1$ for $\tau \leq -1$. The condition $\tau = 1$ reproduces the onset $\omega_{\rm th}$. For $\tau \leq -1$, we recover the untilted values of Eq.~\eqref{eq:epsdopedvnr}, and the same holds for the $\bs{\hat y}$ channel.

The real part of the dielectric function is $\Re \varepsilon^{\rm RPA} = 1 - \omega_{p,i}^2/\omega^2 - 4\,\pi\,e^2\,\Pi_{\rm inter}/q^2$. Below the onset, $\epsilon_{b,i} (\omega)$ increases with $\omega$ and the Drude term $-\,\omega_{p,i}^2/\omega^2$ increases as well. The real part is therefore monotonic, its zero is unique, and the static screened frequency $\tilde\omega_{p,i}^2 = \omega_{p,i}^2/\epsilon_{b,i}$ is an upper bound of it. At the onset, $r_- \to 0$, and $\Re \varepsilon^{\rm RPA}$ stays finite. This differs from the untilted case, where $\Re \varepsilon^{\rm RPA}$ tends to $+\infty$ at $\omega = 2\,\mu$ and a zero below the onset exists for every coupling. For $\eta_z \neq 0$, a zero below the onset is no longer guaranteed. A sufficient condition for its existence is that the static screened frequency lies below $\omega_{\rm th}$. The tilt lowers the onset linearly in $|\eta_z|$ but raises the screened frequencies only at order $\eta_z^2$. The plasmon is therefore pushed into the interband continuum as $|\eta_z|$ grows. The steps behind Eqs.~\eqref{eq:pinterzvnr}--\eqref{eq:imzvnr}, and behind the statements on the zero of $\Re \varepsilon^{\rm RPA}$, are given in Appendix~\ref{app:vnrzscreen}.

\begin{figure*}[t!]
\centering
\includegraphics[width=0.75 \textwidth]{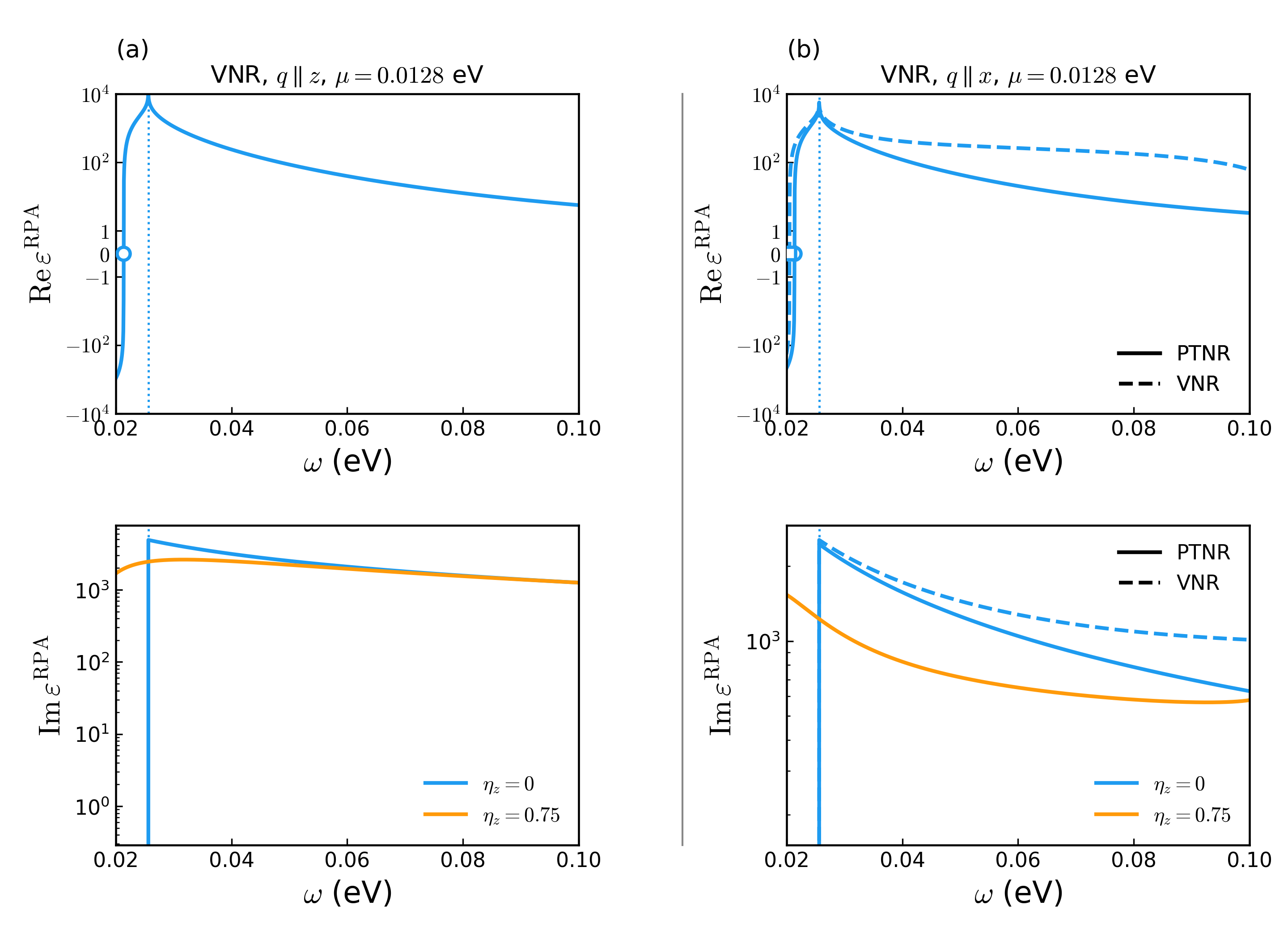}
\caption{\label{fig:epszvnr} RPA dielectric function of the VNR under an axial tilt. Panel (a) shows the $\bs{\hat z}$ channel, which coincides with that of the PTNR. Panel (b) shows the $\bs{\hat x}$ channel, which coincides with the $\bs{\hat y}$ channel. The upper and lower rows show $\Re \varepsilon^{\rm RPA}$ on a symmetric logarithmic scale and $\Im \varepsilon^{\rm RPA}$, respectively. The blue and orange colours correspond to $\eta_z = 0$ and $\eta_z = 0.75$, respectively. The solid lines are the terms that do not involve $\Lambda$. The dashed lines in (b) include the second term of the overlap factor for $\eta_z = 0$ with the cut-off $\Lambda = k_0$, and they stop where $\omega$ approaches $2\,v_0\,\Lambda$. The dotted vertical lines mark the interband onset. The circles mark the zeros of $\Re \varepsilon^{\rm RPA}$ below the onset, and the square marks the zero with the second term of the overlap factor. The parameters are those of Sec.~\ref{secvalidity}.}
\end{figure*}

For $\mu = 0.1\,v_0\,k_0$ and $0.2\,v_0\,k_0$, the background constants are much larger than unity. Without tilt, the zero of $\Re \varepsilon^{\rm RPA}$ lies below the onset $2\,\mu$ for both channels. Evaluating Eq.~\eqref{eq:pinterzvnr} numerically, we find that the zero disappears above a critical tilt, which is smaller along $\bs{\hat z}$ than along $\bs{\hat x}$ and $\bs{\hat y}$. For a large tilt, the onset is lowered to $2\,\mu/(1 + |\eta_z|)$, and the static screened frequencies $\tilde\omega_{p,z}$ and $\tilde\omega_{p,x}$ both lie above it. The sufficient condition for the existence of a zero below the onset is therefore not met at large tilt. The interband screening is essential, since the bare frequencies greatly exceed the onset. Since the tilted case lies at the edge of validity of the expansions for the larger $\mu$, as discussed above, we regard the conclusions at large tilt as indicative. They agree with the PTNR results at the same $\mu$, since the $\bs{\hat z}$ channel and the leading in-plane terms are identical. Figs.~\ref{fig:plasmazvnr} and~\ref{fig:epszvnr} show the plasma frequencies and the dielectric function, respectively.

%%%%%%%%%%%%%%%%%%%%%%%%%%%
%%%%%%%%%%%%%%%%%%%%%%%%%%%

%%%%%%%%%%%%%%%%%%%%%%%%%%%
\subsection{Toroidal-angle dependence of the overlap factor at finite tilt}
\label{secazvortex}

The dependence of the overlap factor on $\phi$ enters only through the second term of Eq.~\eqref{eq:fsplit}. For $\bs{q} = q\,\bs{\hat z}$, it vanishes, and the $\bs{\hat z}$ channel is identical to that of the PTNR, as stated above. For $\bs{q} = q\,\bs{\hat x}$ and $\bs{q} = q\,\bs{\hat y}$, the shifts are $\phi' - \phi \simeq -\,q\sin\phi/k_0$ and $\phi' - \phi \simeq q\cos\phi/k_0$. The second term then carries the weights $\sin^2\phi$ and $\cos^2\phi$, respectively, and each of them averages to $1/2$ over $\phi$. Since the Pauli-blocking radius $\kappa_F (\gamma)$ does not depend on $\phi$ for the axial tilt, the two channels coincide. This is in contrast with the in-plane tilt, where the blocking arcs depend on $\phi$. The tilt does not enter the energy denominators of the in-plane channels, because $v_0\,\bs{\eta} \cdot \bs{q} = 0$. It enters only through the blocking radius $\kappa_F (\gamma)$. Inserting the second term of Eq.~\eqref{eq:fsplit} into Eq.~\eqref{eq:polinter} and performing the integral over $\phi$, we obtain
\begin{align}
\label{eq:dpizvnr}
\delta\Pi (q\,\bs{\hat x}, \omega) = \delta\Pi (q\,\bs{\hat y}, \omega) \simeq \frac{g\,v_0\,q^2}{8\,\pi^2\,k_0}
\int_0^{2\pi} d\gamma\, \sin^2\gamma \int_{\kappa_F (\gamma)}^{\Lambda} \frac{\kappa^2\,d\kappa}{(\omega + i\,0^+)^2 - 4\,v_0^2\,\kappa^2}\,,
\end{align}
where $\Lambda \lesssim k_0$ is the same cut-off as in Eq.~\eqref{eq:dpivnr}. For $\eta_z \to 0$, the integral over $\gamma$ gives $\pi$, and Eq.~\eqref{eq:dpivnr} is recovered. The expansion requires $\kappa_F^{\max} < \Lambda$.

The imaginary part is controlled by the resonant radius $\kappa_0 = \omega/(2\,v_0)$. The condition $\kappa_0 > \kappa_F (\gamma)$ is equivalent to $\sin\gamma > \tau$ for $\eta_z > 0$, with $\tau$ as above, and the same holds for $\eta_z < 0$ after $\gamma \to -\gamma$. The integral of $\sin^2\gamma$ over this arc equals $\pi\,\mathcal{G}_- (\tau)$, and we obtain
\begin{align}
\label{eq:imdpizvnr}
\Im \delta\Pi (q\,\bs{\hat x}, \omega) = \Im \delta\Pi (q\,\bs{\hat y}, \omega) \simeq -\, \frac{g\,q^2\,\omega}{128\,v_0^2\,k_0}\, \mathcal{G}_- (\tau)\,.
\end{align}
It reduces to Eq.~\eqref{eq:imdpivnr} for $\tau \leq -1$. Compared with the leading term $-\,g\,k_0\,q^2\,\mathcal{G}_- (\tau)/(32\,\omega)$ of Eq.~\eqref{eq:imzvnr} in the same channel, it is smaller by $(\omega/2\,v_0\,k_0)^2$. This suppression is independent of the tilt, because both terms carry the same function $\mathcal{G}_-$. It vanishes below the onset $\omega_{\rm th}$.

The real part is again a frequency-independent shift that is dominated by $\kappa \sim \Lambda$. For $\omega \ll v_0\,\Lambda$, we have
\begin{align}
\Re \delta\Pi (q\,\bs{\hat x}, \omega) = \Re \delta\Pi (q\,\bs{\hat y}, \omega) \simeq -\, \frac{g\,q^2}{32\,\pi\,v_0\,k_0} \left( \Lambda - \bar\kappa_F \right),
\end{align}
where $\bar\kappa_F$ is the average of the Fermi momentum over the arc weight $\sin^2\gamma$,
\begin{align}
\label{eq:kbarzvnr}
\bar\kappa_F = \frac{1}{\pi} \int_0^{2\pi} d\gamma\, \sin^2\gamma\, \kappa_F (\gamma)
= \frac{2\,\mu}{v_0\,\eta_z^2} \left( \frac{1}{\sqrt{1 - \eta_z^2}} - 1 \right).
\end{align}
We have used $\int_0^{2\pi} d\gamma\, \sin^2\gamma/(1 + \eta_z \sin\gamma) = (2\,\pi/\eta_z^2) \left( 1/\sqrt{1 - \eta_z^2} - 1 \right)$. For $\eta_z \to 0$, $\bar\kappa_F \to \mu/v_0$, which is the untilted value of $\kappa_F$. To order $\eta_z^2$, we have $\bar\kappa_F \simeq (\mu/v_0)\,(1 + 3\,\eta_z^2/4)$. The in-plane background dielectric constants receive
\begin{align}
\label{eq:debzvnr}
\delta\epsilon_{b,x} = \delta\epsilon_{b,y} \simeq \frac{g\,e^2 \left( \Lambda - \bar\kappa_F \right)}{8\,v_0\,k_0}\,, \quad \delta\epsilon_{b,z} = 0\,.
\end{align}
This reduces to Eq.~\eqref{eq:debvnr} for $\eta_z \to 0$. The axial tilt lowers the shift slightly, because it enlarges the weighted Fermi radius $\bar\kappa_F$. As in the untilted case, the shift is of relative order $\kappa_F\,\Lambda/k_0^2$ with respect to $\epsilon_{b,x} - 1$ of Eq.~\eqref{eq:ebzvnr}. The ratio of the shift to $\epsilon_{b,x} - 1$ is exactly $(\Lambda - \bar\kappa_F)\,\kappa_F/k_0^2$ with $\kappa_F = \mu/v_0$. For $\eta_z = 0$ and $\Lambda = k_0$, this ratio equals $(1 - \kappa_F/k_0)\,\kappa_F/k_0$. Eq.~\eqref{eq:debzvnr} can be used with $\Lambda \lesssim k_0$ only for $|\eta_z| < 1 - \mu/(v_0\,k_0)$, because $\kappa_F^{\max}$ exceeds $k_0$ otherwise. Since it is set by $\Lambda$, it is of the same order as the corrections of relative size $\kappa/k_0$ that were dropped from the leading terms. We therefore regard the results of this section that do not depend on $\Lambda$ as the controlled predictions of the model. The axial tilt leaves the conclusion of Sec.~\ref{secvortexcorr} intact. Relative to the PTNR, the bare plasmon frequencies $\omega_{p,i}$ are the same, the in-plane absorption differs at relative order $(\omega/2\,v_0\,k_0)^2$, and the in-plane screening can only be stronger. The steps behind Eqs.~\eqref{eq:dpizvnr}--\eqref{eq:debzvnr} are given in Appendix~\ref{app:vnrzvortex}.

%%%%%%%%%%%%%%%%%%%%%%%%%%%%%%%%%%%%%%%%%%%%%%%%%%%%%%%%%%%%%%%%%%%%%%%%%%%%%%%%%%%%%%%%%%%%%%%%%%%%%%%%%%%%%%%%%%
\section{Untilted HSM}
\label{sechopf}

The VNR and the HSM are the two simplest topological semimetals whose BC has a dipolar instead of a monopolar structure~\cite{graf-hopf, ips-dipole-vnr, ips-firdous-vnr}. The VNR hosts a continuous ring of BC-dipoles around the nodal loop, as seen in Sec.~\ref{secvortex}, whereas the threefold node of the HSM carries a single BC-dipole at the band-crossing point. The plasmons of the HSM were studied in Ref.~\cite{hopf-plasmon}, which reports that the long-wavelength plasmons are anisotropic, although the band structure, consisting of a Dirac cone and a flat band, is isotropic. The anisotropy was attributed to the dipolar structure of the BC. We derive the DOS, the Drude weights, the interband response, and the plasmon frequencies of the untilted HSM with the local method used for the VNR in Sec.~\ref{secuntilted} and for the untilted WSM in Sec.~\ref{secwsm}, and compare them with those of the untilted VNR and WSM. We do not include any tilt for the HSM.

The effective continuum Hamiltonian of a threefold Hopf node is~\cite{graf-hopf, hopf2, hopf3}
\begin{align}
\label{eq:hhopf}
\mathcal{H}_{\rm H} (\bs{k}) = v_0 \left( k_x\,\mathcal{M}_x + k_y\,\mathcal{M}_y + k_z\,\mathcal{M}_z \right),
\end{align}
where $\mathcal{M}_x$, $\mathcal{M}_y$, and $\mathcal{M}_z$ are the Gell-Mann matrices that couple the pairs of states $\lbrace 1, 2 \rbrace$, $\lbrace 1, 2 \rbrace$, and $\lbrace 1, 3 \rbrace$, respectively,
\begin{align}
\mathcal{M}_x = \begin{pmatrix} 0 & 1 & 0 \\ 1 & 0 & 0 \\ 0 & 0 & 0 \end{pmatrix}, \quad
\mathcal{M}_y = \begin{pmatrix} 0 & -\,i & 0 \\ i & 0 & 0 \\ 0 & 0 & 0 \end{pmatrix}, \quad
\mathcal{M}_z = \begin{pmatrix} 0 & 0 & -\,i \\ 0 & 0 & 0 \\ i & 0 & 0 \end{pmatrix}.
\end{align}
The eigenvalues are
\begin{align}
\label{eq:evhopf}
\varepsilon_s (\bs{k}) =
\begin{cases}
(-1)^s\,v_0\,k & \text{ for } s \in \lbrace 1, 2 \rbrace \\
0 & \text{ for } s = 0
\end{cases},
\end{align}
with $k = |\bs{k}|$. The bands $s = 1$ and $s = 2$ form an isotropic Dirac cone, and $s = 0$ is a nondispersive flat band. Its existence follows from the chiral symmetry $\mathcal{S} = \operatorname{diag} (1, -1, -1)$, which anticommutes with each of the three matrices. The spectrum is therefore symmetric under $\varepsilon \to -\,\varepsilon$, and since the Hilbert space is three-dimensional, at least one state remains at zero energy. The spectrum is the same as that of the isotropic pseudospin-1 triple-point semimetal~\cite{ips-spin1-ph, ips-internode}, and the two systems differ only in their eigenstates.

Writing the Hamiltonian in the block form $\mathcal{H}_{\rm H} = v_0 \begin{pmatrix} 0 & \bs{a}^\dagger \\ \bs{a} & 0 \end{pmatrix}$ with the two-component vector $\bs{a} = (k_+, i\,k_z)^T$ and $k_\pm = k_x \pm i\,k_y$, we obtain $\bs{a}^\dagger\bs{a} = k^2$. The normalised eigenstates are
\begin{align}
\label{eq:statehopf}
|\psi_s (\bs{k}) \rangle = \frac{1}{\sqrt{2}} \begin{pmatrix} 1 \\ (-1)^s\,\bs{a}/k \end{pmatrix}
\text{ for } s \in \lbrace 1, 2 \rbrace\,, \quad
|\psi_0 (\bs{k}) \rangle = \frac{1}{k} \begin{pmatrix} 0 \\ i\,k_z \\ k_- \end{pmatrix}.
\end{align}
The Hamiltonian is not rotationally symmetric in 3d. A rotation of the momentum by an angle $\alpha$ about the $\bs{\hat z}$-axis, $\bs{k} \to R_\alpha\,\bs{k}$, is compensated by the unitary transformation $U_\alpha = \operatorname{diag} (e^{-\,i\,\alpha}, 1, e^{-\,i\,\alpha})$, in the sense that $\mathcal{H}_{\rm H} (R_\alpha\,\bs{k}) = U_\alpha\,\mathcal{H}_{\rm H} (\bs{k})\,U_\alpha^\dagger$. This is the only continuous rotation symmetry of the model, and it singles out the $\bs{\hat z}$-direction. We use the spherical-polar coordinates
\begin{align}
\label{eq:sphhopf}
k_x = k\,\sin\vartheta\,\cos\varphi\,, \quad
k_y = k\,\sin\vartheta\,\sin\varphi\,, \quad
k_z = k\,\cos\vartheta\,,
\end{align}
with $\vartheta \in [0, \pi]$ and $\varphi \in [0, 2\,\pi)$, instead of the toroidal coordinates used for the VNR. The symbols $\vartheta$ and $\varphi$ are distinct from the toroidal angles $\gamma$ and $\phi$ of Sec.~\ref{secmodel}, and the matrices $\mathcal{M}_i$ are distinct from the Pauli matrices of the VNR.

\subsection{BC and orbital magnetic moment}
\label{sechopfgeom}

The HSM is the gapless descendant of a 3d Hopf insulator~\cite{graf-hopf, hopf2, hopf3, hopf-ins-surface-states}. The BC and the orbital magnetic moment (OMM) of the three bands are~\cite{graf-hopf, ips-dipole-vnr}
\begin{align}
\label{eq:bcommhopf}
\bs{\Omega}_s (\bs{k}) =
\begin{cases}
\dfrac{k_z\,\bs{k}}{k^4} & \text{ for } s \in \lbrace 1, 2 \rbrace \\[6pt]
-\,\dfrac{2\,k_z\,\bs{k}}{k^4} & \text{ for } s = 0
\end{cases}
\text{ and } \quad
\bs{m}_s (\bs{k}) =
\begin{cases}
\dfrac{(-1)^{s+1}\,e\,v_0\,k_z\,\bs{k}}{2\,k^3} & \text{ for } s \in \lbrace 1, 2 \rbrace \\[6pt]
0 & \text{ for } s = 0
\end{cases},
\end{align}
respectively, where the overall sign of the BC depends on convention. The two dispersive bands carry the same BC, and the flat band carries the opposite BC with twice the magnitude, so that $\sum_s \bs{\Omega}_s = 0$. In the coordinates of Eq.~\eqref{eq:sphhopf}, $\bs{\Omega}_{1,2} = \cos\vartheta\,\bs{\hat k}/k^2$. The BC points outward for $k_z > 0$, inward for $k_z < 0$, and vanishes in the plane $k_z = 0$. This is the BC-dipole of the node. The flux through any sphere of radius $k$ around the node vanishes,
\begin{align}
\oint \bs{\Omega}_{1,2} \cdot d\bs{S} = \int_0^{2\,\pi} \! d\varphi \int_0^\pi \! d\vartheta\,\sin\vartheta\,\cos\vartheta = 0\,,
\end{align}
and the Chern number is zero on every enclosing surface, in contrast to the Weyl node and the triple-point node, for which the flux is quantised and nonzero. The vanishing takes place at each Hopf node individually, and not by a cancellation between nodes of opposite chirality. The OMM of the dispersive bands has the same angular dependence as the BC, and it is opposite for $s = 1$ and $s = 2$. The flat band has no OMM.

\subsection{Overlap factors and quantum metric}
\label{sechopfqm}

The density response is given by Eq.~\eqref{eq:polgeneral} with $s, s' \in \lbrace 0, 1, 2 \rbrace$ and the overlap factors $\mathcal{F}_{ss'} (\bs{k}, \bs{k}') = |\langle \psi_{s'} (\bs{k}') | \psi_s (\bs{k}) \rangle|^2$, built from Eq.~\eqref{eq:statehopf}. For $\bs{q} \to 0$, the intraband factors tend to unity, and the interband factors are $\mathcal{F}_{ss'} \simeq q_i\,q_j\,Q^{ss'}_{ij} (\bs{k})$ with the band-resolved quantum metric
\begin{align}
Q^{ss'}_{ij} = \frac{v_0^2\,\Re \left[ \langle \psi_s | \mathcal{M}_i | \psi_{s'} \rangle \langle \psi_{s'} | \mathcal{M}_j | \psi_s \rangle \right]}{\left( \varepsilon_s - \varepsilon_{s'} \right)^2}\,.
\end{align}
It satisfies $\sum_{s' \neq s} Q^{ss'}_{ij} = g^{(s)}_{ij}$, the quantum metric of the band $s$.

We write $\mathcal{H}_{\rm H} = v_0 \begin{pmatrix} 0 & \bs{a}^\dagger \\ \bs{a} & 0 \end{pmatrix}$ with $\bs{a} = (k_+, i\,k_z)^T$. The derivatives are $\partial_x \bs{a} = (1, 0)^T$, $\partial_y \bs{a} = (i, 0)^T$, and $\partial_z \bs{a} = (0, i)^T$. With the states of Eq.~\eqref{eq:statehopf}, the matrix elements of $\mathcal{M}_i = \partial_{k_i} \mathcal{H}_{\rm H}/v_0$ between the flat band and the upper band are $\langle \psi_0 | \mathcal{M}_i | \psi_2 \rangle = \bs{y}_0^\dagger\,\partial_i \bs{a}/\sqrt{2}$ with $\bs{y}_0 = (i\,k_z, k_-)^T/k$, namely
\begin{align}
\langle \psi_0 | \mathcal{M}_x | \psi_2 \rangle = -\,\frac{i\,k_z}{\sqrt{2}\,k}\,, \quad
\langle \psi_0 | \mathcal{M}_y | \psi_2 \rangle = \frac{k_z}{\sqrt{2}\,k}\,, \quad
\langle \psi_0 | \mathcal{M}_z | \psi_2 \rangle = \frac{i\,k_+}{\sqrt{2}\,k}\,.
\end{align}
Between the lower and the upper band, $\langle \psi_1 | \mathcal{M}_i | \psi_2 \rangle = i\,\Im \left[ (\partial_i \bs{a}^\dagger)\,\bs{a} \right]/k$, namely
\begin{align}
\langle \psi_1 | \mathcal{M}_x | \psi_2 \rangle = \frac{i\,k_y}{k}\,, \quad
\langle \psi_1 | \mathcal{M}_y | \psi_2 \rangle = -\,\frac{i\,k_x}{k}\,, \quad
\langle \psi_1 | \mathcal{M}_z | \psi_2 \rangle = 0\,.
\end{align}
Dividing by the squared energy differences $v_0^{-2} \left( \varepsilon_s - \varepsilon_{s'} \right)^2$, which equal $k^2$ for the pair $(0, 2)$ and $4\,k^2$ for the pair $(1, 2)$, gives Eq.~\eqref{eq:qhopf}. As a check, the sum $Q^{02}_{ij} + Q^{12}_{ij}$ is the quantum metric of the upper band, for example with $g^{(2)}_{zz} = \sin^2\vartheta/(2\,k^2)$. The pairs that matter below are
\begin{align}
\label{eq:qhopf}
& Q^{02}_{zz} = \frac{\sin^2\vartheta}{2\,k^2}\,, \quad
Q^{02}_{xx} = Q^{02}_{yy} = \frac{\cos^2\vartheta}{2\,k^2}\,, \quad
Q^{02}_{xy} = 0\,, \nn &
Q^{12}_{zz} = 0\,, \quad
Q^{12}_{xx} = \frac{\sin^2\vartheta\,\sin^2\varphi}{4\,k^2}\,, \quad
Q^{12}_{yy} = \frac{\sin^2\vartheta\,\cos^2\varphi}{4\,k^2}\,, \quad
Q^{12}_{xy} = -\,\frac{\sin^2\vartheta\,\sin\varphi\,\cos\varphi}{4\,k^2}\,.
\end{align}
The dependence of $Q^{12}_{ij}$ on $\varphi$ is the analogue of the dependence on the toroidal angle in the VNR, and it vanishes after the integration over $\varphi$ for the in-plane directions considered below. The metric $Q^{02}_{ij}$ is not isotropic, and it is the origin of the anisotropy of the interband response.

\subsection{Fermi surface, DOS, and Drude weights}
\label{sechopfdrude}

For $\mu > 0$, only the band $s = 2$ crosses the Fermi level, and the FS is the sphere $k = k_F = \mu/v_0$. The flat band lies below the Fermi level and is completely filled. Even if it were partially filled, it would not contribute to the intraband response, since its velocity vanishes identically and the Drude weight of Eq.~\eqref{eqdrude} is proportional to the square of the velocity. The chiral symmetry of the Hamiltonian maps $\mu$ to $-\mu$ and exchanges the bands $s = 2$ and $s = 0$. The results below, which are derived for $\mu > 0$, therefore hold for $\mu < 0$ with $\mu$ replaced by $|\mu|$, and the FS is then the sphere of the band $s = 0$. The DOS and the density, measured from $\mu = 0$, are
\begin{align}
\label{eq:doshopf}
\rho (\mu) = \frac{g}{(2\,\pi)^3} \int d\Omega\,\frac{k_F^2}{v_0} = \frac{g\,\mu^2}{2\,\pi^2\,v_0^3}\,, \quad
\tilde{n} = \frac{g\,\mu^3}{6\,\pi^2\,v_0^3}\,.
\end{align}
%%%%%%%%%%%%%%%%%%%%
The band-velocity is $\bs{V}^{(2)} = v_0\,\bs{\hat k}$. It depends only on the direction of $\bs{k}$ and not on the eigenstates, and hence the pseudospin texture and the BC-dipole do not enter the Drude weights. The Drude weights of Eq.~\eqref{eqdrude} are therefore isotropic,
\begin{align}
\label{eq:dhopf}
D_{ij} = \pi\,e^2\,\rho\,\langle V_i\,V_j \rangle = \frac{g\,e^2\,\mu^2}{6\,\pi\,v_0}\,\delta_{ij}\,,
\end{align}
because $\langle \hat k_i\,\hat k_j \rangle = \delta_{ij}/3$ over the sphere. By Eq.~\eqref{eqplasmon}, which gives $\omega_{p,i}^2 = 4\,D_{ii}$, the bare plasmon frequencies are
\begin{align}
\label{eq:wphopf}
\omega_{p,x}^2 = \omega_{p,y}^2 = \omega_{p,z}^2 = \frac{2\,g\,e^2\,\mu^2}{3\,\pi\,v_0}\,.
\end{align}
The off-diagonal components $D_{ij}$ with $i \neq j$ vanish, so that the Drude part of the dielectric function takes the same form $1 - \omega_p^2/\omega^2$ for $\bs{q}$ along $\bs{\hat x}$, $\bs{\hat y}$, and $\bs{\hat z}$. Here and below, $\omega_p^2$ denotes the common value of Eq.~\eqref{eq:wphopf}. These frequencies are identical to those of the untilted WSM, which has the same FS and velocity. In the HSM, the Drude part of the response is thus isotropic, and any anisotropy of the long-wavelength plasmons must come from the interband part, i.e., from the quantum geometry of the eigenstates. This contrasts with the VNR, where the bare frequencies are already anisotropic with $\omega_{p,z}/\omega_{p,x} = \omega_{p,z}/\omega_{p,y} = \sqrt{2}$ owing to the toroidal FS. For the HSM, the dependence $\omega_p \propto \mu \propto \tilde n^{1/3}$ follows from Eqs.~\eqref{eq:doshopf} and~\eqref{eq:wphopf}.

\subsection{Interband response and screened plasmons}
\label{sechopfscreen}

We begin with the intrinsic point, where the FS shrinks to the node and there is no Drude weight. We take the limit $\mu \to 0^+$, so that the flat band and the band $s = 1$ are filled and the band $s = 2$ is empty. The interband transitions are then those from the filled to the empty states of a pair of bands. The transitions $s = 0 \to 2$ have the energy $v_0\,k$, and the transitions $s = 1 \to 2$ have the energy $2\,v_0\,k$. The transitions $s = 1 \to 0$ are blocked, since both bands are filled. The two allowed transitions are unblocked at every $k > 0$, and hence the onset is $\omega_{\rm th} = 0$.

For $q \to 0$ and $v_0\,q \ll \omega \ll v_0\,\Lambda$, with the cut-off $\Lambda$ of the linearised description, we use $\mathcal{F}_{ss'} \simeq q_i\,q_j\,Q^{ss'}_{ij}$. A pair of bands $p$ with the gap $\Delta_p = c_p\,v_0\,k$ contributes the terms $1/(\omega - \Delta_p) - 1/(\omega + \Delta_p) = 2\,\Delta_p/(\omega^2 - \Delta_p^2)$ from the two orderings of the bands. We have $c_p = 1$ for the pair $(0, 2)$ and $c_p = 2$ for the pair $(1, 2)$. Defining $A^p_i (\vartheta, \varphi) = k^2\,Q^p_{ii}$ for $\bs{q} = q\,\bs{\hat e}_i$, we obtain
\begin{align}
\label{eq:pinterhopfgen}
\Pi_{\rm inter} (q\,\bs{\hat e}_i, \omega) 
= \frac{g\,q^2}{(2\,\pi)^3} \sum_p \int d\Omega\,A^p_i 
\int_{k_F}^{\Lambda} dk\,\frac{2\,c_p\,v_0\,k}{\omega^2 - c_p^2\,v_0^2\,k^2}\,,
\end{align}
with $k_F = \mu/v_0 = 0$ at the intrinsic point. The radial integral equals $-\left( c_p\,v_0 \right)^{-1} \ln \left[ (c_p^2\,v_0^2\,\Lambda^2 - \omega^2)/|c_p^2\,\mu^2 - \omega^2| \right]$, which tends to $-\left( c_p\,v_0 \right)^{-1} \ln \left( c_p^2\,v_0^2\,\Lambda^2/\omega^2 \right)$ for $\mu = 0$ and $v_0\,\Lambda \gg \omega$. The angular integrals are $\int d\Omega\,A^{(0,2)}_z = 4\,\pi/3$, $\int d\Omega\,A^{(0,2)}_x = 2\,\pi/3$, $\int d\Omega\,A^{(1,2)}_z = 0$, and $\int d\Omega\,A^{(1,2)}_x = \pi/3$. The last one uses $\int_0^{2\pi} \sin^2\varphi\,d\varphi = \pi$, and the channel $\bs{\hat y}$ equals the channel $\bs{\hat x}$ because $\int_0^{2\pi} \cos^2\varphi\,d\varphi = \pi$ as well. Hence
\begin{align}
\Pi_{\rm inter} (q\,\bs{\hat z}, \omega) = -\,\frac{g\,q^2}{6\,\pi^2\,v_0}\,\ln \!\left( \frac{v_0^2\,\Lambda^2}{\omega^2} \right), \quad
%%%%%%%%%%%%%%%%
\Pi_{\rm inter} (q\,\bs{\hat x}, \omega) 
= \Pi_{\rm inter} (q\,\bs{\hat y}, \omega) = -\,\frac{g\,q^2}{24\,\pi^2\,v_0} 
\left[ \frac{5}{2}\,\ln \!\left( \frac{v_0^2\,\Lambda^2}{\omega^2} \right) + \ln 2 \right].
\end{align}
The first logarithm in each channel is due to the transitions $0 \to 2$. The pair $(1, 2)$ contributes only along $\bs{\hat x}$ and $\bs{\hat y}$, because $Q^{12}_{zz} = 0$, and it gives the term $\ln 2$ together with a part of the coefficient $5/2$.

The dielectric function follows from $\varepsilon^{\rm RPA} = 1 - 4\,\pi\,e^2 \left( \Pi_{\rm intra} + \Pi_{\rm inter} \right)/q^2$, with $\Pi_{\rm intra} = 0$ at the intrinsic point. For the imaginary part, the causal prescription $\omega \to \omega + i\,0^+$ gives $\Im \left[ 2\,\Delta_p/(\omega^2 - \Delta_p^2) \right] = -\,\pi\,\delta (\omega - \Delta_p)$ for $\omega > 0$. The delta function fixes $k = \omega/(c_p\,v_0)$, which lies in the unblocked region for every $\omega > 0$. For $\omega \ll v_0\,\Lambda$, we thus find
\begin{align}
\label{eq:intrhopf}
\Re \varepsilon^{\rm RPA} (q\,\bs{\hat z}, \omega) &= 1 
+ \frac{2\,g\,e^2}{3\,\pi\,v_0}\,\ln \! \left(  \frac{v_0^2\,\Lambda^2}{\omega^2} \right )\,, \nn
\Re \varepsilon^{\rm RPA} (q\,\bs{\hat x}, \omega) &= \Re \varepsilon^{\rm RPA} (q\,\bs{\hat y}, \omega) = 1 + \frac{5\,g\,e^2}{12\,\pi\,v_0}\,\ln \! \left(  \frac{v_0^2\,\Lambda^2}{\omega^2} \right ) + \frac{g\,e^2\,\ln 2}{6\,\pi\,v_0}\,, \nn
\Im \varepsilon^{\rm RPA} (q\,\bs{\hat z}, \omega) &= \frac{2\,g\,e^2}{3\,v_0}\,, \quad
\Im \varepsilon^{\rm RPA} (q\,\bs{\hat x}, \omega) = \Im \varepsilon^{\rm RPA} (q\,\bs{\hat y}, \omega) = \frac{5\,g\,e^2}{12\,v_0}\,.
\end{align}
The real parts exceed unity for $\omega < v_0\,\Lambda$ in every channel, and the imaginary parts are nonzero for all $\omega > 0$. The dielectric function therefore has no zero, and the intrinsic HSM supports no undamped RPA plasmon, in the same way as the intrinsic VNR, the intrinsic PTNR, and undoped graphene. The ratio of the imaginary parts along $\bs{\hat z}$ and $\bs{\hat x}$ is $8/5$.

For $\mu > 0$, the Fermi momentum $k_F = \mu/v_0$ is finite, and the interband transitions with $k < k_F$ are blocked. The transitions $s = 0 \to 2$ and $s = 1 \to 2$ remain allowed for $k > k_F$, and the transitions $s = 1 \to 0$ remain blocked. The interband onset is therefore set by the flat band,
\begin{align}
\label{eq:thhopf}
\omega_{\rm th} = v_0\,k_F = \mu\,,
\end{align}
which is half of the onset $2\,\mu$ of the VNR and the WSM. The lower limit $k_F$ in Eq.~\eqref{eq:pinterhopfgen} turns the argument $\omega^2$ of each logarithm into $|c_p^2\,\mu^2 - \omega^2|$. The Drude term is $\Pi_{\rm intra} = q^2\,D/(\pi\,e^2\,\omega^2)$, which adds $-\,\omega_p^2/\omega^2$ to $\varepsilon^{\rm RPA}$, with $\omega_p^2$ of Eq.~\eqref{eq:wphopf}. Along $\bs{\hat z}$, $\bs{\hat x}$, and $\bs{\hat y}$, the interband polarisability is
\begin{align}
\label{eq:pinterhopf}
& \Pi_{\rm inter} (q\,\bs{\hat z}, \omega) = 
-\,\frac{g\,q^2}{6\,\pi^2\,v_0}\,
\ln \!\left( \frac{v_0^2\,\Lambda^2}{\left| \mu^2 - \omega^2 \right| } \right)\,, \quad
%%%%%%%%%%%%%%%%
\Pi_{\rm inter} (q\,\bs{\hat x}, \omega) = -\,\frac{g\,q^2}{24\,\pi^2\,v_0} 
\left[ 2\,\ln \!\left( \frac{v_0^2\,\Lambda^2}{\left| \mu^2 - \omega^2 \right| } \right) + \frac{1}{2}\,
\ln \! \left(  \frac{4\,v_0^2\,\Lambda^2}{\left| 4\,\mu^2 - \omega^2 \right|} \right )\right],
%%%%%%%%%
\nn & \Pi_{\rm inter} (q\,\bs{\hat y}, \omega) = \Pi_{\rm inter} (q\,\bs{\hat x}, \omega)\,.
\end{align}
The first logarithm in each channel is due to the transitions $0 \to 2$, and the second one in the $\bs{\hat x}$ channel is due to the transitions $1 \to 2$. The real part of the dielectric function is
%%%%%%%%%%%%%
\begin{align}
\label{eq:epshopf}
& \Re \varepsilon^{\rm RPA} (q\,\bs{\hat z}, \omega) 
= 1 - \frac{\omega_p^2}{\omega^2} + \frac{2\,g\,e^2}{3\,\pi\,v_0}\,\ln \!\left( \frac{v_0^2\,\Lambda^2}{\left| \mu^2 - \omega^2 \right| } \right)\,, \nn
%%%%%%%
& \Re \varepsilon^{\rm RPA} (q\,\bs{\hat x}, \omega) 
= 1 - \frac{\omega_p^2}{\omega^2} + \frac{g\,e^2}{6\,\pi\,v_0} \left[ 2\,
\ln \!\left( \frac{v_0^2\,\Lambda^2}{\left| \mu^2 - \omega^2 \right| } \right) 
+ \frac{1}{2}\,\ln \! 
\left(  \frac{4\,v_0^2\,\Lambda^2}{\left| 4\,\mu^2 - \omega^2 \right|} \right )\right], \nn
%%%%%%%%%%
 & \Re \varepsilon^{\rm RPA} (q\,\bs{\hat y}, \omega) =
  \Re \varepsilon^{\rm RPA} (q\,\bs{\hat x}, \omega)\,.
\end{align}
%%%%%%%%%%%
In the imaginary part, the delta function fixes $k = \omega/(c_p\,v_0)$, which lies in the unblocked region for $\omega > c_p\,\mu$. The imaginary part is independent of $\omega$ above the onsets,
\begin{align}
\label{eq:imhopf}
& \Im \varepsilon^{\rm RPA} (q\,\bs{\hat z}, \omega) = 
\frac{2\,g\,e^2}{3\,v_0}\,\Theta (\omega - \mu)\,, \quad
\Im \varepsilon^{\rm RPA} (q\,\bs{\hat x}, \omega) 
= \frac{g\,e^2}{3\,v_0}\,\Theta (\omega - \mu) + \frac{g\,e^2}{12\,v_0}\,\Theta (\omega - 2\,\mu)\,, \nn &
\Im \varepsilon^{\rm RPA} (q\,\bs{\hat y}, \omega) = 
\Im \varepsilon^{\rm RPA} (q\,\bs{\hat x}, \omega)\,.
\end{align}
These reduce to Eq.~\eqref{eq:intrhopf} for $\mu \to 0$.

We write $\epsilon_{b,i} (\omega)$ for the part of $\Re \varepsilon^{\rm RPA}$ along $\bs{\hat e}_i$ that remains after the Drude term of Eq.~\eqref{eq:epshopf} is removed. Its static limit $\omega \to 0$ gives the background dielectric constants
\begin{align}
\label{eq:ebhopf}
\epsilon_{b,z} = 1 + \frac{4\,g\,e^2}{3\,\pi\,v_0}\,\ln \! \left(  \frac{v_0\,\Lambda}{\mu} \right )\,, \quad
\epsilon_{b,x} = \epsilon_{b,y} = 1 + \frac{5\,g\,e^2}{6\,\pi\,v_0}\,
\ln \! \left(  \frac{v_0\,\Lambda}{\mu} \right )\,.
\end{align}
The dependence on $\Lambda$ is logarithmic and, as for the WSM, only the dependence on $\mu$ is independent of the regularisation. The background is anisotropic, with the ratio $8/5$ of the coefficients of the logarithm along $\bs{\hat z}$ and along $\bs{\hat x}$ or $\bs{\hat y}$. This is the ratio of the imaginary parts at the intrinsic point.

For $\omega < \omega_{\rm th}$, both $\epsilon_{b,i} (\omega)$ and the Drude term $-\,\omega_p^2/\omega^2$ increase with $\omega$. The real part is therefore monotonic. It tends to $-\,\infty$ for $\omega \to 0$ and to $+\,\infty$ for $\omega \to \mu^-$, and it has a unique zero below the onset in each channel for every coupling. The static screened frequencies
\begin{align}
\label{eq:wpscrhopf}
\tilde\omega_{p,i}^2 = \frac{\omega_p^2}{\epsilon_{b,i}}
\end{align}
are upper bounds of these zeros, since $\epsilon_{b,i} (\omega) \geq \epsilon_{b,i}$. At the zero $\omega_*$, the relation $\omega_p^2/\omega_*^2 = \epsilon_{b,i} (\omega_*)$ holds, and hence $\omega_* \leq \tilde\omega_{p,i}$. Since the zero also lies below the onset, $\omega_* < \min (\tilde\omega_{p,i}, \mu)$, and the plasmon is undamped. The static screened frequencies are anisotropic, with $\tilde\omega_{p,z}^2/\tilde\omega_{p,x}^2 = \tilde\omega_{p,z}^2/\tilde\omega_{p,y}^2 = \epsilon_{b,x}/\epsilon_{b,z}$. This ratio decreases monotonically with the logarithm of $v_0\,\Lambda/\mu$ from unity towards $5/8$. The long-wavelength plasmons of the HSM are thus anisotropic, although the FS and the bare frequencies are isotropic, in line with the finding of Ref.~\cite{hopf-plasmon}.

%%%%%%%%%%%%%%%%%%%%%%%%%5
\section{Comparison across different systems}
\label{seccompare}

In this section, we compare the results of Secs.~\ref{secuntilted}--\ref{secaxial} with those of Ref.~\cite{ips-nr-plasmon}. That work treats the PTNR, its gapped counterpart, the GNR, and the WSMs. The VNR is a distinct system, with a Hamiltonian, a pseudospin texture, and a distribution of the Berry curvature that differ from those of the PTNR and of the GNR. Its linearised spectrum nevertheless coincides with that of the PTNR, whereas the mass gap $2\,\Delta$ of the GNR changes the spectrum itself. The comparison therefore shows how much the pseudospin texture matters for the collective modes when the linearised bands are the same. We first list which results coincide with those of the PTNR in Sec.~\ref{secinherit}, and then describe how the pseudospin texture of the VNR makes the others differ in Sec.~\ref{secvortexeffects}. The VNR is contrasted with the GNR in Sec.~\ref{secgnr} and with the WSM in Sec.~\ref{secwsm}. The last subsection, Sec.~\ref{sechopfvnr}, compares the HSM of Sec.~\ref{sechopf} with the VNR and the WSM. Tables~\ref{tab:vnr-ptnr}, \ref{tab:vnr-ptnr-gnr}, \ref{tab:untilted-all}, and~\ref{tab:vnr-wsm-tilted} collect the main entries for all the systems.

%%%%%%%%%%%%%%%%%%%%%%%%%%%
\subsection{VNR versus PTNR: coinciding results}
\label{secinherit}

The VNR Hamiltonian of Eq.~\eqref{eq:hamvnr} is different from that of the PTNR. Its pseudospin vector has three nonvanishing components and rotates with the toroidal angle $\phi$, whereas the PTNR pseudospin lies in a fixed plane. The linearised eigenvalues $\varepsilon_s$, the band velocities $\bs{V}^{(s)}$, and the integration measure nevertheless coincide in the two systems. Every quantity that is fixed by these three ingredients is therefore the same for the VNR and the PTNR, for each of the three configurations. This holds for the FS, which is a ring torus without tilt, a ring-cyclide or horn-cyclide for the in-plane tilt, and a deformed torus for the axial tilt. It also holds for the DOS, the Drude weights, the bare plasmon frequencies, and their anisotropy ratios. Examples are $\omega_{p,z}/\omega_{p,x} = \sqrt{2}$ without tilt, $\omega_{p,z}^2/\omega_{p,x}^2 = 2\,(1 + \eta_x^2/16)$ for the in-plane tilt in the ring-cyclide regime, and $\omega_{p,z}^2/\omega_{p,x}^2 = 2\,\sqrt{1 - \eta_z^2}$ for the axial tilt. The interband onsets $2\,\mu$, $2\,(\mu - |\tilde{\eta}|)/(1 + |\eta_x|)$, and $2\,\mu/(1 + |\eta_z|)$ are the same as well. The same is true of the static background dielectric constants that follow from the first term of the overlap factor, which are independent of $\eta_z$ and depend on $\eta_x$ only through $\tilde{\eta}$.

The overlap factor of Eq.~\eqref{eq:fsplit} also takes the PTNR form for $\bs{q} = q\,\bs{\hat z}$. The whole $\bs{\hat z}$ channel is therefore the same in the two systems, namely the polarisability, the dielectric function, and the damping. This holds in every regime of Secs.~\ref{secuntilted}--\ref{secaxial} and is exact in the linearised model. The qualitative conclusions of Ref.~\cite{ips-nr-plasmon} for the PTNR therefore apply to the VNR as well. The intrinsic system supports no undamped plasmon. An in-plane tilt at $\mu = 0$ produces the symmetric horn-cyclide, which has a finite Drude weight, but the resulting plasmon is always damped because the interband continuum has no finite threshold. In the doped untilted system, the zero of $\Re \varepsilon^{\rm RPA}$ lies below the onset at every coupling. For the axial tilt, the onset decreases linearly with $|\eta_z|$ while the plasmon frequencies change only at order $\eta_z^2$, and the screened plasmon enters the interband continuum at large tilt.

For $\mu = 0.1\,v_0\,k_0$ and $0.2\,v_0\,k_0$, this entry takes place at a smaller tilt along $\bs{\hat z}$ than along $\bs{\hat x}$ and $\bs{\hat y}$. The large tilt used in Ref.~\cite{ips-nr-plasmon} lies beyond both. This statement inherits the caveat of Sec.~\ref{secaxial} on the validity of the expansions at large tilt.

The intrinsic point with a small in-plane tilt provides a further check. The damping of Eq.~\eqref{eq:imintrxvnr} agrees with the corresponding PTNR expressions of Ref.~\cite{ips-nr-plasmon} in the $\bs{\hat z}$ channel, which has no dependence on $\eta_x$, and in the $\bs{\hat x}$ channel, which carries the factor $1 + v_0\,\eta_x\,q/\omega$. The PTNR derivation obtains the frequency shift $\Omega = \omega - v_0\,\eta_x\,q$ directly from the tilt contribution to the transition energy. In the VNR, the same shift arises as the sum of a local part and a part due to the displacement along the ring, as shown in Appendix~\ref{app:vnrxintr}. The $\bs{\hat y}$ channel is independent of $\eta_x$, and the vortex term of Sec.~\ref{secxintr} is the only difference. Below the continuum, the pocket correction of Eq.~\eqref{eq:dpipocketmainvnr} also coincides with the PTNR result of Ref.~\cite{ips-nr-plasmon}, because the $\bs{\hat z}$ channel is the same in the two systems. The correction is of relative order $(\tilde{\eta}/v_0\,q)^3$ and leaves the absence of a zero below the continuum unchanged in both systems. This result is parametric. It requires $\eta_x\,k_0 \ll q \ll k_0$ and holds to leading order in $\eta_x$. At the parameters of Sec.~\ref{secvalidity}, the ratio $\tilde{\eta}/(v_0\,k_0)$ is not small. The window is therefore empty, and the result is not a prediction for those parameters.

%%%%%%%%%%%%%%%%%%%%%%%%%%%
\subsection{VNR versus PTNR: effects of the pseudospin texture}
\label{secvortexeffects}

The pseudospin texture of the VNR differs from that of the PTNR through the toroidal-angle dependence of the overlap factor. This dependence matters only for wavevectors with an in-plane component. Table~\ref{tab:vnr-ptnr} compares the VNR results with those of the PTNR for the quantities discussed in this paper.

\begin{table*}[t!]
\centering
\small
\renewcommand{\arraystretch}{1.8}
\setlength{\tabcolsep}{6pt}
\begin{tabular}{|l|c|c|}
\hline
Quantity & Difference from the PTNR & Size \\
\hline
\makecell[l]{$\varepsilon_s$, $\bs{V}^{(s)}$, FS, DOS, \\ $D_{ij}$, $\omega_{p,i}$, $\omega_{\rm th}$} & none & exact in the linearised model \\
\hline
$\Pi (q\,\bs{\hat z}, \omega)$ for all $\bs{\eta}$ and $\mu$ & none & exact in the linearised model \\
\hline
\makecell[l]{$\Im \Pi (q\,\bs{\hat x}, \omega)$ and \\ $\Im \Pi (q\,\bs{\hat y}, \omega)$} & \makecell{extra contribution with the weights \\ $\sin^2\phi$ and $\cos^2\phi$} & \makecell{relative order \\ $(\omega/2\,v_0\,k_0)^2$} \\
\hline
\makecell[l]{in-plane background \\ dielectric constants \\ $\epsilon_{b,x}$, $\epsilon_{b,y}$} & \makecell{positive shift \\ $\delta\epsilon_{b} \simeq g\,e^2\,(\Lambda - \kappa_F)/(8\,v_0\,k_0)$, \\ with $\kappa_F \to \bar\kappa_F$ for the axial tilt} & \makecell{relative order $\kappa_F\,\Lambda/k_0^2$ \\ with respect to $\epsilon_{b,x} - 1$} \\
\hline
quantum metric $g_{ij}$ & \makecell{second term \\ $\sin^2\gamma\,\hat\phi_i\,\hat\phi_j/(4\,k_\perp^2)$} & relative order $(\kappa/k_0)^2$ \\
\hline
\makecell[l]{Berry curvature \\ for $k_z \neq 0$} & \makecell{$|\bs{\Omega}| = |\sin\gamma|/(2\,\kappa\,k_\perp)$ \\ instead of zero} & \makecell{zero flux through every \\ torus $\kappa = \text{const}$} \\
\hline
\end{tabular}
\caption{\label{tab:vnr-ptnr} Comparison of the VNR with the PTNR of Ref.~\cite{ips-nr-plasmon}. The VNR results are from Secs.~\ref{secuntilted}--\ref{secaxial}, and $\Lambda \lesssim k_0$ is the momentum scale at which the linearised Hamiltonian ceases to hold.}
\end{table*}

The absorption in the in-plane channels differs between the two systems in a way that is easily stated. Let $F_0$ and $F_2$ be the Pauli-blocking functions of Eq.~\eqref{eq:fnvnr}, which describe the unblocked arcs of the ring, and let $r_\omega = (\omega/2\,v_0\,k_0)^2$. For the in-plane tilt, Eqs.~\eqref{eq:imhornvnr} and~\eqref{eq:imdpixvnr} combine into
%%%%%%%%%%%%%%%%%%%%%%%%%%%%%%%
\begin{align}
\label{eq:mixvnr}
& \Im \Pi (q\,\bs{\hat x}, \omega) \simeq -\, \frac{g\,k_0\,q^2}{16\,\omega} 
\left[ F_2 + r_\omega\,(F_0 - F_2) \right], \quad
\Im \Pi (q\,\bs{\hat y}, \omega) \simeq -\, \frac{g\,k_0\,q^2}{16\,\omega} 
\left[ F_0 - F_2 + r_\omega\,F_2 \right],\nn &
%%%%%%%%%%
\Im \Pi (q\,\bs{\hat z}, \omega) = \frac{-\, g\,k_0\,q^2\,F_0} {16\,\omega}\,.
\end{align}
%%%%%%%%%%%%%%%%%%%%%%%%%%
Note that the last one is the same as for the PTNR. The second term of the overlap factor has the weights $\sin^2\phi$ and $\cos^2\phi$, which are interchanged relative to the PTNR weights. It therefore mixes the two in-plane channels. The difference between them, which the tilt creates, is reduced by the factor $1 - r_\omega$ with respect to the PTNR value $(2\,F_2 - F_0)$ in units of the common prefactor. Without tilt, $F_0 = 1$ and $F_2 = 1/2$, and the intrinsic damping anisotropy between $\bs{\hat z}$ and $\bs{\hat x}$ becomes $2/(1 + r_\omega)$ instead of the factor of two found for the PTNR. For the axial tilt, the in-plane absorption of the VNR is that of the PTNR multiplied by $1 + r_\omega$ at every tilt. In all cases the second term vanishes below the onset $\omega_{\rm th}$ of the doped system. The criterion that the zero of $\Re \varepsilon^{\rm RPA}$ lies below the onset is therefore not affected by the imaginary part. At $\mu = 0$ with an in-plane tilt, the second term has no threshold either, and the plasmon remains damped.

The real part differs through the shift $\delta\epsilon_{b}$ of the in-plane background dielectric constants. For $\omega < \omega_{\rm th}$, the integrand of Eq.~\eqref{eq:dpizvnr} is negative, and its magnitude grows with $\omega$. The shift is therefore positive and does not decrease with $\omega$. The static screened in-plane frequencies $\tilde\omega_{p,i}^2 = \omega_{p,i}^2/(\epsilon_{b,i} + \delta\epsilon_{b,i})$ of the VNR then lie below the PTNR values of Ref.~\cite{ips-nr-plasmon}. For the axial tilt, a zero of $\Re \varepsilon^{\rm RPA}$ below the onset exists if and only if $\Re \varepsilon^{\rm RPA}$ is positive at $\omega_{\rm th}$, since $\Re \varepsilon^{\rm RPA}$ is monotonic below the onset and tends to $-\infty$ for $\omega \to 0$. A positive shift keeps this condition satisfied at every tilt for which it holds in the PTNR. The in-plane plasmon of the VNR therefore does not enter the continuum at a smaller $|\eta_z|$ than that of the PTNR. The size of the displacement depends on $\Lambda$ and is not fixed within the linearised model. A numerical evaluation with a hard cut-off would not be controlled either. Moreover, Eq.~\eqref{eq:dpizvnr} requires $\kappa_F^{\max} < \Lambda \lesssim k_0$. Since $\kappa_F^{\max} = \mu/[v_0\,(1 - |\eta_z|)]$ grows with the tilt, this restricts the tilt to $|\eta_z| < 1 - \mu/(v_0\,\Lambda)$. We therefore regard the results that do not depend on $\Lambda$ as the controlled predictions for the VNR. The $\bs{\hat z}$ channel carries no such uncertainty.

The geometric quantities differ more visibly than the response. In the PTNR, the BC vanishes away from the ring and the quantum metric has only the first term of Eq.~\eqref{eq:qmetric}. In the VNR, the winding of the pseudospin gives rise to the second term of the metric and to a finite BC. Its radial direction in the poloidal cross-section and its zero net flux make it a dipole-like source~\cite{ips-dipole-vnr}. The RPA plasmon frequencies depend on the pseudospin texture only through the interband screening, which differs between the VNR and the PTNR at relative order $\kappa_F\,\Lambda/k_0^2$. The geometry of the eigenstates therefore differs between the two systems much more than the plasmons do.

%%%%%%%%%%%%%%%%%%%%%%%%%%%
\subsection{VNR versus GNR}
\label{secgnr}

Table~\ref{tab:vnr-ptnr-gnr} compares the VNR with the PTNR and the GNR of Ref.~\cite{ips-nr-plasmon}. Its last column lists the corresponding entries of the HSM, which are derived in Sec.~\ref{sechopf} and compared with the VNR in Sec.~\ref{sechopfvnr}.

\begin{table*}[t!]
\centering
\small
\renewcommand{\arraystretch}{1.8}
\setlength{\tabcolsep}{6pt}
\begin{tabular}{|l|c|c|c|c|}
\hline
Quantity & VNR & PTNR & GNR & HSM \\
\hline
$\varepsilon_{1,2}$ for $\bs{\eta} = 0$ & $\mp\, v_0\,\kappa$ & $\mp\, v_0\,\kappa$ & $\mp \sqrt{v_0^2\,\kappa^2 + \Delta^2}$ & \makecell{$\mp\, v_0\,k$ \\ (and a flat band at $0$)} \\
\hline
\makecell[l]{Berry curvature \\ for $k_z \neq 0$} & \makecell{radial, $\dfrac{|\sin\gamma|}{2\,\kappa\,k_\perp}$, \\ zero net flux} & zero & finite & \makecell{radial, $\dfrac{|\cos\vartheta|}{k^2}$, \\ zero net flux} \\
\hline
$D_{zz}$ for $\bs{\eta} = 0$ & $\dfrac{g\,e^2\,k_0\,\mu}{4}$ & $\dfrac{g\,e^2\,k_0\,\mu}{4}$ & $\dfrac{g\,e^2\,k_0\,(\mu^2 - \Delta^2)}{4\,\mu}$ & $\dfrac{g\,e^2\,\mu^2}{6\,\pi\,v_0}$ \\
\hline
$D_{xx} = D_{yy}$ for $\bs{\eta} = 0$ & $\dfrac{g\,e^2\,k_0\,\mu}{8}$ & $\dfrac{g\,e^2\,k_0\,\mu}{8}$ & $\dfrac{g\,e^2\,k_0\,(\mu^2 - \Delta^2)}{8\,\mu}$ & $\dfrac{g\,e^2\,\mu^2}{6\,\pi\,v_0}$ \\
\hline
$\omega_{p,z}^2/\omega_{p,x}^2$ for $\bs{\eta} = 0$ & $2$ & $2$ & $2$ & $1$ \\
\hline
$\omega_{\rm th}$ for $\bs{\eta} = 0$ ($\mu > \Delta$) & $2\,\mu$ & $2\,\mu$ & $2\,\mu$ & $\mu$ \\
\hline
$\omega_{p,z}^2/\omega_{p,x}^2$ for the $k_z$-tilt & $2\,\sqrt{1 - \eta_z^2}$ & $2\,\sqrt{1 - \eta_z^2}$ & $\dfrac{2\,\nu\,\mu}{M}$ & -- \\
\hline
$\omega_{\rm th}$ for the $k_z$-tilt ($\mu \ge \Delta$) & $\dfrac{2\,\mu}{1 + |\eta_z|}$ & $\dfrac{2\,\mu}{1 + |\eta_z|}$ & $\dfrac{2\,(\mu - |\eta_z|\,\sqrt{K})}{\nu^2}$ & -- \\
\hline
\makecell[l]{$\omega_{\rm th}$ for the $k_x$-tilt, \\ ring-cyclide regime} & $\dfrac{2\,(\mu - |\tilde{\eta}|)}{1 + |\eta_x|}$ & $\dfrac{2\,(\mu - |\tilde{\eta}|)}{1 + |\eta_x|}$ & \makecell{$2\,\Delta$ \\ (gapped regimes)} & -- \\
\hline
$\Pi (q\,\bs{\hat z}, \omega)$ & same as PTNR & gapless local cones & gapped local cones & \makecell{Dirac cone \\ and flat band} \\
\hline
\makecell[l]{plasmon at $\mu = 0$, \\ $\bs{\eta} = 0$ or $k_z$-tilt} & damped or absent & damped or absent & \makecell{absent \\ (insulating)} & damped or absent \\
\hline
\makecell[l]{plasmon at $\mu = 0$, \\ $k_x$-tilt} & always damped & always damped & \makecell{undamped \\ (twin-crescent regime)} & -- \\
\hline
\end{tabular}
\caption{\label{tab:vnr-ptnr-gnr} Comparison of the VNR with the PTNR and the GNR of Ref.~\cite{ips-nr-plasmon} and with the HSM of Sec.~\ref{sechopf}. The VNR results are from Secs.~\ref{secuntilted}--\ref{secaxial}. For the GNR, $\Delta$ is the mass gap, $\nu = \sqrt{1 - \eta_z^2}$, $K = \mu^2 - \Delta^2\,\nu^2$, and $M = \sqrt{\mu^2 + \eta_z^2\,\Delta^2}$. For the HSM, $k = |\bs{k}|$ and $\vartheta$ is the polar angle of $\bs{k}$, and a dash marks the entries for which the HSM is not studied, since no tilt is included for it. The entry ``damped or absent'' means that the intrinsic system has no undamped plasmon.}
\end{table*}

In the GNR of Ref.~\cite{ips-nr-plasmon}, the mass term gives the pseudospin a constant out-of-plane component. This opens the gap $2\,\Delta$ at the ring and produces a finite BC in the bulk~\cite{ips-magnus, ips-nlsm-ph, ips-gnr-strain}. Its local cones are those of gapped graphene~\cite{pyatkovskiy-gapped-graphene}. The gap, unlike the pseudospin winding of the VNR, enters the energies. For $\bs{\eta} = 0$, the Drude weights follow from those of the PTNR by the replacement $\mu \to (\mu^2 - \Delta^2)/\mu$, the FS acquires several topologically distinct regimes for the in-plane tilt, and the interband onset is set by $2\,\Delta$ in the gapped regimes. None of these features has a VNR counterpart, because the linearised VNR spectrum is that of the gapless PTNR. In particular, the fully-insulating, twin-crescent, and crescent-cyclide regimes of the in-plane tilted GNR have no analogue here. The VNR has only the ring-cyclide and horn-cyclide regimes, as does the PTNR.

The GNR and the PTNR are related in the limit of vanishing gap, because the GNR Hamiltonian reduces to the PTNR one for $\Delta \to 0$. The VNR has no such connection with either system, since no parameter turns the PTNR Hamiltonian into that of the VNR. Nevertheless, for $\Delta \to 0$, the GNR results coincide with those of Secs.~\ref{secuntilted}--\ref{secaxial} for $\bs{q} = q\,\bs{\hat z}$, because the latter coincide with the PTNR ones. An example is $D_{zz} = g\,e^2\,k_0\,(\mu^2 - \Delta^2)/(4\,\mu) \to g\,e^2\,k_0\,\mu/4$. The interband screening provides a second point of contact. In the GNR, the static background dielectric constant is bounded by its insulating value, which diverges as $\Delta \to 0$. In the gapless VNR and PTNR, the divergence is instead controlled by the doping, as seen in Eq.~\eqref{eq:ebvnr}, and the interband contribution grows as $1/\mu$.

The VNR and the GNR thus differ from the PTNR in different ways. The gap of the GNR changes the spectrum and reshapes the FS and the Drude weights, for example by the factor $1 - \Delta^2/\mu^2$ for $\bs{\eta} = 0$. The VNR has the same linearised spectrum as the PTNR but a different pseudospin texture. Its response differs from that of the PTNR only through the eigenstates, at relative order $(\kappa/k_0)^2$ in the absorption.

\subsection{VNR versus WSM}
\label{secwsm}

We now contrast the VNR with the WSM, which is the simplest 3d semimetal with a point-like band-touching. The WSM node is a point-like BC-monopole, whereas the VNR carries a ring of BC-dipoles. The long-wavelength response of an untilted and of a type-I tilted WSM node was derived with the same local-FS method in Ref.~\cite{ips-nr-plasmon}. The node is described by $H_{\rm W} = v_0\,\bs{k} \cdot \bs{\sigma} + v_0\,\bs{\eta} \cdot \bs{k}\;\mathbb{I}_{2\times 2}$ with $|\bs{\eta}| = \eta < 1$, and its FS is an ellipsoid with the node at a focus. We quote the results that are needed here for the same Fermi velocity $v_0$ and the same degeneracy $g$. Since the quantities listed below coincide for the VNR and the PTNR, the VNR entries are those of Secs.~\ref{secuntilted} and~\ref{secaxial}. Table~\ref{tab:untilted-all} collects the untilted results, together with those of the HSM of Sec.~\ref{sechopf}, and Table~\ref{tab:vnr-wsm-tilted} collects the tilted results.

\begin{table*}[t!]
\centering
\small
\renewcommand{\arraystretch}{1.8}
\setlength{\tabcolsep}{6pt}
\begin{tabular}{|l|c|c|c|}
\hline
Quantity & Untilted VNR & Untilted WSM & Untilted HSM \\
\hline
FS & torus & sphere & sphere \\
\hline
DOS $\rho$ & $\dfrac{g\,k_0\,\mu}{2\,\pi\,v_0^2}$ & $\dfrac{g\,\mu^2}{2\,\pi^2\,v_0^3}$ & $\dfrac{g\,\mu^2}{2\,\pi^2\,v_0^3}$ \\
\hline
Density $\tilde n$ & $\dfrac{g\,k_0\,\mu^2}{4\,\pi\,v_0^2}$ & $\dfrac{g\,\mu^3}{6\,\pi^2\,v_0^3}$ & $\dfrac{g\,\mu^3}{6\,\pi^2\,v_0^3}$ \\
\hline
Drude weight $D_{ij}$ & \makecell{$\dfrac{g\,e^2\,k_0\,\mu}{4}\,\delta_{iz}\,\delta_{jz}$ \\ $+ \dfrac{g\,e^2\,k_0\,\mu}{8}\,(\delta_{ix}\,\delta_{jx} + \delta_{iy}\,\delta_{jy})$} & $\dfrac{g\,e^2\,\mu^2}{6\,\pi\,v_0}\,\delta_{ij}$ & $\dfrac{g\,e^2\,\mu^2}{6\,\pi\,v_0}\,\delta_{ij}$ \\
\hline
$\omega_{p,i}^2 = 4\,D_{ii}$ & \makecell{$g\,e^2\,k_0\,\mu$ ($z$) \\ half of it ($x, y$)} & \makecell{$\dfrac{2\,g\,e^2\,\mu^2}{3\,\pi\,v_0}$ \\ ($x, y, z$)} & \makecell{$\dfrac{2\,g\,e^2\,\mu^2}{3\,\pi\,v_0}$ \\ ($x, y, z$)} \\
\hline
$\omega_p \propto$ & ${\tilde n}^{1/4}$ & ${\tilde n}^{1/3}$ & ${\tilde n}^{1/3}$ \\
\hline
$\omega_{\rm th}$ & $2\,\mu$ & $2\,\mu$ & $\mu$ \\
\hline
$\omega_p^2/\omega_{\rm th}^2$ & \makecell{$\dfrac{g\,e^2\,k_0}{4\,\mu}$ ($z$) \\ $\dfrac{g\,e^2\,k_0}{8\,\mu}$ ($x, y$)} & $\dfrac{g\,e^2}{6\,\pi\,v_0}$ & $\dfrac{2\,g\,e^2}{3\,\pi\,v_0}$ \\
\hline
$\epsilon_b - 1$ & \makecell{$\dfrac{g\,e^2\,k_0}{4\,\mu}$ ($z$) \\ $\dfrac{g\,e^2\,k_0}{8\,\mu}$ ($x, y$)} & $\dfrac{g\,e^2}{3\,\pi\,v_0} \ln \dfrac{v_0\,\Lambda}{\mu}$ & \makecell{$\dfrac{4\,g\,e^2}{3\,\pi\,v_0} \ln \dfrac{v_0\,\Lambda}{\mu}$ ($z$) \\ $\dfrac{5\,g\,e^2}{6\,\pi\,v_0} \ln \dfrac{v_0\,\Lambda}{\mu}$ ($x, y$)} \\
\hline
$\Im \varepsilon^{\rm RPA}$ above the onset & $\propto 1/\omega$ & independent of $\omega$ & independent of $\omega$ \\
\hline
Anisotropy $\omega_{p,z}^2/\omega_{p,x}^2$ & $2$ & $1$ & $1$ \\
\hline
\makecell[l]{Origin of the anisotropy \\ of the screened plasmons} & \makecell{FS (bare) \\ and interband part} & none & interband part only \\
\hline
\end{tabular}
\caption{\label{tab:untilted-all} Comparison of the long-wavelength response of the untilted VNR, WSM, and HSM with the same Fermi velocity $v_0$. The density is $\tilde n = \int_0^\mu \rho\,d\mu'$, the Drude weights are $D_{ij} = \pi\,e^2\,\rho\,\langle V_i\,V_j \rangle$, and $\Lambda$ is the momentum cut-off of the linearised description. The VNR results are from Sec.~\ref{secuntilted}, the WSM results are from Ref.~\cite{ips-nr-plasmon}, and the HSM results are derived in Sec.~\ref{sechopf}.}
\end{table*}

\begin{table*}[t!]
\centering
\small
\renewcommand{\arraystretch}{1.8}
\setlength{\tabcolsep}{6pt}
\begin{tabular}{|l|c|c|}
\hline
Quantity & Axially-tilted VNR & Tilted WSM \\
\hline
FS & deformed torus & ellipsoid \\
\hline
DOS $\rho$ & $\dfrac{g\,k_0\,\mu}{2\,\pi\,v_0^2 \left( 1 - \eta_z^2 \right)^{3/2}}$ & $\dfrac{g\,\mu^2}{2\,\pi^2\,v_0^3 \left( 1 - \eta^2 \right)^2}$ \\
\hline
$\omega_p^2$ & \makecell{$\dfrac{2\,g\,e^2\,k_0\,\mu}{\eta_z^2} \left( 1 - \sqrt{1 - \eta_z^2} \right)$ ($z$) \\ $\dfrac{g\,e^2\,k_0\,\mu}{\eta_z^2} \left( \dfrac{1}{\sqrt{1 - \eta_z^2}} - 1 \right)$ ($x, y$)} & \makecell{$\dfrac{2\,g\,e^2\,\mu^2}{\pi\,v_0}\,\dfrac{\operatorname{arctanh} \eta - \eta}{\eta^3}$ ($\parallel$) \\ $\dfrac{g\,e^2\,\mu^2}{\pi\,v_0\,\eta^3} \left( \dfrac{\eta}{1 - \eta^2} - \operatorname{arctanh} \eta \right)$ ($\perp$)} \\
\hline
$\omega_{\rm th}$ & $\dfrac{2\,\mu}{1 + |\eta_z|}$ & $\dfrac{2\,\mu}{1 + \eta}$ \\
\hline
$\epsilon_b - 1$ & \makecell{$\dfrac{g\,e^2\,k_0}{4\,\mu}$ ($z$) \\ $\dfrac{g\,e^2\,k_0}{8\,\mu}$ ($x, y$)} & \makecell{$\dfrac{g\,e^2}{3\,\pi\,v_0} \ln\!\left( \dfrac{v_0\,\Lambda}{\mu} \right)$ \\ $+ \dfrac{g\,e^2}{4\,\pi\,v_0}\,\mathcal{B}_i (\eta)$} \\
\hline
\makecell[l]{Anisotropy $\omega_{p,z}^2/\omega_{p,x}^2$ \\ or $\omega_{p,\parallel}^2/\omega_{p,\perp}^2$} & $2\,\sqrt{1 - \eta_z^2}$ & $1 - \dfrac{3\,\eta^2}{5} + \mathcal{O} (\eta^4)$ \\
\hline
\end{tabular}
\caption{\label{tab:vnr-wsm-tilted} Comparison of the long-wavelength response of the axially-tilted VNR and the tilted WSM with the same Fermi velocity $v_0$. The symbols $\parallel$ and $\perp$ refer to the tilt direction of the WSM. The functions in the WSM background are $\mathcal{B}_\parallel = \int_{-1}^{1} dc\,(1 - c^2) \ln (1 + \eta\,c)$ and $\mathcal{B}_\perp = \tfrac{1}{2} \int_{-1}^{1} dc\,(1 + c^2) \ln (1 + \eta\,c)$, and $\Lambda$ is the momentum cut-off of the linearised description. The VNR results are from Sec.~\ref{secaxial}, and the WSM results are from Ref.~\cite{ips-nr-plasmon}.}
\end{table*}

We first compare the untilted systems. For both, the Drude weight can be written as $D_{ii} = \pi\,e^2\,\rho\,\langle V_i^2 \rangle$, where $\langle V_i^2 \rangle$ is the average of the squared Fermi velocity over the FS. In the WSM, the velocity $v_0\,\bs{\hat k}$ is distributed uniformly over the sphere, and hence $\langle V_i^2 \rangle = v_0^2/3$ for every $i$. In the VNR at $\mu \ll v_0\,k_0$, the weight is uniform in $\gamma$ and $\phi$ instead of in the solid angle. This gives $\langle V_z^2 \rangle = v_0^2/2$ and $\langle V_x^2 \rangle = \langle V_y^2 \rangle = v_0^2/4$. In both cases, the three averages add up to $v_0^2$, because the speed on the FS is constant. The two untilted responses therefore differ for two reasons. The first is the DOS, which is linear in $\mu$ for the VNR and quadratic in $\mu$ for the WSM, since the ring supplies a factor $k_0$ to the phase space. The second is the anisotropic velocity distribution of the VNR, which is absent in the WSM. Consequently, $\omega_p \propto \tilde n^{1/4}$ for the VNR and $\omega_p \propto \tilde n^{1/3}$ for the WSM. Both $\omega_p$ and $\omega_{\rm th}$ are proportional to $\mu$ in the WSM, and the ratio $\omega_p^2/\omega_{\rm th}^2$ is then independent of $\mu$ and fixed by the coupling alone. For the VNR, the ratio $g\,e^2\,k_0/(4\,\mu)$ along $\bs{\hat z}$ grows as $\mu \to 0$.

The background screening differs as well. For the VNR, it is finite and proportional to $1/\mu$, because the integral $\int d^2\kappa/\kappa^3$ converges in the ultraviolet. For the WSM, it contains a cut-off-dependent logarithm, and only the dependence on $\mu$ is independent of the regularisation. As a result, the screened frequency $\tilde\omega_{p,z}$ of the VNR approaches the onset $2\,\mu$ from below for $g\,e^2\,k_0 \gg \mu$, instead of growing with the coupling.

The tilt modifies the two systems in a qualitatively similar way. In both, the interband onset decreases linearly with the tilt, whereas the Drude weights change only at order $\eta^2$, and their relative growth along the tilt is smaller than across it. The coefficients of the leading correction differ. For the WSM, the Drude weights are enhanced by the factors $1 + 3\,\eta^2/5$ along the tilt and $1 + 6\,\eta^2/5$ across it. For the axially-tilted VNR, the squared frequencies of Eq.~\eqref{eq:wpzvnr} are enhanced by $1 + \eta_z^2/4$ along $\bs{\hat z}$ and by $1 + 3\,\eta_z^2/4$ in the plane, and the coefficients of the in-plane tilt of Eq.~\eqref{eq:wpxvnr} are $9/8$, $17/16$, and $27/16$. The ring thus has two inequivalent tilt orientations with different responses, whereas the isotropic WSM depends on $\eta$ alone.

The difference is sharper near the type-II transition. The DOS diverges as $\left( 1 - \eta^2 \right)^{-2}$ for the WSM and as $\left( 1 - \eta_z^2 \right)^{-3/2}$ for the axially-tilted VNR. The WSM Drude weights diverge there, logarithmically for $D_\parallel$ and as $\left( 1 - \eta^2 \right)^{-1}$ for $D_\perp$. In the axially-tilted VNR, $D_{zz}$ stays finite and $D_{xx}$ diverges only as $\left( 1 - \eta_z^2 \right)^{-1/2}$. The measure $k_0\,\kappa_F$ of the ring carries one power of the Fermi momentum fewer than the measure $k_F^2$ of the WSM, which softens the divergence at the tip of the FS. Finally, a type-I WSM has no FS at $\mu = 0$ for any tilt and hence no Drude weight. In contrast, the in-plane-tilted VNR hosts pockets along the ring at the intrinsic point (Sec.~\ref{secxintr}), with a finite Drude weight and a damped plasmon. The pseudospin texture of the VNR does not alter any of these conclusions, because they follow from the spectrum, the velocities, and the integration measure.

\subsection{HSM versus VNR and WSM}
\label{sechopfvnr}

The HSM of Sec.~\ref{sechopf} has a point-like BC-dipole, in contrast to the ring of BC-dipoles of the VNR and the point-like BC-monopole of the WSM. Table~\ref{tab:untilted-all} compares the three untilted systems, and the last column of Table~\ref{tab:vnr-ptnr-gnr} lists the HSM entries for the spectrum, the BC, and the Drude weights.

The VNR and the HSM differ in the geometry of the FS, which is a torus for the former and a sphere for the latter. The DOS is linear in $\mu$ for the VNR and quadratic in $\mu$ for the HSM, and hence $\omega_p \propto \tilde n^{1/4}$ for the VNR and $\omega_p \propto \tilde n^{1/3}$ for the HSM, as for the WSM. The anisotropy of the bare plasmon frequencies is a property of the toroidal FS of the VNR and is absent in the HSM. The interband parts of the two systems are anisotropic for different reasons. In the VNR, the anisotropy comes from the weights of the local Dirac cones along the ring, and the background is finite and proportional to $1/\mu$. In the HSM, it comes from the anisotropic quantum metric of the transitions between the flat band and the Dirac cone, and the background contains a logarithm in $v_0\,\Lambda/\mu$. The flat band also lowers the onset to $\mu$, since the transitions from the flat band to the upper band cost only $v_0\,k$. These transitions give a constant imaginary part of the dielectric function, whereas the transitions between the two Dirac bands of the VNR give $\Im \varepsilon^{\rm RPA} \propto 1/\omega$. In both systems, the real part of the dielectric function tends to $+\,\infty$ at the onset, and a zero below the onset exists at every coupling.

The HSM and the WSM have the same FS, DOS, and Drude weights, and hence the same bare plasmon frequencies. The flat band of the HSM changes only the interband part. It lowers the onset from $2\,\mu$ to $\mu$. It also enhances the coefficient of the logarithm in the background constant by the factors $4$ along $\bs{\hat z}$ and $5/2$ in the plane, relative to the isotropic coefficient of the WSM, and it makes the background anisotropic.

%%%%%%%%%%%%%%%%%%%%%%%%%%%%%%%%%%%%%%
\section{Summary and outlook}
\label{secsum}

We have studied the plasmons of the VNR without tilt and with in-plane and axial band tilts, and we have compared them with those of the PTNR, the GNR, the WSM, and the HSM. The main outcome is that the vortex texture of the VNR leaves the plasmons of the nodal ring almost untouched, although it changes the quantum geometry markedly. The reason is structural. After linearisation, the VNR and the PTNR~\cite{ips-nr-plasmon} share their eigenvalues, band velocities, and integration measures, and at low doping the toroidal FS of both maps onto a continuum of 2d Dirac cones. The noninteracting polarisability then follows from the graphene result~\cite{hwang-dassarma-graphene}, and the long-wavelength limit of the dielectric function gives the RPA plasmon frequencies. Everything fixed by these ingredients is common to the two systems. This covers the FS, the DOS, the Drude weights, the bare plasmon frequencies, and the interband onsets in every configuration. It also covers the complete response to a wavevector along the axial direction in every regime, because the momentum transfer does not shift the toroidal angle. The mass gap of the GNR, in contrast, changes the spectrum itself. It reshapes the FS and the Drude weights, and it sets the interband onset in the gapped regimes.

The differences appear for in-plane wavevectors. The pseudospin vortex enters through the overlap factor and the quantum metric, both of which depend on the toroidal angle, and it produces a finite BC at all $k_z \neq 0$. The dependence on the toroidal angle adds to the polarisability a contribution with no PTNR counterpart. Its imaginary part is parametrically small and vanishes below the interband onset of the doped system. Its real part is a frequency-independent upward shift of the in-plane background dielectric constants. This shift depends on the cut-off of the linearised description and stays small compared with the interband screening at low doping. The bare plasmon frequencies are therefore unchanged, and the in-plane screening can only be stronger than in the PTNR. Under an in-plane tilt, the contribution also mixes the two in-plane channels, because its angular weights are interchanged relative to those of the PTNR.

Consequently, the qualitative picture established for the PTNR carries over. Without doping, the dielectric function stays above unity below the interband continuum, and no undamped plasmon exists. An in-plane tilt creates electron- and hole-pockets even at zero doping. The resulting symmetric horn-cyclide has a finite Drude weight and hence a zero of the real part of the dielectric function. The plasmon is nevertheless always damped, because the interband continuum has no finite threshold. In the doped untilted system, the zero of the real part lies below the interband onset at every coupling. Under an axial tilt, the onset decreases with the tilt while the plasmon frequencies change much more slowly, and the screened plasmon is eventually pushed into the interband continuum. The upward shift of the in-plane background dielectric constants delays this entry rather than advancing it. The interband screening is essential throughout, since the background dielectric constants are large at low doping and lower the bare frequencies substantially. Taken together, these comparisons lead to a simple verdict. A plasmon measurement at low doping is unlikely to distinguish a VNR from a PTNR, because the two systems share their long-wavelength plasmon response to a remarkable degree. Probes of the quantum geometry, such as the magnetoelectric conductivity~\cite{ips-dipole-vnr, ips-firdous-vnr}, are better suited for this purpose.

The contrast with the WSM separates the features that come from the toroidal FS from those that come from the pseudospin texture. The WSM has point-like BC-monopoles, whereas the VNR has a ring of BC-dipoles. The dependence of the plasmon frequency on the carrier density, the anisotropy of the response without tilt, and the nature of the interband background screening are properties of the toroidal FS. They differ between the VNR and the WSM, and they are not affected by the vortex texture. The tilt modifies the two systems in a qualitatively similar way. The VNR has two inequivalent tilt orientations, whereas the isotropic WSM depends on the magnitude of the tilt alone. An in-plane tilt also leaves the VNR with a finite Drude weight at the intrinsic point, unlike a type-I WSM.

The HSM provides a complementary case. Its BC is a point-like dipole, as for the VNR, but its FS is a sphere. Its bare plasmons are therefore isotropic, as for the WSM. The anisotropy arises only from the interband part, through the quantum geometry of the transitions between the flat band and the Dirac cone. The flat band also lowers the interband onset relative to that of the Dirac cone alone. In the VNR, by comparison, the anisotropy is already present in the bare frequencies owing to the toroidal FS. Without doping, the HSM supports no undamped plasmon, as does the VNR.

Three caveats apply to the calculational framework used here. The linearised description requires low doping and, in the ring-cyclide regime, a small in-plane tilt. The statements on the intrinsic point under an in-plane tilt are parametric in the same sense, since they assume a small tilt, and the tilt used in our numerical estimates lies at the edge of this window. Those estimates are therefore indicative only. Next, the toroidal-angle contribution to the real part of the polarisability depends on a cut-off that the linearised model cannot fix, and the same holds for the background screening of the WSM and the HSM. Only the results independent of the cut-off are controlled predictions. Finally, the local mapping onto 2d Dirac cones requires two separate conditions, namely $\kappa_F \ll k_0$ for the validity of the linearised Hamiltonian and $q \ll k_0$ for the decoupling of the copies at different toroidal angles. The first keeps the internal momenta that contribute to each copy well below the ultraviolet scale set by the ring radius. The second ensures that the component of $\bs{q}$ along the ring is negligible, so that the copies at different toroidal angles do not mix. Both hold in the low-doping long-wavelength regime considered here, and they are independent of each other. In particular, the physically relevant window $q \sim \kappa_F$ is accessible, since $\kappa_F \ll k_0$. Neither condition extends to wavevectors comparable to the ring radius, where a lattice model or a full numerical treatment would be required.

Several extensions suggest themselves. A lattice model would allow us to study the momentum dependence of the response beyond the long-wavelength limit, the cut-off dependence of the in-plane response, and the validity of the local mapping at wavevectors comparable to the ring radius. For the axial tilt, the real part of the dielectric function stays finite at the interband onset, and the coefficient of the cusp there remains to be determined. The effect of a tilt on the plasmons of the HSM is also left open. Interactions beyond the RPA, disorder, and finite temperature offer further directions. A comparison with 3d Luttinger semimetals would show how different band-touching geometries and quantum-geometric structures enter the collective modes. The approach extends to other topological semimetals with toroidal FSs, and the results may be relevant to materials in which nodal rings have been identified~\cite{balents-nodal, fu_nlsm, yang_review_nlsm, schnyder_nodal}.

%%%%%%%%%%%%%%%%%%%%%%%%%%%%%%%%%
\appendix

%%%%%%%%%%%%%%%
\section{Untilted VNR}
\label{app:untilted}

This appendix gives the intermediate steps behind the results of Sec.~\ref{secuntilted}, where $\bs{\eta} = 0$. The subsections follow the order of Secs.~\ref{secreduction} to~\ref{secvortexcorr}. Throughout, $\epsilon_k = v_0\,\kappa$ denotes the energy of the band $s = 2$ and $\kappa_F = \mu/v_0$ denotes the Fermi momentum.

%%%%%%%%%%%%%%%%%%
\subsection{Mapping onto two-dimensional Dirac cones}
\label{app:vnrdirac}

We give the steps behind Eq.~\eqref{eq:ringmapvnr}. At fixed $\phi$ we use the local coordinates $u = \kappa \cos\gamma = k_\perp - k_0$ and $w = \kappa \sin\gamma = k_z$. Since $d k_x\, d k_y = k_\perp\, d k_\perp\, d\phi$, the measure becomes
\begin{align}
g \int \frac{d^3 \bs{k}}{(2\,\pi)^3} = g\, k_0 \int_0^{2\pi} \frac{d\phi}{2\,\pi} \int \frac{du\, dw}{(2\,\pi)^2} \left( 1 + \frac{u}{k_0} \right) .
\end{align}
The term $u/k_0$ is of order $\kappa/k_0$ and is dropped. This is the replacement $k_0 + \kappa \cos\gamma \to k_0$ of Sec.~\ref{secreduction}.

In terms of $u$ and $w$, Eq.~\eqref{eqham} reads $\mathcal{H}_\ell = v_0 \left( u\, \sigma_z - w\, \sigma_\phi \right)$ with $\sigma_\phi = \cos\phi\, \sigma_x + \sin\phi\, \sigma_y$. The unitary matrix $U_\phi = e^{-i\,\phi\,\sigma_z/2}$ commutes with $\sigma_z$ and satisfies $U_\phi\, \sigma_x\, U_\phi^\dagger = \sigma_\phi$. Hence
\begin{align}
\mathcal{H}_\ell = U_\phi\, h (u, w)\, U_\phi^\dagger \,, \quad
h (u, w) = v_0 \left( u\, \sigma_z - w\, \sigma_x \right) .
\end{align}
The matrix $h$ describes a massless 2d Dirac cone of velocity $v_0$. It does not depend on $\phi$. Its eigenvalues are $\pm\, v_0\, \kappa$, and its eigenstates are the spin-$1/2$ coherent states along $\pm\, \bs{n} (\gamma, 0)$. The eigenstates of $\mathcal{H}_\ell$ are obtained by applying $U_\phi$ to those of $h$. The overlap between two states with the same $\phi$ is therefore the overlap of the corresponding states of $h$, because $U_\phi^\dagger\, U_\phi = 1$. This gives $\mathcal{F} = \left[ 1 - \cos (\gamma - \gamma') \right]/2$, as stated after Eq.~\eqref{eq:formfactor}. For different angles, the product $U_\phi^\dagger\, U_{\phi'} = e^{-i\, (\phi' - \phi)\, \sigma_z/2}$ does not cancel. This is the origin of the second term of Eq.~\eqref{eq:fsplit}, which is treated in Appendix~\ref{app:vnrvortex}.

For $\bs{q} = q\, \bs{\hat z}$ we have exactly $\phi' = \phi$, $u' = u$, and $w' = w + q$. For $\bs{q} = q\, \bs{\hat x}$ we have $k_x' = k_x + q$ and $k_y' = k_y$. This gives
\begin{align}
k_\perp' = \sqrt{k_\perp^2 + 2\, k_\perp\, q \cos\phi + q^2} = k_\perp + q \cos\phi + \frac{q^2 \sin^2\phi}{2\, k_\perp} + \mathcal{O} \left( \frac{q^3}{k_0^2} \right) ,
\quad
\sin (\phi' - \phi) = -\, \frac{q \sin\phi}{k_\perp'} \,,
\end{align}
where the second relation follows from $k_\perp'\, \sin (\phi' - \phi) = k_y' \cos\phi - k_x' \sin\phi = -\, q \sin\phi$. Hence $\phi' - \phi \simeq -\, q \sin\phi/k_0$, and the local coordinates change as $u' = u + q \cos\phi$ and $w' = w$ up to corrections of order $q^2/k_0$. The component of $\bs{q}$ along the ring therefore shifts the energies only at order $v_0\, q^2/k_0$. It also shifts $\phi$ by an amount of order $q/k_0$. If the dependence of $\mathcal{F}$ on $\phi' - \phi$ is set aside, the copy at the angle $\phi$ is a 2d problem with the local transfer $\bs{q}_{\rm loc}$. It equals $q\, \bs{\hat w}$ for $\bs{q} = q\, \bs{\hat z}$ and $q \cos\phi\, \bs{\hat u}$ for $\bs{q} = q\, \bs{\hat x}$.

Inserting the measure into Eq.~\eqref{eq:polgeneral}, the integrand at fixed $\phi$ is $\Pi_{\rm 2D} (\bs{q}_{\rm loc}, \omega)$. It is defined with $g \int d^2 \bs{\kappa}/(2\,\pi)^2$ and the cone $h$, where $\bs{\kappa} = (u, w)$. A rotation of the $(u, w)$ plane shifts $\gamma$ and $\gamma'$ by the same angle. The overlap factor depends only on their difference, and the energies are unchanged. Hence $\Pi_{\rm 2D}$ depends only on $|\bs{q}_{\rm loc}|$, and Eq.~\eqref{eq:ringmapvnr} follows. For $\bs{q} = q\, \bs{\hat z}$ the steps above are exact up to the replacement of the Jacobian. For $\bs{q} = q\, \bs{\hat x}$ the neglected shifts are of relative order $q/k_0$. A response proportional to $|\bs{q}_{\rm loc}|^n$ acquires, for $\bs{q} = q\, \bs{\hat x}$, the factor
\begin{align}
\int_0^{2\pi} \frac{d\phi}{2\,\pi}\, \cos^2\phi = \frac{1}{2} \,, \quad
\int_0^{2\pi} \frac{d\phi}{2\,\pi}\, |\cos\phi| = \frac{2}{\pi}\,,
\end{align}
for $n = 2$ and $n = 1$, respectively. The first value gives the factor of two between the $\bs{\hat z}$ and $\bs{\hat x}$ channels of $\Im \Pi$ and $\Pi_{\rm inter}$. The second value relates the two expressions of Eq.~\eqref{eq:pistaticvnr}.

%%%%%%%%%%%%%%%%%%
\subsection{Polarisability of the intrinsic VNR}
\label{app:vnrintrinsic}

We give the steps behind Eqs.~\eqref{eq:polmu0vnr}, \eqref{eq:imvnr}, \eqref{eq:pistaticvnr}, and \eqref{eq:epsintrvnr}. By Appendix~\ref{app:vnrdirac}, it is sufficient to evaluate $\Pi_{\rm 2D} (Q, \omega)$ for a single undoped cone with the local transfer $Q = |\bs{q}_{\rm loc}|$. At $\mu = 0$ the band $s = 1$ is filled and the band $s = 2$ is empty. Combining the two terms of Eq.~\eqref{eq:polinter}, we obtain
\begin{align}
\Pi_{\rm 2D} (Q, \omega) = g \int \frac{d^2 \bs{\kappa}}{(2\,\pi)^2}\, \mathcal{F}\, \frac{2\, (E + E')}{(\omega + i\, 0^+)^2 - (E + E')^2} \,, \quad
E = v_0\, \kappa \,, \quad E' = v_0\, \kappa' \,,
\end{align}
with $\bs{\kappa}' = \bs{\kappa} + \bs{q}_{\rm loc}$. The scalar product of the pseudospin vectors of the cone is $\cos (\gamma - \gamma') = \bs{\kappa} \cdot \bs{\kappa}'/(\kappa\, \kappa')$, and $\bs{\kappa} \cdot \bs{\kappa}' = \left( \kappa^2 + \kappa'^2 - Q^2 \right)/2$. Hence
\begin{align}
\mathcal{F} = \frac{1 - \cos (\gamma - \gamma')}{2} = \frac{Q^2 - (\kappa - \kappa')^2}{4\, \kappa\, \kappa'} \,.
\end{align}

The integrand depends on $\bs{\kappa}$ only through $\kappa$ and $\kappa'$, which are the distances to the two foci $\bs{\kappa} = 0$ and $\bs{\kappa} = -\, \bs{q}_{\rm loc}$. This suggests the elliptic coordinates
\begin{align}
\kappa + \kappa' = Q\, \xi \,, \quad \kappa - \kappa' = Q\, \chi \,, \quad \xi \in [1, \infty) \,, \quad \chi \in [-1, 1] \,.
\end{align}
Each pair $(\xi, \chi)$ corresponds to two points, which are mirror images with respect to the line through the foci. We obtain
\begin{align}
d^2 \bs{\kappa} = 2 \cdot \frac{Q^2}{4}\, \frac{\xi^2 - \chi^2}{\sqrt{\left( \xi^2 - 1 \right) \left( 1 - \chi^2 \right)}}\, d\xi\, d\chi \,, \quad
\kappa\, \kappa' = \frac{Q^2}{4} \left( \xi^2 - \chi^2 \right) , \quad
\mathcal{F} = \frac{1 - \chi^2}{\xi^2 - \chi^2} \,,
\end{align}
where the factor of two in the first expression counts the two points. The form factor cancels the factor $\xi^2 - \chi^2$ of the area element. The integral over $\chi$ is then elementary, and for any function $\tilde{G} (\xi)$ we have
\begin{align}
\int d^2 \bs{\kappa}\, \mathcal{F}\, \tilde{G} (\xi) = \frac{Q^2}{2} \int_1^\infty \frac{d\xi\, \tilde{G} (\xi)}{\sqrt{\xi^2 - 1}} \int_{-1}^{1} d\chi \sqrt{1 - \chi^2} = \frac{\pi\, Q^2}{4} \int_1^\infty \frac{d\xi\, \tilde{G} (\xi)}{\sqrt{\xi^2 - 1}} \,.
\end{align}
With $E + E' = v_0\, Q\, \xi$ and the substitution $t = \sqrt{\xi^2 - 1}$, which gives $\xi\, d\xi/\sqrt{\xi^2 - 1} = dt$, the polarisability becomes
\begin{align}
\Pi_{\rm 2D} (Q, \omega) = \frac{g\, Q^2}{16\, \pi} \int_0^\infty dt\, \frac{2\, v_0\, Q}{\omega^2 - v_0^2\, Q^2 - v_0^2\, Q^2\, t^2 + i\, 0^+} \,.
\end{align}

For $\omega < v_0\, Q$ the denominator is negative and real. With $\int_0^\infty dt/(a^2 + b^2\, t^2) = \pi/(2\, a\, b)$ for $a^2 = v_0^2\, Q^2 - \omega^2$ and $b = v_0\, Q$, we find
\begin{align}
\Pi_{\rm 2D} (Q, \omega) = -\, \frac{g\, Q^2}{16\, \sqrt{v_0^2\, Q^2 - \omega^2}} \,, \quad \omega < v_0\, Q \,.
\end{align}
For $\omega > v_0\, Q$ the denominator vanishes at $t_0 = \sqrt{\omega^2 - v_0^2\, Q^2}/(v_0\, Q)$. The principal value of $\int_0^\infty dt/(t_0^2 - t^2)$ is zero, and the delta function gives $-\, i\, \pi \int_0^\infty dt\, \delta (t_0^2 - t^2) = -\, i\, \pi/(2\, t_0)$. Hence
\begin{align}
\Pi_{\rm 2D} (Q, \omega) = -\, i\, \frac{g\, Q^2}{16\, \sqrt{\omega^2 - v_0^2\, Q^2}} \,, \quad \omega > v_0\, Q \,.
\end{align}
For $\bs{q} = q\, \bs{\hat z}$ we have $Q = q$, and multiplication by $k_0$ gives Eq.~\eqref{eq:polmu0vnr} together with its continuation above the threshold. For $\omega \gg v_0\, q$ the square root reduces to $\omega$, which gives Eq.~\eqref{eq:imvnr}. In the static limit, $\Pi_{\rm 2D} (Q, 0) = -\, g\, Q/(16\, v_0)$. For $\bs{q} = q\, \bs{\hat z}$ this gives the first expression of Eq.~\eqref{eq:pistaticvnr}. For $\bs{q} = q\, \bs{\hat x}$, we have $Q = q\, |\cos\phi|$, and the average of $|\cos\phi|$ in Appendix~\ref{app:vnrdirac} gives
\begin{align}
\Pi (q\, \bs{\hat x}, 0) = -\, \frac{g\, k_0\, q}{16\, v_0} \int_0^{2\pi} \frac{d\phi}{2\,\pi}\, |\cos\phi| = -\, \frac{g\, k_0\, q}{8\, \pi\, v_0} \,.
\end{align}
In the same way, the average of $\cos^2\phi$ halves $\Im \Pi$ in the window $v_0\, q \ll \omega \ll v_0\, k_0$, which gives $\Im \Pi (q\, \bs{\hat x}, \omega) \simeq -\, g\, k_0\, q^2/(32\, \omega)$.

The dielectric function follows from Eq.~\eqref{eq:epsrpa} with $-\, 4\, \pi\, e^2\, \Pi/q^2$ and the result of Eq.~\eqref{eq:polmu0vnr}. This gives Eq.~\eqref{eq:epsintrvnr}. The second term is positive for $0 \leq \omega < v_0\, q$, and therefore $\varepsilon^{\rm RPA} > 1$ and no zero exists.

For $\bs{q} = q\, \bs{\hat x}$ the exact local result gives
\begin{align}
\Im \Pi (q\, \bs{\hat x}, \omega) = -\, \frac{g\, k_0\, q^2}{16} \int_0^{2\pi} \frac{d\phi}{2\,\pi}\, \frac{\cos^2\phi\, \Theta \left( \omega - v_0\, q\, |\cos\phi| \right)}{\sqrt{\omega^2 - v_0^2\, q^2 \cos^2\phi}} \,.
\end{align}
The set of angles with $v_0\, q\, |\cos\phi| < \omega$ has a finite measure for every $\omega > 0$, and the integrand is positive there. The imaginary part therefore never vanishes, and the continuum extends down to $\omega = 0$. The decoupled description fails near $\phi = \pi/2$. There the local transfer $q\, |\cos\phi|$ becomes comparable to the neglected shift $q^2 \sin^2\phi/(2\, k_\perp)$ of $k_\perp'$, which happens for $|\cos\phi| \lesssim q/k_0$. The weight of this window in the average over $\phi$ is of order $q/k_0$, and the conclusion above is unchanged.

%%%%%%%%%%%%%%%%%%
\subsection{Density-of-states, Drude weights, and bare plasmon frequencies}
\label{app:vnrdrude}

We give the steps behind Eqs.~\eqref{eq:dosvnr}, \eqref{eq:dvnr}, and \eqref{eq:wpvnr}. In the coordinates of Eqs.~\eqref{eqtrs} and~\eqref{eqtor} the exact measure is
\begin{align}
d^3 \bs{k} = k_\perp\, d k_\perp\, d\phi\, d k_z = \kappa \left( k_0 + \kappa \cos\gamma \right) d\kappa\, d\gamma\, d\phi \,.
\end{align}
For $\mu > 0$ only the band $s = 2$ has a FS, because $\varepsilon_1 = -\, v_0\, \kappa \leq 0$. The delta function is $\delta (\varepsilon_2 - \mu) = \delta (\kappa - \kappa_F)/v_0$, and Eq.~\eqref{eqdos} gives
\begin{align}
\rho (\mu) = \frac{g}{(2\,\pi)^3\, v_0} \int_0^{2\pi} d\phi \int_0^{2\pi} d\gamma\, \kappa_F \left( k_0 + \kappa_F \cos\gamma \right) = \frac{g\, k_0\, \kappa_F}{2\,\pi\, v_0} = \frac{g\, k_0\, \mu}{2\,\pi\, v_0^2} \,.
\end{align}
The term proportional to $\cos\gamma$ integrates to zero. No expansion in $\kappa_F/k_0$ has been made, and the result holds as long as $\kappa_F < k_0$, which is the condition for the FS to be a ring torus.

With Eq.~\eqref{eqev} and $\epsilon_k = v_0\, \kappa$, the velocity of the band $s = 2$ is $\bs{V}^{(2)} = v_0\, \bs{\hat V}$ with $\bs{\hat V} = \left\lbrace \cos\gamma \cos\phi,\, \cos\gamma \sin\phi,\, \sin\gamma \right\rbrace$. Using $v_0\, \kappa_F = \mu$, Eq.~\eqref{eqdrude} becomes
\begin{align}
D_{ij} = \frac{\pi\, g\, e^2}{(2\,\pi)^3}\, \frac{v_0^2}{v_0}\, k_0\, \kappa_F \int_0^{2\pi} d\phi \int_0^{2\pi} d\gamma\, \hat V_i\, \hat V_j = \frac{g\, e^2\, k_0\, \mu}{8\, \pi^2} \int_0^{2\pi} d\phi \int_0^{2\pi} d\gamma\, \hat V_i\, \hat V_j \,.
\end{align}
The piece of the measure proportional to $\kappa_F \cos\gamma$ gives $\int_0^{2\pi} d\gamma \sin^2\gamma \cos\gamma = \int_0^{2\pi} d\gamma \cos^3\gamma = 0$ and drops out. The off-diagonal entries are proportional to $\int_0^{2\pi} d\phi \cos\phi \sin\phi$, $\int_0^{2\pi} d\phi \cos\phi$, or $\int_0^{2\pi} d\phi \sin\phi$, and they vanish. For the diagonal entries,
\begin{align}
\int d\phi\, d\gamma\, \sin^2\gamma = 2\, \pi^2 \,, \quad
\int d\phi\, d\gamma\, \cos^2\gamma \cos^2\phi = \int d\phi\, d\gamma\, \cos^2\gamma \sin^2\phi = \pi^2 \,,
\end{align}
which gives $D_{zz} = g\, e^2\, k_0\, \mu/4$ and $D_{xx} = D_{yy} = g\, e^2\, k_0\, \mu/8$, as in Eq.~\eqref{eq:dvnr}. The ratio of two comes from the two integrals above.

We now derive the intraband polarisability of Eq.~\eqref{eqintra}. Only the band $s = 2$ is partially occupied, and the occupation difference of the filled band $s = 1$ vanishes. We take $q \to 0$ at fixed $\omega$ and expand the denominator in powers of $1/\omega$. With $\bs{k}' = \bs{k} + \bs{q}$,
\begin{align}
\Pi_{\rm intra} (\bs{q}, \omega) \simeq g \int \frac{d^3 \bs{k}}{(2\,\pi)^3}\, \mathcal{F}_{22}\, \left[ f_0 (\varepsilon_2 (\bs{k})) - f_0 (\varepsilon_2 (\bs{k}')) \right] \left[ \frac{1}{\omega} + \frac{\varepsilon_2 (\bs{k}') - \varepsilon_2 (\bs{k})}{\omega^2} + \cdots \right] .
\end{align}
The overlap factor is $\mathcal{F}_{22} = 1 - q_i\, q_j\, g_{ij}$. For $\mathcal{F}_{22} = 1$ the first term vanishes, because the integral of $f_0 (\varepsilon_2 (\bs{k})) - f_0 (\varepsilon_2 (\bs{k}'))$ is zero after a shift of the integration variable. The correction from $\mathcal{F}_{22} - 1$ is of higher order in $q$. In the second term we use $f_0 (\varepsilon_2 (\bs{k})) - f_0 (\varepsilon_2 (\bs{k}')) \simeq -\, q_i\, \partial f_0/\partial k_i$ and $\varepsilon_2 (\bs{k}') - \varepsilon_2 (\bs{k}) \simeq q_j\, V_j^{(2)}$. At $T = 0$ we have $\partial f_0/\partial k_i = -\, \delta (\varepsilon_2 - \mu)\, V_i^{(2)}$. Hence
\begin{align}
\Pi_{\rm intra} (\bs{q}, \omega) \simeq \frac{q_i\, q_j}{\omega^2}\, g \int \frac{d^3 \bs{k}}{(2\,\pi)^3}\, \delta (\varepsilon_2 - \mu)\, V_i^{(2)}\, V_j^{(2)} = \frac{q_i\, q_j\, D_{ij}}{\pi\, e^2\, \omega^2} \,,
\end{align}
which is Eq.~\eqref{eqintra}. For $\bs{q}$ along a principal axis $\bs{\hat e}_i$, Eq.~\eqref{eq:epsrpa} gives
\begin{align}
\varepsilon^{\rm RPA} (q\, \bs{\hat e}_i, \omega) = 1 - \frac{4\, \pi\, e^2}{q^2}\, \frac{q^2\, D_{ii}}{\pi\, e^2\, \omega^2} = 1 - \frac{4\, D_{ii}}{\omega^2} \,.
\end{align}
The zero gives $\omega_{p,i}^2 = 4\, D_{ii}$, which is Eq.~\eqref{eqplasmon}. The values of Eq.~\eqref{eq:wpvnr} follow from Eq.~\eqref{eq:dvnr}, and $\omega_{p,z}/\omega_{p,x} = \sqrt{2}$.

The bare frequencies can be written as $\omega_{p,z} = 2\, \mu \sqrt{\lambda}$ with $\lambda = g\, e^2\, k_0/(4\, \mu)$. Hence $\omega_{p,z} > 2\, \mu$ if and only if $\lambda > 1$. For $\mu = 0.1\, v_0\, k_0$, $g = e = 1$, and $v_0 = 0.0004$, we have $\lambda = 1/(0.4\, v_0) = 6250$. In addition, $\omega_{p,z}^2/(v_0\, k_0)^2 = 0.1\, g\, e^2/v_0 = 250$. Thus $\omega_{p,z}$ exceeds both $2\, \mu$ and $v_0\, k_0$ for these parameters, as stated in Sec.~\ref{secdopedvnr}.

%%%%%%%%%%%%%%%%%%
\subsection{Interband screening and the screened plasmons}
\label{app:vnrscreen}

We give the steps behind Eqs.~\eqref{eq:pinterdopedvnr} to~\eqref{eq:wpscreenedvnr}, together with the statements on the zero of $\Re \varepsilon^{\rm RPA}$ made below them. We take $\mu > 0$ and $\omega \gg v_0\, q$. The denominators of Eq.~\eqref{eq:polinter} are evaluated at $q = 0$, where $\epsilon_{k'} = \epsilon_k$. The bracket then becomes
\begin{align}
\frac{1}{\omega - 2\, \epsilon_k + i\, 0^+} - \frac{1}{\omega + 2\, \epsilon_k + i\, 0^+} = \frac{4\, \epsilon_k}{(\omega + i\, 0^+)^2 - 4\, \epsilon_k^2} \,.
\end{align}
The band $s = 1$ is filled, and the transitions are allowed only into the empty states of the band $s = 2$, which have $\kappa > \kappa_F$. The overlap factor is $\mathcal{F} \simeq q_i\, q_j\, g_{ij}$, which requires $q \ll \kappa_F$. We keep only the first term of Eq.~\eqref{eq:qmetric} in this subsection. The second term is treated in Appendix~\ref{app:vnrvortex}. For $\bs{q} = q\, \bs{\hat z}$ the second term vanishes, because $\hat\phi_z = 0$, and the first term gives $q^2 \cos^2\gamma/(4\, \kappa^2)$. With the measure $k_0\, \kappa\, d\kappa\, d\gamma\, d\phi$ we obtain
\begin{align}
\Pi_{\rm inter} (q\, \bs{\hat z}, \omega) = \frac{g\, k_0}{(2\,\pi)^3} \int_0^{2\pi} d\phi \int_0^{2\pi} d\gamma \int_{\kappa_F}^\infty \kappa\, d\kappa\, \frac{q^2 \cos^2\gamma}{4\, \kappa^2}\, \frac{4\, v_0\, \kappa}{(\omega + i\, 0^+)^2 - 4\, v_0^2\, \kappa^2} = \frac{g\, k_0\, q^2}{4\, \pi} \int_\mu^\infty \frac{d\epsilon}{(\omega + i\, 0^+)^2 - 4\, \epsilon^2} \,,
\end{align}
where we have used $\int_0^{2\pi} d\phi \int_0^{2\pi} d\gamma \cos^2\gamma = 2\, \pi^2$ and the substitution $\epsilon = v_0\, \kappa$. The factors of $\kappa$ cancel, and the radial integral is elementary. The derivative of $\ln \left| (\omega + 2\, \epsilon)/(\omega - 2\, \epsilon) \right|$ with respect to $\epsilon$ equals $4\, \omega/(\omega^2 - 4\, \epsilon^2)$, and this function vanishes as $\epsilon \to \infty$. The real part of the integral is therefore $-\, \ln \left| (\omega + 2\, \mu)/(\omega - 2\, \mu) \right|/(4\, \omega)$. The imaginary part is $-\, \pi \int_\mu^\infty d\epsilon\, \delta (\omega^2 - 4\, \epsilon^2) = -\, \pi\, \Theta (\omega - 2\, \mu)/(4\, \omega)$, because $|d (4\, \epsilon^2)/d\epsilon| = 4\, \omega$ at $\epsilon = \omega/2$. Hence
\begin{align}
\Pi_{\rm inter} (q\, \bs{\hat z}, \omega) = -\, \frac{g\, k_0\, q^2}{16\, \pi\, \omega} \left[ \ln \left| \frac{\omega + 2\, \mu}{\omega - 2\, \mu} \right| + i\, \pi\, \Theta (\omega - 2\, \mu) \right] ,
\end{align}
which is Eq.~\eqref{eq:pinterdopedvnr}. At $\mu = 0$ it reduces to the imaginary part of Eq.~\eqref{eq:imvnr}. For $\bs{q} = q\, \bs{\hat x}$, the first term of Eq.~\eqref{eq:qmetric} gives $q^2 \sin^2\gamma \cos^2\phi/(4\, \kappa^2)$. The angular integral is $\int_0^{2\pi} d\phi \int_0^{2\pi} d\gamma \sin^2\gamma \cos^2\phi = \pi^2$, which is half of the value for $\bs{\hat z}$. Therefore $\Pi_{\rm inter} (q\, \bs{\hat x}, \omega)$ is half of the expression above.

At small frequencies, $\ln \left[ (2\, \mu + \omega)/(2\, \mu - \omega) \right] \simeq \omega/\mu$, and the static value is $\Pi_{\rm inter} (q\, \bs{\hat z}, 0) = -\, g\, k_0\, q^2/(16\, \pi\, \mu)$. Adding Eq.~\eqref{eqintra} with $D_{zz}$ of Eq.~\eqref{eq:dvnr}, Eq.~\eqref{eq:epsrpa} gives
\begin{align}
\varepsilon^{\rm RPA} (q\, \bs{\hat z}, \omega) = 1 - \frac{g\, e^2\, k_0\, \mu}{\omega^2} + \frac{g\, e^2\, k_0}{4\, \omega} \left[ \ln \left| \frac{\omega + 2\, \mu}{\omega - 2\, \mu} \right| + i\, \pi\, \Theta (\omega - 2\, \mu) \right] ,
\end{align}
which is Eq.~\eqref{eq:epsdopedvnr}. Both the Drude weight and $\Pi_{\rm inter}$ are halved along $\bs{\hat x}$, and this is the replacement of $g\, e^2\, k_0$ by $g\, e^2\, k_0/2$ stated in the main text. Since $\ln \left[ (1 + x)/(1 - x) \right] = 2 \operatorname{arctanh} x$, the real part of the interband term is $g\, e^2\, k_0 \operatorname{arctanh} (\omega/2\, \mu)/(2\, \omega)$. This is the background constant $\epsilon_{b,z} (\omega) - 1$. Its limit $\operatorname{arctanh} x \to x$ gives the static values of Eq.~\eqref{eq:ebvnr}.

With $\lambda = g\, e^2\, k_0/(4\, \mu)$ and $\tilde{x} = \omega/(2\, \mu)$, the Drude term is $g\, e^2\, k_0\, \mu/\omega^2 = \lambda/\tilde{x}^2$ and the interband term is $\lambda \operatorname{arctanh} \tilde{x}/\tilde{x}$. Hence, below the onset,
\begin{align}
\varepsilon^{\rm RPA} (q\, \bs{\hat z}, \omega) = 1 + \lambda\, \tilde{h} (\tilde{x}) \,, \quad
\tilde{h} (\tilde{x}) = \frac{\operatorname{arctanh} \tilde{x}}{\tilde{x}} - \frac{1}{\tilde{x}^2} \,, \quad 0 < \tilde{x} < 1 \,.
\end{align}
The function $\operatorname{arctanh} \tilde{x}/\tilde{x} = \sum_{n \geq 0} \tilde{x}^{2\, n}/(2\, n + 1)$ increases on $(0, 1)$ and diverges as $\tilde{x} \to 1^-$. The function $-\, 1/\tilde{x}^2$ increases as well and tends to $-\, \infty$ as $\tilde{x} \to 0$. Therefore $\tilde{h}$ is strictly increasing from $-\, \infty$ to $+\, \infty$, and $1 + \lambda\, \tilde{h} (\tilde{x}) = 0$ has exactly one root for every $\lambda > 0$. The $\bs{\hat x}$ channel has the same form with $\lambda \to \lambda/2$.

The static screened frequency is an upper bound of this root. Since $\operatorname{arctanh} \tilde{x}/\tilde{x} \geq 1$, we have $\Re \varepsilon^{\rm RPA} \geq 1 + \lambda - \lambda/\tilde{x}^2$, and the right-hand side vanishes at $\tilde{x}^2 = \lambda/(1 + \lambda)$. Because $\Re \varepsilon^{\rm RPA}$ increases with $\tilde{x}$, the root cannot exceed this value. In terms of the frequency, $\omega^2 = 4\, \mu^2\, \lambda/(1 + \lambda)$, which is $\tilde\omega_{p,z}^2$ of Eq.~\eqref{eq:wpscreenedvnr}. The same steps with $\lambda \to \lambda/2$ and the bare value $\omega_{p,x}^2 = 2\, \mu^2\, \lambda$ give $\tilde\omega_{p,x}^2 = 2\, \mu^2\, \lambda/(1 + \lambda/2)$. Both values are below $4\, \mu^2$ and hence below the onset. The ratio of the two is
\begin{align}
\frac{\tilde\omega_{p,z}^2}{\tilde\omega_{p,x}^2} = \frac{2\, (1 + \lambda/2)}{1 + \lambda} \,,
\end{align}
which is smaller than two for $\lambda > 0$ and tends to unity for $\lambda \gg 1$.

For $\lambda \ll 1$ the root lies at small $\tilde{x}$, where $\tilde{h} (\tilde{x}) \simeq 1 - 1/\tilde{x}^2$. The condition $1 + \lambda\, \tilde{h} = 0$ gives $\tilde{x}^2 \simeq \lambda$, that is $\omega^2 \simeq 4\, \mu^2\, \lambda = g\, e^2\, k_0\, \mu$. This is the bare frequency of Eq.~\eqref{eq:wpvnr}. For $\lambda \to \infty$ the root approaches the zero $\tilde{x}_*$ of $\tilde{h}$, which solves $\tilde{x}_* \operatorname{arctanh} \tilde{x}_* = 1$, or equivalently $\operatorname{arctanh} \tilde{x}_* = 1/\tilde{x}_*$. It is unique and lies below unity, because $\tilde{h}$ is strictly increasing and diverges at $\tilde{x} = 1$. Numerically $\tilde{x}_* \simeq 0.834$. For finite $\lambda$ we have $\tilde{h} (\tilde{x}) = -\, 1/\lambda < 0 = \tilde{h} (\tilde{x}_*)$, and the root lies below $\tilde{x}_*$. The root of the $\bs{\hat x}$ channel obeys $\tilde{h} (\tilde{x}) = -\, 2/\lambda$ and approaches the same $\tilde{x}_*$. The two exact zeros therefore have a ratio that tends to unity, and both are below the static values, which tend to $2\, \mu$ from below.

%%%%%%%%%%%%%%%%%%
\subsection{Toroidal-angle dependence of the overlap factor for in-plane wavevectors}
\label{app:vnrvortex}

We give the steps behind Eqs.~\eqref{eq:fsplit} to~\eqref{eq:debvnr}. Writing $\cos (\phi - \phi') = 1 - \left[ 1 - \cos (\phi - \phi') \right]$ in Eq.~\eqref{eq:formfactor} gives
\begin{align}
\bs{n} \cdot \bs{n}' = \cos (\gamma - \gamma') - \sin\gamma \sin\gamma' \left[ 1 - \cos (\phi - \phi') \right] .
\end{align}
Inserting this into $\mathcal{F} = \left( 1 - \bs{n} \cdot \bs{n}' \right)/2$ gives Eq.~\eqref{eq:fsplit}. The second term vanishes for $\phi' = \phi$. For $\bs{q} = q\, \bs{\hat x}$, Appendix~\ref{app:vnrdirac} gives $\phi' - \phi \simeq -\, q \sin\phi/k_0$. To leading order, $\sin\gamma \sin\gamma' \to \sin^2\gamma$ and $1 - \cos (\phi - \phi') \simeq (\phi' - \phi)^2/2 = q^2 \sin^2\phi/(2\, k_0^2)$. The second term becomes
\begin{align}
\delta\mathcal{F} = \frac{q^2 \sin^2\gamma \sin^2\phi}{4\, k_0^2} \,.
\end{align}
This agrees with the second term of Eq.~\eqref{eq:qmetric}, since $\hat\phi_x = -\, \sin\phi$ and $k_\perp \to k_0$.

The corresponding contribution to the polarisability follows from Eq.~\eqref{eq:polinter}, with the bracket $4\, \epsilon_k/[(\omega + i\, 0^+)^2 - 4\, \epsilon_k^2]$ of Appendix~\ref{app:vnrscreen} and the same blocking $\kappa > \kappa_F$. Using the measure $k_0\, \kappa\, d\kappa\, d\gamma\, d\phi$,
\begin{align}
\delta\Pi (q\, \bs{\hat x}, \omega) = \frac{g\, k_0}{(2\,\pi)^3}\, \frac{q^2}{4\, k_0^2} \int_0^{2\pi} d\phi \sin^2\phi \int_0^{2\pi} d\gamma \sin^2\gamma \int_{\kappa_F}^{\Lambda} \kappa\, d\kappa\, \frac{4\, v_0\, \kappa}{(\omega + i\, 0^+)^2 - 4\, v_0^2\, \kappa^2} \,.
\end{align}
Each of the two angular integrals equals $\pi$, and the prefactor becomes $g\, v_0\, q^2/(8\, \pi\, k_0)$. This is Eq.~\eqref{eq:dpivnr}. The integrand tends to $-\, 1/(4\, v_0^2)$ at large $\kappa$. The integral then grows linearly, and the cut-off $\Lambda \lesssim k_0$ of the linearised description is required.

The imaginary part is controlled by the resonant radius $\kappa_0 = \omega/(2\, v_0)$. Using $|d (4\, v_0^2\, \kappa^2)/d\kappa| = 8\, v_0^2\, \kappa_0$ at $\kappa = \kappa_0$,
\begin{align}
\Im \int_{\kappa_F}^{\Lambda} \frac{\kappa^2\, d\kappa}{(\omega + i\, 0^+)^2 - 4\, v_0^2\, \kappa^2} = -\, \pi\, \frac{\kappa_0^2}{8\, v_0^2\, \kappa_0}\, \Theta (\kappa_0 - \kappa_F) = -\, \frac{\pi\, \omega}{16\, v_0^3}\, \Theta (\omega - 2\, \mu) \,.
\end{align}
The step function holds for $\kappa_F < \kappa_0 < \Lambda$, and the second condition is $\omega < 2\, v_0\, \Lambda$. Multiplying by $g\, v_0\, q^2/(8\, \pi\, k_0)$ gives Eq.~\eqref{eq:imdpivnr}. The leading term in the same channel is $-\, g\, k_0\, q^2/(32\, \omega)$, and the ratio is
\begin{align}
\frac{g\, q^2\, \omega/(128\, v_0^2\, k_0)}{g\, k_0\, q^2/(32\, \omega)} = \frac{\omega^2}{4\, v_0^2\, k_0^2} = \left( \frac{\omega}{2\, v_0\, k_0} \right)^2 .
\end{align}
This is the factor $(\kappa/k_0)^2$ of the quantum metric at $\kappa = \kappa_0$.

For the real part we use the identity
\begin{align}
\frac{\kappa^2}{\omega^2 - 4\, v_0^2\, \kappa^2} = -\, \frac{1}{4\, v_0^2} + \frac{\omega^2}{4\, v_0^2} \, \frac{1}{\omega^2 - 4\, v_0^2\, \kappa^2} \,.
\end{align}
Together with the antiderivative of Appendix~\ref{app:vnrscreen}, with $\epsilon = v_0\, \kappa$, it gives
\begin{align}
\Re \int_{\kappa_F}^{\Lambda} \frac{\kappa^2\, d\kappa}{(\omega + i\, 0^+)^2 - 4\, v_0^2\, \kappa^2} = -\, \frac{\Lambda - \kappa_F}{4\, v_0^2} + \frac{\omega}{16\, v_0^3} \left[ \ln \left| \frac{\omega + 2\, v_0\, \kappa}{\omega - 2\, v_0\, \kappa} \right| \right]_{\kappa_F}^{\Lambda} .
\end{align}
The second term is smaller than the first by a factor of order $\omega/(v_0\, \Lambda)$, up to a logarithm that is singular only at $\omega = 2\, \mu$. For $\omega \ll v_0\, \Lambda$ we drop it. Multiplying the first term by $g\, v_0\, q^2/(8\, \pi\, k_0)$ gives
\begin{align}
\Re \delta\Pi (q\, \bs{\hat x}, \omega) \simeq -\, \frac{g\, q^2 \left( \Lambda - \kappa_F \right)}{32\, \pi\, v_0\, k_0} \,.
\end{align}
The same value follows from the static limit of Eq.~\eqref{eq:polinterstatic}. Its second term of the quantum metric gives $-\, g \int d^3 \bs{k}/(2\,\pi)^3\, \delta\mathcal{F}/\epsilon_k$ over $\kappa > \kappa_F$, and the integrand $k_0\, \kappa \cdot q^2 \sin^2\gamma \sin^2\phi/(4\, k_0^2\, v_0\, \kappa)$ gives the same result after the angular integrals $\pi^2$.

The shift of the background constant follows from $\delta\epsilon_{b,x} = -\, 4\, \pi\, e^2\, \Re \delta\Pi/q^2$, which gives Eq.~\eqref{eq:debvnr}. For $\bs{q} = q\, \bs{\hat y}$ the shift of $\phi$ is $\phi' - \phi \simeq q \cos\phi/k_0$, and the weight $\sin^2\phi$ is replaced by $\cos^2\phi$. The angular integral is again $\pi$, and $\delta\epsilon_{b,y} = \delta\epsilon_{b,x}$, as required by the combined rotation of Sec.~\ref{secmodel}. For $\bs{q} = q\, \bs{\hat z}$ the shift vanishes. Relative to $\epsilon_{b,x} - 1 = g\, e^2\, k_0/(8\, \mu)$ of Eq.~\eqref{eq:ebvnr}, the ratio is
\begin{align}
\frac{\delta\epsilon_{b,x}}{\epsilon_{b,x} - 1} = \frac{g\, e^2 \left( \Lambda - \kappa_F \right)/(8\, v_0\, k_0)}{g\, e^2\, k_0/(8\, \mu)} = \frac{\left( \Lambda - \kappa_F \right) \kappa_F}{k_0^2} \,,
\end{align}
where $\mu/v_0 = \kappa_F$ has been used. For $\Lambda = k_0$ it equals $\left( 1 - \kappa_F/k_0 \right) \kappa_F/k_0$. The shift is positive for $\Lambda > \kappa_F$, and it therefore raises the in-plane background constants and lowers the screened in-plane frequencies $\tilde\omega_{p,i}^2 = \omega_{p,i}^2/\epsilon_{b,i}$.

%%%%%%%%%%%%%%%%%%%%%%%%%
%%%%%%%%%%%%%%%
\section{VNR tilted with respect to the \texorpdfstring{$k_x$}{kx}-axis}
\label{app:xtilt}

This appendix gives the intermediate steps behind the results of Sec.~\ref{secinplane}, where $\bs{\eta} = \eta_x\,\bs{\hat x}$. The subsections follow the order of Secs.~\ref{secxintr} to~\ref{secxvortex}, and they are preceded by the steps behind Eq.~\eqref{eq:cyclidevnr}. We take $\eta_x > 0$ without loss of generality, because $\eta_x \to -\,\eta_x$ is equivalent to $\phi \to \phi + \pi$. We use the shorthand $c_\gamma = \cos\gamma$, $s_\gamma = \sin\gamma$, $c_\phi = \cos\phi$, and $s_\phi = \sin\phi$. The local chemical potential at the toroidal angle $\phi$ is $\mu_{\rm eff} (\phi) = \mu - \tilde{\eta}\,c_\phi$, with $\tilde{\eta} = v_0\,\eta_x\,k_0$. We also define $\zeta_\pm = 1 \pm \eta_x\,c_\gamma\,c_\phi$.

%%%%%%%%%%%%%%%%%%
\subsection{Fermi momentum, delta function, and transition energies}
\label{app:vnrxfs}

We give the steps behind Eq.~\eqref{eq:cyclidevnr} and the statements on the transition energies made after it. With $k_x = (k_0 + \kappa\,c_\gamma)\,c_\phi$, the eigenvalues of Sec.~\ref{secinplane} can be written as
\begin{align}
\varepsilon_s (\bs{k}) = (-1)^s\,v_0\,\kappa + v_0\,\eta_x\,(k_0 + \kappa\,c_\gamma)\,c_\phi
= v_0\,\kappa\,\zeta_s (\phi, \gamma) + \tilde{\eta}\,c_\phi\,,
\quad \zeta_s = (-1)^s + \eta_x\,c_\gamma\,c_\phi\,.
\end{align}
At fixed angles, this is linear in $\kappa$. The condition $\varepsilon_s = \mu$ therefore has the single solution
\begin{align}
\label{eq:axkappas}
\kappa_s (\phi, \gamma) = \frac{\mu_{\rm eff} (\phi)}{v_0\,\zeta_s (\phi, \gamma)}\,,
\end{align}
which is Eq.~\eqref{eq:cyclidevnr}. Since $\partial \varepsilon_s/\partial \kappa = v_0\,\zeta_s$, the delta function becomes $\delta (\varepsilon_s - \mu) = \delta (\kappa - \kappa_s)/(v_0\,|\zeta_s|)$. It contributes only if $\kappa_s > 0$.

For $|\eta_x| < 1$, we have $\zeta_2 = \zeta_+ > 0$ and $\zeta_1 = -\,\zeta_- < 0$. The band $s = 2$ therefore has a FS at the angles with $\mu_{\rm eff} > 0$, and the band $s = 1$ has one at the angles with $\mu_{\rm eff} < 0$. For $|\tilde{\eta}| < \mu$, the local chemical potential is positive for every $\phi$, and only the electron surface exists. This is the ring-cyclide. For $|\tilde{\eta}| > \mu$, the local chemical potential changes sign at $\cos\phi_b = \mu/\tilde{\eta}$. We then have an electron arc $\phi \in (\phi_b, 2\,\pi - \phi_b)$ and a hole arc $|\phi| < \phi_b$. They meet where $\mu_{\rm eff} = 0$ and $\kappa_s = 0$, which is at $\phi = \pm\,\pi/2$ for $\mu = 0$. The band velocities follow from Eq.~\eqref{eqev} and the tilt. We write them as $\bs{V}^{(s)} = \pm\,v_0\,\bs{\hat V}^{(s)}$, with
\begin{align}
\bs{\hat V}^{(2)} = \left\lbrace c_\gamma\,c_\phi + \eta_x,\, c_\gamma\,s_\phi,\, s_\gamma \right\rbrace,
\quad
\bs{\hat V}^{(1)} = \left\lbrace c_\gamma\,c_\phi - \eta_x,\, c_\gamma\,s_\phi,\, s_\gamma \right\rbrace.
\end{align}
The Drude weight is quadratic in the velocity, and the overall sign is therefore irrelevant. The vector $\bs{\hat V}^{(1)}$ follows from $\bs{\hat V}^{(2)}$ by $\eta_x \to -\,\eta_x$.

The exact measure is $\kappa\,(k_0 + \kappa\,c_\gamma)\,d\kappa\,d\gamma\,d\phi$. Inserting Eq.~\eqref{eq:axkappas} into Eqs.~\eqref{eqdos} and~\eqref{eqdrude}, we find that the band $s = 2$ contributes $\kappa_2\,(k_0 + \kappa_2\,c_\gamma)/\zeta_+ = \mathcal{W}_+/v_0$. The band $s = 1$ contributes the same expression with $\zeta_+ \to \zeta_-$ and $\mu_{\rm eff} \to |\mu_{\rm eff}|$ in the linear term. The weight is
\begin{align}
\mathcal{W}_\pm (\phi, \gamma) = \frac{k_0\,|\mu_{\rm eff}|}{\zeta_\pm^2} + \frac{\mu_{\rm eff}^2\,c_\gamma}{v_0\,\zeta_\pm^3}\,,
\end{align}
and we obtain
\begin{align}
\label{eq:axmaster}
& \rho (\mu) = \frac{g}{(2\,\pi)^3\,v_0^2} \left[ \int_{\mu_{\rm eff} > 0} \! d\phi\, d\gamma\; \mathcal{W}_+ + \int_{\mu_{\rm eff} < 0} \! d\phi\, d\gamma\; \mathcal{W}_- \right], \nn &
%%%%%%%%%%%%%%%%%%%%%%%%%%%%%
D_{ij} = \frac{g\,e^2}{8\,\pi^2} \left[ \int_{\mu_{\rm eff} > 0} 
\! d\phi\, d\gamma\; \hat V_i^{(2)}\,\hat V_j^{(2)}\,\mathcal{W}_+ + \int_{\mu_{\rm eff} < 0} \! d\phi\, d\gamma\; \hat V_i^{(1)}\,\hat V_j^{(1)}\,\mathcal{W}_- \right].
\end{align}
The prefactor of $D_{ij}$ is $\pi\,g\,e^2\,v_0^2/[(2\,\pi)^3\,v_0^2]$. The off-diagonal entries vanish, as argued in Sec.~\ref{secresponse}, and we need only $D_{xx}$, $D_{yy}$, and $D_{zz}$.

We now turn to the transition energies. The tilt enters Eq.~\eqref{eq:polinter} only through $v_0\,\bs{\eta} \cdot \bs{q} = v_0\,\eta_x\,q_x$. For $\bs{q} = q\,\bs{\hat y}$ and $\bs{q} = q\,\bs{\hat z}$, this vanishes because $k_x' = k_x$. For $\bs{q} = q\,\bs{\hat x}$, the difference $k_x' - k_x = q$ holds at every $\bs{k}$, and the shift equals $v_0\,\eta_x\,q$ for all angles. The denominators of Eq.~\eqref{eq:polinter} then become $\Omega \mp (\epsilon_k + \epsilon_{k'})$, with $\Omega = \omega - v_0\,\eta_x\,q$. The overlap factor depends only on $\bs{n}$ and $\bs{n}'$, and it is unchanged, because the tilt term is proportional to the identity.

For the parameters of Sec.~\ref{secvalidity}, $v_0\,k_0 = 0.064$ eV. The tilt $\tilde{\eta} = 0.03$ eV then gives $\eta_x = \tilde{\eta}/(v_0\,k_0) = 15/32$, which obeys $|\eta_x| < 1$. The ring-cyclide condition $|\tilde{\eta}| < \mu$ is equivalent to $\eta_x < \mu/(v_0\,k_0)$, which equals $0.1$ or $0.2$ for the two values of $\mu$. It is not met, and the FS is in the horn-cyclide regime.

%%%%%%%%%%%%%%%%%%
\subsection{Interband response at the intrinsic point}
\label{app:vnrxintr}

We give the steps behind Eq.~\eqref{eq:imintrxvnr} and the statements made after it. At fixed $\phi$, we use the local coordinates $u = \kappa\,c_\gamma$ and $w = \kappa\,s_\gamma$ of Appendix~\ref{app:vnrdirac}, for which $k_x = (k_0 + u)\,c_\phi$. The eigenvalues of Appendix~\ref{app:vnrxfs} then read
\begin{align}
\varepsilon_s = (-1)^s\,v_0\,\kappa + \tilde{\eta}\,c_\phi + v_0\,\eta_x\,c_\phi\,u\,.
\end{align}
The first term is the massless cone of Appendix~\ref{app:vnrdirac}. The second is a constant offset, proportional to the identity in band space. It acts as the local chemical potential $\mu_{\rm eff} = -\,\tilde{\eta}\,c_\phi$ at $\mu = 0$, and it creates a local pocket of radius $|\tilde{\eta}\,c_\phi|/v_0$. The third term is a tilt of the local cone along $u$. We drop it inside the cone, which modifies the Pauli-blocking boundary at relative order $\eta_x$. At $q \to 0$, the resonance $v_0\,(\kappa + \kappa') = \Omega$ selects $\kappa = \Omega/(2\,v_0)$. The transition at this angle is blocked only if $\Omega < 2\,|\tilde{\eta}\,c_\phi|$. For $\Omega > 2\,|\tilde{\eta}|$, no angle is blocked, and the local cones are the undoped ones. This holds for $\omega \gg |\tilde{\eta}|$ and $\omega \gg v_0\,\eta_x\,q$.

For $\bs{q} = q\,\bs{\hat x}$, the shift $v_0\,\eta_x\,q$ splits into two parts. The first is the local tilt acting on the local transfer $q\,c_\phi$ along $u$, namely $v_0\,\eta_x\,c_\phi\,(q\,c_\phi)$. The second is the change of the offset when $\phi$ moves to $\phi' \simeq \phi - q\,s_\phi/k_0$, which is
\begin{align}
\tilde{\eta}\,(\cos\phi' - \cos\phi) \simeq -\,\tilde{\eta}\,s_\phi\,(\phi' - \phi) = \frac{\tilde{\eta}\,q\,s_\phi^2}{k_0} = v_0\,\eta_x\,q\,s_\phi^2\,.
\end{align}
The two parts add up to $v_0\,\eta_x\,q\,(c_\phi^2 + s_\phi^2) = v_0\,\eta_x\,q$, in agreement with Appendix~\ref{app:vnrxfs}. The second part is of leading order in $\eta_x$, because $\tilde{\eta}/k_0 = v_0\,\eta_x$. It is the reason why the shift is the same for every $\phi$.

The local mapping of Appendix~\ref{app:vnrdirac} then gives
\begin{align}
\Im \Pi (q\,\bs{\hat e}, \omega) \simeq k_0 \int_0^{2\pi} 
\frac{d\phi}{2\,\pi}\, \Im \Pi_{\rm 2D} \left( Q_e (\phi), \Omega_e \right),
\quad \Im \Pi_{\rm 2D} (Q, \Omega) \simeq -\, \frac{g\,Q^2}{16\,\Omega}\,.
\end{align}
%%%%%%%%%%%%%%%%%%%
Here $Q_z = q$, $Q_x = q\,|c_\phi|$, and $Q_y = q\,|s_\phi|$ are the local transfers, with $\Omega_x = \Omega$ and $\Omega_y = \Omega_z = \omega$. The second relation follows from Appendix~\ref{app:vnrintrinsic} for $\Omega \gg v_0\,Q$. Using $\int_0^{2\pi} d\phi\, c_\phi^2/(2\,\pi) = \int_0^{2\pi} d\phi\, s_\phi^2/(2\,\pi) = 1/2$ and $1/\Omega \simeq (1/\omega)\,(1 + v_0\,\eta_x\,q/\omega)$, we obtain
\begin{align}
\label{eq:axintr}
\Im \Pi (q\,\bs{\hat z}, \omega) \simeq -\, \frac{g\,k_0\,q^2}{16\,\omega}\,, \quad
\Im \Pi (q\,\bs{\hat x}, \omega) \simeq -\, \frac{g\,k_0\,q^2}{32\,\Omega} 
\simeq -\, \frac{g\,k_0\,q^2}{32\,\omega} \left( 1 + \frac{v_0\,\eta_x\,q}{\omega} \right), \quad
\Im \Pi (q\,\bs{\hat y}, \omega) \simeq -\, \frac{g\,k_0\,q^2}{32\,\omega}\,,
\end{align}
which is Eq.~\eqref{eq:imintrxvnr}.

For $\bs{q} = q\,\bs{\hat z}$ and $\omega \gg v_0\,q$, Eq.~\eqref{eq:axintr} gives a nonzero imaginary part, and $\varepsilon^{\rm RPA}$ is complex throughout the window.
For $\omega < v_0\,q$, the interband part is real, and the pockets add a correction to Eq.~\eqref{eq:polmu0vnr}. The local mapping of Appendix~\ref{app:vnrdirac} is exact for the $\bs{\hat z}$ channel. We have $\phi' = \phi$, the local transfer is $Q = q$ along $w$, and the tilt does not shift the transition energies, because $\bs{\eta} \cdot \bs{q} = 0$. Hence
\begin{align}
\Pi (q\,\bs{\hat z}, \omega) = k_0 \int_0^{2\pi} \frac{d\phi}{2\,\pi}\, \Pi_{\rm 2D} \left( q, \omega; \mu_{\rm eff} (\phi) \right),
\quad \mu_{\rm eff} = -\,\tilde{\eta}\,c_\phi\,,
\end{align}
where $\Pi_{\rm 2D} (Q, \omega; \mu_{\rm eff})$ is the polarisability of the massless Dirac cone $h (u, w)$ with the local chemical potential $\mu_{\rm eff}$. We drop the local tilt along $u$, as above. The cone has the symmetries $\varepsilon_1 = -\,\varepsilon_2$, $\mathcal{F}_{11} = \mathcal{F}_{22}$, and $\mathcal{F}_{12} = \mathcal{F}_{21}$. They map a hole-pocket of the band $s = 1$ onto an electron-pocket of the band $s = 2$ and reverse the sign of $\omega$. The result below is even in $\omega$, and it therefore depends on $|\mu_{\rm eff}|$ only. We take $\mu_{\rm eff} > 0$ and write $\Pi_{\rm 2D} = \Pi_{\rm 2D}^{(0)} + \delta\Pi_{\rm 2D}$. The undoped part is $\Pi_{\rm 2D}^{(0)} = -\,g\,Q^2/\big( 16\,\sqrt{v_0^2\,Q^2 - \omega^2} \big)$, which gives Eq.~\eqref{eq:polmu0vnr} after the average over $\phi$.

The band $s = 2$ is occupied for $\kappa < \kappa_F = \mu_{\rm eff}/v_0$, and the band $s = 1$ is filled. With $n_{\bs{\kappa}} = \Theta (\kappa_F - \kappa)$, the occupations of Eq.~\eqref{eq:polgeneral} differ from the undoped ones only through $n_{\bs{\kappa}}$. The term with $s = s' = 1$ does not change. The terms $(s, s') = (1, 2)$ and $(2, 1)$ change by $-\,n_{\bs{\kappa}'}\,\mathcal{F}_{12}/(\omega - \epsilon_{\kappa} - \epsilon_{\kappa'})$ and $n_{\bs{\kappa}}\,\mathcal{F}_{12}/(\omega + \epsilon_{\kappa} + \epsilon_{\kappa'})$, respectively. The term $(2, 2)$ is the intraband response of the pocket carriers, with the weight $n_{\bs{\kappa}} - n_{\bs{\kappa}'}$ and $\mathcal{F}_{22}$. After a shift of the integration variable in the terms with $n_{\bs{\kappa}'}$, all four contributions are integrated over the pocket only. We define $\epsilon_\kappa = v_0\,\kappa$, $\epsilon_\kappa^\pm = v_0\,|\bs{\kappa} \pm \bs{Q}|$, and $\mathcal{F}_{ss'}^\pm$ for the overlap between $\bs{\kappa}$ and $\bs{\kappa} \pm \bs{Q}$. Then
\begin{align}
\delta\Pi_{\rm 2D} = g \int_{\kappa < \kappa_F} \frac{d^2 \bs{\kappa}}{(2\,\pi)^2}\, \mathcal{T} (\bs{\kappa})\,, \quad
\mathcal{T} = \frac{\mathcal{F}_{12}^+}{\omega + \epsilon_\kappa + \epsilon_\kappa^+} - \frac{\mathcal{F}_{12}^-}{\omega - \epsilon_\kappa - \epsilon_\kappa^-}
+ \frac{\mathcal{F}_{22}^+}{\omega + \epsilon_\kappa - \epsilon_\kappa^+} - \frac{\mathcal{F}_{22}^-}{\omega - \epsilon_\kappa + \epsilon_\kappa^-}\,.
\end{align}
The overlaps are $\mathcal{F}_{12} = (1 - \cos\alpha)/2$ and $\mathcal{F}_{22} = (1 + \cos\alpha)/2$, with $\alpha$ the angle between the two momenta. All denominators are nonzero for $\omega < v_0\,Q - 2\,v_0\,\kappa_F$, which is the lower edge of the intraband continuum of the pocket.

We expand $\mathcal{T}$ in $\kappa/Q$ at fixed $\theta$, the angle between $\bs{\kappa}$ and $\bs{Q}$. We write $c_\theta = \cos\theta$, $s_\theta = \sin\theta$, $\omega_+ = \omega + v_0\,Q$, and $\omega_- = \omega - v_0\,Q$. To first order, $\epsilon_\kappa^\pm = v_0\,(Q \pm \kappa\,c_\theta)$, $\cos\alpha_+ = c_\theta + \tilde{\delta}_\theta$, and $\cos\alpha_- = -\,c_\theta + \tilde{\delta}_\theta$, with $\tilde{\delta}_\theta = \kappa\,s_\theta^2/Q$. At $\kappa \to 0$, the four terms are $(1 - c_\theta)/(2\,\omega_+)$, $-\,(1 + c_\theta)/(2\,\omega_-)$, $(1 + c_\theta)/(2\,\omega_-)$, and $-\,(1 - c_\theta)/(2\,\omega_+)$. They cancel in pairs, and $\mathcal{T}$ has no term of order $\kappa^0$. The terms of order $\kappa$ combine to
\begin{align}
\mathcal{T} = s_\theta^2\,\kappa \left[ -\,\frac{1}{Q\,\omega_+} + \frac{1}{Q\,\omega_-} - \frac{v_0}{\omega_+^2} - \frac{v_0}{\omega_-^2} \right]
= -\,\frac{4\,v_0^3\,Q^2\,\sin^2\theta\;\kappa}{\left( v_0^2\,Q^2 - \omega^2 \right)^2}\,,
\end{align}
where we have used $A - B = 2\,v_0\,Q$ and $A\,B = \omega^2 - v_0^2\,Q^2$. The integral over the pocket uses $\int_0^{2\pi} d\theta\,\sin^2\theta = \pi$ and $\int_0^{\kappa_F} \kappa^2\,d\kappa = \kappa_F^3/3$. With $v_0\,\kappa_F = |\mu_{\rm eff}|$, we obtain
\begin{align}
\delta\Pi_{\rm 2D} (Q, \omega) = -\,\frac{g\,|\mu_{\rm eff}|^3\,Q^2}{3\,\pi \left( v_0^2\,Q^2 - \omega^2 \right)^2}\,.
\end{align}
This is a relative correction $(16/3\,\pi)\,|\mu_{\rm eff}|^3/(v_0^2\,Q^2 - \omega^2)^{3/2}$ to $\Pi_{\rm 2D}^{(0)}$. It is cubic in $|\mu_{\rm eff}|$, because the intraband response of the pocket carriers, of order $|\mu_{\rm eff}|^2/(v_0\,Q)$, cancels against the Pauli-blocking of the interband transitions at that order.

We now insert $\mu_{\rm eff} = -\,\tilde{\eta}\,c_\phi$ and use $\int_0^{2\pi} d\phi\,|c_\phi|^3/(2\,\pi) = 4/(3\,\pi)$. The pocket correction of the $\bs{\hat z}$ channel is
\begin{align}
\label{eq:dpipocketvnr}
\delta\Pi (q\,\bs{\hat z}, \omega) = -\,\frac{4\,g\,k_0\,q^2\,|\tilde{\eta}|^3}{9\,\pi^2 \left( v_0^2\,q^2 - \omega^2 \right)^2}\,,
\quad \omega < v_0\,q - 2\,|\tilde{\eta}|\,.
\end{align}
Its ratio to $\Pi (q\,\bs{\hat z}, \omega)$ of Eq.~\eqref{eq:polmu0vnr} is $64\,|\tilde{\eta}|^3/[9\,\pi^2\,(v_0^2\,q^2 - \omega^2)^{3/2}]$. The correction is negative, as is Eq.~\eqref{eq:polmu0vnr}. From Eq.~\eqref{eq:epsrpa}, the dielectric function becomes
\begin{align}
\varepsilon^{\rm RPA} (q\,\bs{\hat z}, \omega) = 1 + \frac{\pi\,g\,e^2\,k_0}{4\,\sqrt{v_0^2\,q^2 - \omega^2}} \left[ 1 + \frac{64\,|\tilde{\eta}|^3}{9\,\pi^2 \left( v_0^2\,q^2 - \omega^2 \right)^{3/2}} \right].
\end{align}
The pockets therefore add a positive term to Eq.~\eqref{eq:epsintrvnr}. Hence $\varepsilon^{\rm RPA} > 1$ persists, and there is no zero below the continuum.

For $\bs{q} = q\,\bs{\hat x}$, the local threshold $v_0\,q\,|c_\phi|$ vanishes at $\phi = \pi/2$. As in Appendix~\ref{app:vnrintrinsic}, the angular integrand of $\Im \Pi$ is positive on a set of angles of finite measure for every $\Omega > 0$. The continuum therefore extends down to the lower edge of the window $\omega \gg |\tilde{\eta}|$, and $\varepsilon^{\rm RPA}$ is complex throughout it.

The vortex term $\delta\Pi$ is derived in Appendix~\ref{app:vnrxvortex}. At the intrinsic point, no angle is blocked in the window above, and Eq.~\eqref{eq:imdpivnr} applies with $\omega \to \Omega$ and without the step function. The tilt thus changes it by the relative amount $v_0\,\eta_x\,q/\omega$. This is a correction to a term that is already suppressed by $(\omega/2\,v_0\,k_0)^2$, and we drop it.

%%%%%%%%%%%%%%%%%%
\subsection{Ring-cyclide regime: density-of-states, Drude weights, and bare plasmons}
\label{app:vnrxring}

We give the steps behind Eqs.~\eqref{eq:dosxvnr}, \eqref{eq:dxvnr}, and~\eqref{eq:wpxvnr}, together with the ratios below them. For $|\tilde{\eta}| < \mu$, only the first terms of Eq.~\eqref{eq:axmaster} contribute, and the angular integrals run over the full range. With $\mu_{\rm eff} = \mu - v_0\,\eta_x\,k_0\,c_\phi$ and the expansions
\begin{align}
\zeta_+^{-2} = 1 - 2\,\eta_x\,c_\gamma\,c_\phi + 3\,\eta_x^2\,c_\gamma^2\,c_\phi^2 + \mathcal{O} (\eta_x^3)\,,
\quad
\zeta_+^{-3} = 1 - 3\,\eta_x\,c_\gamma\,c_\phi + 6\,\eta_x^2\,c_\gamma^2\,c_\phi^2 + \mathcal{O} (\eta_x^3)\,,
\end{align}
the weight $\mathcal{W}_+$ becomes a polynomial in $\eta_x$. Its second term contains $\mu^2/v_0$ times $c_\gamma$, $c_\gamma^2\,c_\phi$, and $c_\gamma^3\,c_\phi^2$ at the orders $\eta_x^0$, $\eta_x^1$, and $\eta_x^2$, respectively. Each of these is odd under $\gamma \to \pi - \gamma$ or under $\phi \to \pi - \phi$ once multiplied by $1$, $s_\gamma^2$, $\left( c_\gamma\,c_\phi + \eta_x \right)^2$, or $c_\gamma^2\,s_\phi^2$, up to the order we keep. These terms integrate to zero and are dropped. The remaining terms are the first term of $\mathcal{W}_+$ and the pieces of the second term that contain $\tilde{\eta}/v_0 = \eta_x\,k_0$. In particular, the cross-term between the offset $\tilde{\eta}\,c_\phi$ and the factor $c_\gamma$ of the measure is kept. It gives
\begin{align}
\label{eq:axwring}
\mathcal{W}_+ \simeq k_0\,\mu \left( 1 - 4\,\eta_x\,c_\gamma\,c_\phi + 9\,\eta_x^2\,c_\gamma^2\,c_\phi^2 \right)
- v_0\,\eta_x\,k_0^2\,c_\phi \left( 1 - 3\,\eta_x\,c_\gamma\,c_\phi \right).
\end{align}
The coefficient $4$ is the sum of $2$ from $\zeta_+^{-2}$ and $2$ from the cross-term. The coefficient $9$ is the sum of $3$ from $\zeta_+^{-2}$ and $6$ from the cross-term with $\zeta_+^{-3}$. The last bracket collects the pieces proportional to $v_0\,\eta_x\,k_0^2$.

The angular integrals needed below are
\begin{align}
\label{eq:axangint}
& \int d\phi\, d\gamma\, c_\gamma^2\,c_\phi^2 = \pi^2\,, \quad
\int d\phi\, d\gamma\, s_\gamma^2 = 2\,\pi^2\,, \quad
\int d\phi\, d\gamma\, s_\gamma^2\,c_\gamma^2\,c_\phi^2 = \frac{\pi^2}{4}\,, \quad
\int d\phi\, d\gamma\, c_\gamma^4\,c_\phi^4 = \frac{9\,\pi^2}{16}\,, \nn &
\int d\phi\, d\gamma\, c_\gamma^4\,c_\phi^2\,s_\phi^2 = \frac{3\,\pi^2}{16}\,.
\end{align}
%%%%%%%%%%%%%%%%%%%%%%%%%%%%%%%%%%%
Terms odd in $c_\gamma$ or in $c_\phi$ vanish. Equation~\eqref{eq:axwring} then gives
\begin{align}
\label{eq:axringint}
& \int d\phi\, d\gamma\, \mathcal{W}_+ = k_0\,\mu 
\left( 4\,\pi^2 + 9\,\pi^2\,\eta_x^2 \right), \quad
\int d\phi\, d\gamma\, s_\gamma^2\,\mathcal{W}_+ = k_0\,\mu 
\left( 2\,\pi^2 + \frac{9\,\pi^2\,\eta_x^2}{4} \right), \nn
%%%%%%%%%%%%
 & \int d\phi\, d\gamma\, \left( c_\gamma\,c_\phi + \eta_x \right)^2 \mathcal{W}_+ 
= k_0\,\mu\,\pi^2 \left( 1 + 4\,\eta_x^2 - 8\,\eta_x^2 + \frac{81\,\eta_x^2}{16} \right),
%%%%%%%%%%%
\quad \int d\phi\, d\gamma\, c_\gamma^2\,s_\phi^2\,\mathcal{W}_+ &= k_0\,\mu \left( \pi^2 + \frac{27\,\pi^2\,\eta_x^2}{16} \right).
\end{align}
In the third line, the three corrections come from $\eta_x^2$ in the velocity multiplied by the leading weight, from the cross-term $2\,\eta_x\,c_\gamma\,c_\phi$ in the velocity multiplied by $-\,4\,\eta_x\,c_\gamma\,c_\phi$ in the weight, and from $c_\gamma^2\,c_\phi^2$ in the velocity multiplied by the $\eta_x^2$ term of the weight. The terms proportional to $v_0\,\eta_x\,k_0^2$ in Eq.~\eqref{eq:axwring} drop out of all four integrals. They carry an odd power of $c_\phi$ or of $c_\gamma$ in every product.

The first line of Eq.~\eqref{eq:axringint} gives $\rho (\mu)$ with the prefactor $g/[(2\,\pi)^3\,v_0^2]$, and we recover Eq.~\eqref{eq:dosxvnr}. The other lines have the prefactor $g\,e^2/(8\,\pi^2)$ of Eq.~\eqref{eq:axmaster}. They give
\begin{align}
D_{zz} = \frac{g\,e^2\,k_0\,\mu}{4} \left( 1 + \frac{9\,\eta_x^2}{8} \right), \quad
D_{xx} = \frac{g\,e^2\,k_0\,\mu}{8} \left( 1 + \frac{17\,\eta_x^2}{16} \right), \quad
D_{yy} = \frac{g\,e^2\,k_0\,\mu}{8} \left( 1 + \frac{27\,\eta_x^2}{16} \right),
\end{align}
which is Eq.~\eqref{eq:dxvnr}. The correction to $D_{yy}$ arises only from the deformation of the FS, since $\hat V_y^{(2)}$ does not involve $\eta_x$. The plasmon frequencies of Eq.~\eqref{eq:wpxvnr} follow from $\omega_{p,i}^2 = 4\,D_{ii}$. For the ratios, we expand to the order $\eta_x^2$ and obtain
\begin{align}
\frac{\omega_{p,z}^2}{\omega_{p,x}^2} = 2\,\frac{1 + 9\,\eta_x^2/8}{1 + 17\,\eta_x^2/16} 
\simeq 2 \left( 1 + \frac{\eta_x^2}{16} \right), \quad
\frac{\omega_{p,x}^2}{\omega_{p,y}^2} \simeq 1 - \frac{5\,\eta_x^2}{8}\,, \quad
\frac{\omega_{p,z}^2}{\omega_{p,y}^2} \simeq 2 \left( 1 - \frac{9\,\eta_x^2}{16} \right).
\end{align}
Taking the square root gives the ratios of the frequencies quoted below Eq.~\eqref{eq:wpxvnr}. The condition $|\tilde{\eta}| < \mu$ is equivalent to $\eta_x < \mu/(v_0\,k_0) \equiv \kappa_F/k_0$ with $\kappa_F = \mu/v_0$, and it makes the $\eta_x^2$ corrections of order $(\kappa_F/k_0)^2$.

%%%%%%%%%%%%%%%%%%
\subsection{Ring-cyclide regime: interband screening}
\label{app:vnrxringscreen}

We give the steps behind Eqs.~\eqref{eq:pinterxvnr} and~\eqref{eq:ebxvnr}, and the statements on the screened frequencies. At fixed $\phi$, the local cone has the chemical potential $\mu_{\rm eff} (\phi)$. The blocked states are those with $v_0\,\kappa < |\mu_{\rm eff}|$, as in the discussion of Appendix~\ref{app:vnrxintr}. For $v_0\,q \ll \omega \ll v_0\,k_0$, the doped result of Appendix~\ref{app:vnrscreen} holds in each copy with $\mu \to |\mu_{\rm eff}|$ and with the local transfer $Q_e (\phi)$. Averaging over $\phi$ with the weights $\check w_z = 1$, $\check w_x = c_\phi^2$, and $\check w_y = s_\phi^2$ gives Eq.~\eqref{eq:pinterxvnr}. The shift $v_0\,\eta_x\,q$ of $\omega$ in the $\bs{\hat x}$ channel is neglected for $\omega \gg v_0\,q$.

The interband onset is the smallest value of $2\,v_0\,\kappa_2 (\phi, \gamma) = 2\,\mu_{\rm eff}/\zeta_+$, since the band $s = 1$ is completely filled. For fixed $\phi$ with $c_\phi > 0$, the minimum over $\gamma$ is at $c_\gamma = 1$, which gives $(\mu - \tilde{\eta}\,c_\phi)/(1 + \eta_x\,c_\phi)$. Its derivative with respect to $c_\phi$ is $-\left( \tilde{\eta} + \eta_x\,\mu \right)/(1 + \eta_x\,c_\phi)^2 < 0$, and the minimum is at $c_\phi = 1$. For $c_\phi < 0$, the minimum is at $c_\gamma = -1$, which gives $(\mu + \tilde{\eta}\,|c_\phi|)/(1 + \eta_x\,|c_\phi|)$. Its derivative with respect to $|c_\phi|$ is $\eta_x\,(v_0\,k_0 - \mu)/(1 + \eta_x\,|c_\phi|)^2 > 0$, and the minimum is at $c_\phi = 0$, where it equals $\mu$. The global minimum is therefore
\begin{align}
\omega_{\rm th} = \frac{2\,(\mu - \tilde{\eta})}{1 + \eta_x}\,,
\end{align}
at $\phi = \gamma = 0$. For $\eta_x < 0$, it is at $\phi = \pi$ and $\gamma = 0$, and $\tilde{\eta} \to |\tilde{\eta}|$, $\eta_x \to |\eta_x|$. Equation~\eqref{eq:pinterxvnr} has the onset $2\,(\mu - |\tilde{\eta}|)$, because the local tilt along $u$ was dropped. The two differ by the factor $1 + |\eta_x|$.

For $\omega \ll 2\,(\mu - |\tilde{\eta}|)$, the logarithm becomes $\ln [(2\,|\mu_{\rm eff}| + \omega)/(2\,|\mu_{\rm eff}| - \omega)] \simeq \omega/|\mu_{\rm eff}|$, and the static values follow as
\begin{align}
\epsilon_{b,i} = 1 + \frac{g\,e^2\,k_0}{4} \int_0^{2\pi} \frac{d\phi}{2\,\pi}\, \frac{\check w_i (\phi)}{\mu - \tilde{\eta}\,c_\phi}\,.
\end{align}
The prefactor is $4\,\pi\,e^2\,g\,k_0/(16\,\pi)$. With $\tilde w = \sqrt{\mu^2 - \tilde{\eta}^2}$, the standard integral gives $\int_0^{2\pi} d\phi/(2\,\pi)\,(\mu - \tilde{\eta}\,c_\phi)^{-1} = 1/\tilde w$. For the weight $c_\phi^2$, we write $c_\phi = (\mu - \mu_{\rm eff})/\tilde{\eta}$ and use $\langle \mu_{\rm eff} \rangle = \mu$. This gives
\begin{align}
\left\langle \frac{c_\phi^2}{\mu_{\rm eff}} \right\rangle = \frac{\mu^2/\tilde w - 2\,\mu + \mu}{\tilde{\eta}^2} = \frac{\mu\,(\mu - \tilde w)}{\tilde w\,\tilde{\eta}^2} = \frac{\mu}{\tilde w\,(\mu + \tilde w)}\,,
\end{align}
where we have used $\tilde{\eta}^2 = (\mu - \tilde w)\,(\mu + \tilde w)$. The weight $s_\phi^2$ gives the difference $1/\tilde w - \mu/[\tilde w\,(\mu + \tilde w)] = 1/(\mu + \tilde w)$. Multiplying by $g\,e^2\,k_0/4$ reproduces Eq.~\eqref{eq:ebxvnr}. These expressions reduce to Eq.~\eqref{eq:ebvnr} for $\tilde{\eta} \to 0$.

Below the onset, each term of the average has the form $(2/\omega)\,\operatorname{arctanh} [\omega/(2\,|\mu_{\rm eff}|)]$. Its series in $\omega$ has positive coefficients, and it increases with $\omega$ for $\omega < 2\,|\mu_{\rm eff}|$. Hence $\epsilon_{b,i} (\omega)$ increases with $\omega$ below the onset. The Drude term $-\,\omega_{p,i}^2/\omega^2$ increases as well, and $\Re \varepsilon^{\rm RPA}$ is monotonic, tending to $-\,\infty$ for $\omega \to 0$. Its zero is unique. Since $\epsilon_{b,i} (\omega) \geq \epsilon_{b,i} (0)$, we have $\Re \varepsilon^{\rm RPA} \geq \epsilon_{b,i} (0) - \omega_{p,i}^2/\omega^2$, which vanishes at the static screened frequency. The zero therefore cannot exceed $\tilde\omega_{p,i}$. At the onset, the minimum of $|\mu_{\rm eff}|$ is quadratic in $\phi$ near $\phi = 0$, and the logarithm has a singularity of the type $\ln |\phi^2 - \phi_\omega^2|$. This is integrable, and $\Re \varepsilon^{\rm RPA}$ stays finite. If $\tilde\omega_{p,i} < \omega_{\rm th}$, then $\Re \varepsilon^{\rm RPA} (\omega_{\rm th}) \geq \epsilon_{b,i} (0) - \omega_{p,i}^2/\omega_{\rm th}^2 > 0$, and a zero exists below the onset. The functions $\epsilon_{b,z}$ and $\epsilon_{b,x}$ diverge as $1/\tilde w$ for $|\tilde{\eta}| \to \mu$, and $\epsilon_{b,y}$ grows to $1 + g\,e^2\,k_0/(4\,\mu)$. The screening therefore becomes stronger as the ring-cyclide approaches the horn-cyclide.

%%%%%%%%%%%%%%%%%%
\subsection{Horn-cyclide regime: density-of-states, Drude weights, and bare plasmons}
\label{app:vnrxhorn}

We give the steps behind Eqs.~\eqref{eq:horn0vnr}, \eqref{eq:horncycvnr}, and~\eqref{eq:horncorrvnr}. We take $\tilde{\eta} > 0$ and $0 \leq \mu < \tilde{\eta}$. The electron arc is $E = (\phi_b, 2\,\pi - \phi_b)$ and the hole arc is $H = (-\phi_b, \phi_b)$, with $\cos\phi_b = \mu/\tilde{\eta}$, $\sin\phi_b = \varsigma$, and $\bar{w} = \tilde{\eta}\,\varsigma$.

At leading order, we set $\zeta_\pm \to 1$, $\mathcal{W}_\pm \to k_0\,|\mu_{\rm eff}|$, and $\hat V^{(1)} = \hat V^{(2)} = \left\lbrace c_\gamma\,c_\phi,\, c_\gamma\,s_\phi,\, s_\gamma \right\rbrace$. The hole arc therefore adds to the electron arc. With $\int_0^{2\pi} d\gamma = 2\,\pi$, $\int_0^{2\pi} d\gamma\, s_\gamma^2 = \int_0^{2\pi} d\gamma\, c_\gamma^2 = \pi$, and the moments
\begin{align}
\mathcal{J}_n = \int_0^{2\pi} d\phi\, |\mu_{\rm eff}|\,c_\phi^n\,,
\end{align}
Eq.~\eqref{eq:axmaster} gives
\begin{align}
\label{eq:axhornlead}
\rho = \frac{g\,k_0}{4\,\pi^2\,v_0^2}\,\mathcal{J}_0\,, \quad
D_{zz} = \frac{g\,e^2\,k_0}{8\,\pi}\,\mathcal{J}_0\,, \quad
D_{xx} = \frac{g\,e^2\,k_0}{8\,\pi}\,\mathcal{J}_2\,, \quad
D_{yy} = \frac{g\,e^2\,k_0}{8\,\pi} \left( \mathcal{J}_0 - \mathcal{J}_2 \right).
\end{align}
For $\mathcal{J}_0$ we use $\int_E - \int_H = \int_{\rm all} - 2\int_H$, which holds for the integrand $\mu_{\rm eff}\,c_\phi^n$ because $|\mu_{\rm eff}| = \pm\,\mu_{\rm eff}$ on $E$ and $H$. Define $h_n = \int_{-\phi_b}^{\phi_b} c_\phi^n\,d\phi$ and $A_n = \int_0^{2\pi} c_\phi^n\,d\phi$. Then
\begin{align}
& h_0 = 2\,\phi_b\,, \quad h_1 = 2\,\varsigma\,, \quad 
h_2 = \phi_b + \tfrac{1}{2}\,\sin \!\left( 2\,\phi_b \right) \,, \quad
h_3 = 2 \left( \varsigma - \frac{\varsigma^3}{3} \right), \quad
h_4 = \frac{3\,\phi_b}{4} + \frac{\sin \!\left( 2\,\phi_b \right) }{2} + \frac{\sin \!\left( 4\,\phi_b \right) }{16}\,, \quad
\nn & h_5 = 2 \left( \varsigma - \frac{2\,\varsigma^3}{3} + \frac{\varsigma^5}{5} \right),
\end{align}
and $A_0 = 2\,\pi$, $A_2 = \pi$, $A_4 = 3\,\pi/4$, with $A_n = 0$ for odd $n$. We obtain
\begin{align}
\mathcal{J}_0 = \mu\,A_0 - 2\,(\mu\,h_0 - h_1) = 2\,\pi\,\mu - 4\,\mu\,\phi_b + 4\,\tilde{\eta}\,\varsigma\,.
\end{align}
Using $\phi_b = \pi/2 - \arcsin (\mu/\tilde{\eta})$, this equals $4\,[\mu \arcsin (\mu/\tilde{\eta}) + \bar{w}]$, and the first relation of Eq.~\eqref{eq:axhornlead} gives the DOS of Eq.~\eqref{eq:horncycvnr}. The same method for $c_\phi^2$ gives
\begin{align}
\mathcal{J}_2 = \mu\,(A_2 - 2\,h_2) + 2\,\tilde{\eta}\,h_3
= 2\,\mu \arcsin \!\left(  \frac{\mu}{\tilde{\eta}} \right)  
+ \frac{2\,\bar{w}\,(4\,\tilde{\eta}^2 - \mu^2)}{3\,\tilde{\eta}^2}\,.
\end{align}
Here we have used $\sin \!\left( 2\,\phi_b \right)  = 2\,\varsigma\,\mu/\tilde{\eta}$ and $\mu^2/\tilde{\eta}^2 = 1 - \varsigma^2$. Inserting $\mathcal{J}_0$ and $\mathcal{J}_2$ into Eq.~\eqref{eq:axhornlead} reproduces $D_{zz}$, $D_{xx}$, and $D_{yy}$ of Eq.~\eqref{eq:horncycvnr}. The prefactors are $g\,e^2\,k_0/(2\,\pi)$ for $D_{zz}$ and $g\,e^2\,k_0/(4\,\pi)$ for the other two.

At $\mu = 0$, we have $\phi_b = \pi/2$, $\mathcal{J}_0 = 4\,\tilde{\eta}$, and $\mathcal{J}_2 = 8\,\tilde{\eta}/3$, which gives Eq.~\eqref{eq:horn0vnr} and the frequencies $\omega_{p,z}^2 = 2\,g\,e^2\,k_0\,\tilde{\eta}/\pi$, $\omega_{p,x}^2 = 4\,g\,e^2\,k_0\,\tilde{\eta}/(3\,\pi)$, and $\omega_{p,y}^2 = 2\,g\,e^2\,k_0\,\tilde{\eta}/(3\,\pi)$. At $\mu = \tilde{\eta}$, we have $\mathcal{J}_0 = 2\,\pi\,\tilde{\eta}$ and $\mathcal{J}_2 = \pi\,\tilde{\eta}$, which join onto the ring-cyclide values $\rho = g\,k_0\,\tilde{\eta}/(2\,\pi\,v_0^2)$ and $D_{zz} = g\,e^2\,k_0\,\tilde{\eta}/4$. The anisotropy ratios are $D_{zz}/D_{xx} = \mathcal{J}_0/\mathcal{J}_2$ and $D_{yy}/D_{xx} = \mathcal{J}_0/\mathcal{J}_2 - 1$. They equal $3/2$ and $1/2$ at $\mu = 0$, and $2$ and $1$ at $\mu = \tilde{\eta}$. Their monotonic increase in between is seen in Fig.~\ref{fig:drudexvnr}(b).

We now turn to the corrections of order $\eta_x^2$. The hole band follows from the electron band by $\eta_x \to -\,\eta_x$ in $\zeta_\pm$ and in $\hat V$, with $\mu_{\rm eff} \to |\mu_{\rm eff}|$ in the linear term of $\mathcal{W}$. The quantity $\tilde{\eta} = v_0\,\eta_x\,k_0$ is not flipped, since it only enters through $\mu_{\rm eff}$. The expansions of Appendix~\ref{app:vnrxring} apply on each arc, and the terms proportional to $\mu^2/v_0$ again vanish by parity. The $\gamma$-integrals of the band $s = 2$ are
%%%%%%%%%%%%%%%%%%%%
\begin{align}
& \int d\gamma\, \mathcal{W}_+ = 2\,\pi\,k_0\,\mu_{\rm eff} 
\left( 1 + \tfrac{3}{2}\,\eta_x^2\,c_\phi^2 \right) 
- \frac{3\,\pi\,\eta_x}{v_0}\,c_\phi\,\mu_{\rm eff}^2\,, \quad
%%%%%%%%%%%%%%%%%%%%%%%%%%%5
\int d\gamma\, s_\gamma^2\,\mathcal{W}_+ = \pi\,k_0\,\mu_{\rm eff} 
\left( 1 + \tfrac{3}{4}\,\eta_x^2\,c_\phi^2 \right) 
- \frac{3\,\pi\,\eta_x}{4\,v_0}\,c_\phi\,\mu_{\rm eff}^2\,, \nn
%%%%%%%%%%%%%%%%%%%%
& \int d\gamma\, \big( \hat V_x^{(2)} \big)^2 \mathcal{W}_+ = \pi\,k_0\,\mu_{\rm eff} \left[ c_\phi^2 + \eta_x^2 \left( 2 - 4\,c_\phi^2 + \tfrac{9}{4}\,c_\phi^4 \right) \right] + \frac{\pi\,\eta_x}{v_0}\,\mu_{\rm eff}^2 \left( 2\,c_\phi - \tfrac{9}{4}\,c_\phi^3 \right), \nn
%%%%%%%%%%%%
& \int d\gamma\, \big( \hat V_y^{(2)} \big)^2 \mathcal{W}_+ = \pi\,k_0\,\mu_{\rm eff}\,s_\phi^2 \left( 1 + \tfrac{9}{4}\,\eta_x^2\,c_\phi^2 \right) - \frac{9\,\pi\,\eta_x}{4\,v_0}\,c_\phi\,s_\phi^2\,\mu_{\rm eff}^2\,.
\end{align}
For the band $s = 1$, the terms containing $\mu_{\rm eff}^2$ change sign, and $\mu_{\rm eff}$ in the first terms becomes $|\mu_{\rm eff}|$. Summing over the two arcs, the first terms give the moments $\mathcal{J}_n$ and the second terms give
\begin{align}
\mathcal{K}_n = \int_0^{2\pi} d\phi\, \operatorname{sgn} (\mu_{\rm eff})\,\mu_{\rm eff}^2\,c_\phi^n\,.
\end{align}
We also use $\eta_x/v_0 = \eta_x^2\,k_0/\tilde{\eta}$. Then
\begin{align}
\label{eq:axhorncorr}
\rho &= \frac{g\,k_0}{4\,\pi^2\,v_0^2} \left[ \mathcal{J}_0 + \frac{3\,\eta_x^2}{2} \left( \mathcal{J}_2 - \frac{\mathcal{K}_1}{\tilde{\eta}} \right) \right], \quad
%%%%%%%%%%%%%%%%%%%
D_{zz} = \frac{g\,e^2\,k_0}{8\,\pi} \left[ \mathcal{J}_0 + \frac{3\,\eta_x^2}{4} \left( \mathcal{J}_2 - \frac{\mathcal{K}_1}{\tilde{\eta}} \right) \right], \nn
D_{xx} &= \frac{g\,e^2\,k_0}{8\,\pi} \left[ \mathcal{J}_2 
+ \eta_x^2 \left( 2\,\mathcal{J}_0 - 4\,\mathcal{J}_2 + \frac{9}{4}\,\mathcal{J}_4 + \frac{2\,\mathcal{K}_1}{\tilde{\eta}} - \frac{9\,\mathcal{K}_3}{4\,\tilde{\eta}} \right) \right], \nn
%%%%%%%%%%%%%%%%%%%%%%
D_{yy} &= \frac{g\,e^2\,k_0}{8\,\pi} \left[ \mathcal{J}_0 - \mathcal{J}_2 + \frac{9\,\eta_x^2}{4} 
\left( \mathcal{J}_2 - \mathcal{J}_4 - \frac{\mathcal{K}_1 - \mathcal{K}_3}{\tilde{\eta}} \right) \right].
\end{align}
%%%%%%%%%%%%%%%%%%
The new moments are evaluated with the same method. For $\mathcal{J}_4$ we need $\int_{\rm all} - 2\int_H$ of $\mu_{\rm eff}\,c_\phi^4$, which equals $\mu\,A_4 - 2\,(\mu\,h_4 - h_5)$. For $\mathcal{K}_n$ we need the same combination for $\mu_{\rm eff}^2\,c_\phi^n$, with $\mu_{\rm eff}^2\,c_\phi = \mu^2\,c_\phi - 2\,\mu\,\tilde{\eta}\,c_\phi^2 + \tilde{\eta}^2\,c_\phi^3$ and the analogous expression for $c_\phi^3$. We use $\sin \!\left( 4\,\phi_b \right)  = 4\,\varsigma\,(\mu/\tilde{\eta})\,(1 - 2\,\varsigma^2)$ and $\mu^2/\tilde{\eta}^2 = 1 - \varsigma^2$, and we collect the terms in powers of $\varsigma$. The results are
\begin{align}
\mathcal{J}_0 &= 4\,\tilde{\eta}\,\Xi\,, \quad
\mathcal{J}_2 = \tilde{\eta} \left( 2\,\Xi + \frac{2\,\varsigma^3}{3} \right), \quad
\mathcal{J}_4 = \tilde{\eta} \left( \frac{3\,\Xi}{2} + \frac{5\,\varsigma^3}{6} - \frac{\varsigma^5}{5} \right), \nn
\mathcal{K}_1 &= \tilde{\eta}^2 \left( -\,4\,\Xi + \frac{4\,\varsigma^3}{3} \right), \quad
\mathcal{K}_3 = \tilde{\eta}^2 \left( -\,3\,\Xi + \varsigma^3 - \frac{2\,\varsigma^5}{15} \right),
\end{align}
with $\Xi = [\bar{w} + \mu \arcsin (\mu/\tilde{\eta})]/\tilde{\eta}$ as in the main text. The relation $\Xi = \varsigma + (\mu/\tilde{\eta})\,\arcsin (\mu/\tilde{\eta})$ was used. These results agree with direct evaluation at $\mu = 0$ and at $\mu = \tilde{\eta}$. For example, $\mathcal{K}_1 = -\,8\,\tilde{\eta}^2/3$ and $\mathcal{K}_3 = -\,32\,\tilde{\eta}^2/15$ at $\mu = 0$, and $\mathcal{K}_1 = -\,2\,\pi\,\tilde{\eta}^2$ and $\mathcal{K}_3 = -\,3\,\pi\,\tilde{\eta}^2/2$ at $\mu = \tilde{\eta}$.

Substituting these moments into Eq.~\eqref{eq:axhorncorr}, the combinations that appear are
\begin{align}
& \mathcal{J}_2 - \frac{\mathcal{K}_1}{\tilde{\eta}} 
= \tilde{\eta} \left( 6\,\Xi - \frac{2\,\varsigma^3}{3} \right), \quad
2\,\mathcal{J}_0 - 4\,\mathcal{J}_2 + \frac{9}{4}\,\mathcal{J}_4 + \frac{2\,\mathcal{K}_1}{\tilde{\eta}} - \frac{9\,\mathcal{K}_3}{4\,\tilde{\eta}} = \tilde{\eta} \left( \frac{17\,\Xi}{8} - \frac{3\,\varsigma^3}{8} - \frac{3\,\varsigma^5}{20} \right), \nn
%%%%%%%%%%%%
& \mathcal{J}_2 - \mathcal{J}_4 - \frac{\mathcal{K}_1 - \mathcal{K}_3}{\tilde{\eta}} 
= \tilde{\eta} \left( \frac{3\,\Xi}{2} - \frac{\varsigma^3}{2} + \frac{\varsigma^5}{15} \right).
\end{align}
The leading terms are $\mathcal{J}_0 = 4\,\tilde{\eta}\,\Xi$, $\mathcal{J}_2 = \tilde{\eta}\,(2\,\Xi + 2\,\varsigma^3/3)$, and $\mathcal{J}_0 - \mathcal{J}_2 = \tilde{\eta}\,(2\,\Xi - 2\,\varsigma^3/3)$. Inserting everything into Eq.~\eqref{eq:axhorncorr} gives the expressions of Eq.~\eqref{eq:horncorrvnr}. The first combination, multiplied by $3/4$ and divided by the leading factor $4\,\tilde{\eta}$, gives the correction $9\,\Xi/8 - \varsigma^3/8$ of $D_{zz}$. The second and third combinations, multiplied by $1/2$ and $9/8$, respectively, and divided by the leading factor $2\,\tilde{\eta}$, give the corrections $17\,\Xi/16 - 3\,\varsigma^3/16 - 3\,\varsigma^5/40$ of $D_{xx}$ and $27\,\Xi/16 - 9\,\varsigma^3/16 + 3\,\varsigma^5/40$ of $D_{yy}$.

At $\mu = 0$, $\Xi = \varsigma = 1$, and the corrections multiply the leading values by $1 + 2\,\eta_x^2$ for $\rho$, by $1 + \eta_x^2$ for $D_{zz}$, by $1 + 3\,\eta_x^2/5$ for $D_{xx}$, and by $1 + 9\,\eta_x^2/5$ for $D_{yy}$. For $D_{xx}$, the leading term is $\Xi + \varsigma^3/3 = 4/3$ and the correction is $17/16 - 3/16 - 3/40 = 4/5$. For $D_{yy}$, the leading term is $\Xi - \varsigma^3/3 = 2/3$ and the correction is $27/16 - 9/16 + 3/40 = 6/5$. At $\mu = \tilde{\eta}$, $\varsigma = 0$ and $\Xi = \pi/2$, and Eq.~\eqref{eq:horncorrvnr} reduces to the ring-cyclide corrections of Eqs.~\eqref{eq:dosxvnr} and~\eqref{eq:dxvnr}. The relations $\Pi_{\rm intra} (q\,\bs{\hat e}_i, \omega) \simeq q^2\,D_{ii}/(\pi\,e^2\,\omega^2)$ and $\omega_{p,i}^2 = 4\,D_{ii}$ do not depend on the regime, because their derivation in Appendix~\ref{app:vnrdrude} only uses $\delta (\varepsilon_s - \mu)$ and $\bs{V}^{(s)}$.

%%%%%%%%%%%%%%%%%%
\subsection{Horn-cyclide regime: interband response and damping}
\label{app:vnrxhorninter}

We give the steps behind Eqs.~\eqref{eq:fnvnr} to~\eqref{eq:epshornvnr} and the statements on the damping. For $v_0\,q \ll \omega \ll v_0\,k_0$, the local cone at the angle $\phi$ has the imaginary part $-\,g\,Q^2\,\Theta (\omega - 2\,|\mu_{\rm eff}|)/(16\,\omega)$. This follows from the doped result of Appendix~\ref{app:vnrscreen} with $\mu \to |\mu_{\rm eff}|$. Averaging over $\phi$ with the weights $1$, $c_\phi^2$, and $s_\phi^2 = 1 - c_\phi^2$, we obtain Eq.~\eqref{eq:imhornvnr} with $F_n$ of Eq.~\eqref{eq:fnvnr}.

The condition $\omega \geq 2\,|\mu - \tilde{\eta}\,c_\phi|$ is equivalent to $a_\omega \leq c_\phi \leq b_\omega$, with $a_\omega = \max [-1, (\mu - \omega/2)/\tilde{\eta}]$ and $b_\omega = \min [1, (\mu + \omega/2)/\tilde{\eta}]$. The set of angles with $\cos\phi \in [a_\omega, b_\omega]$ has the measure $2\,(\arcsin b_\omega - \arcsin a_\omega)$. Changing the variable to $c = \cos\phi$ gives
\begin{align}
F_0 = \frac{1}{\pi} \int_{a_\omega}^{b_\omega} \frac{dc}{\sqrt{1 - c^2}}\,, \quad
F_2 = \frac{1}{\pi} \int_{a_\omega}^{b_\omega} \frac{c^2\,dc}{\sqrt{1 - c^2}} = \frac{1}{2\,\pi} \left[ \arcsin c - c\,\sqrt{1 - c^2} \right]_{a_\omega}^{b_\omega} \,,
\end{align}
which are the expressions below Eq.~\eqref{eq:imhornvnr}. For $\omega > 2\,(\tilde{\eta} + \mu)$, the window covers $[-1, 1]$, and $F_0 = 1$ and $F_2 = 1/2$. For $\omega \to 0$ with $\mu < \tilde{\eta}$, the window shrinks around $\mu/\tilde{\eta}$, and $\arcsin b_\omega - \arcsin a_\omega \simeq (\omega/\tilde{\eta})/\sqrt{1 - \mu^2/\tilde{\eta}^2} = \omega/\bar{w}$. Hence $F_0 \simeq \omega/(\pi\,\bar{w})$, and $\Im \Pi (q\,\bs{\hat z}, \omega) \to -\,g\,k_0\,q^2/(16\,\pi\,\bar{w})$. The functions $F_0$ and $F_2$ are positive for every $\omega > 0$, because the window has a finite measure and $c^2$ vanishes only at a single point. The interband continuum therefore has no onset.

The imaginary part of $\varepsilon^{\rm RPA}$ follows from $\Im \varepsilon^{\rm RPA} = -\,4\,\pi\,e^2\,\Im \Pi/q^2$, which gives $\pi\,g\,e^2\,k_0\,F_n/(4\,\omega)$. Together with the Drude term, this yields Eq.~\eqref{eq:epshornvnr}, where the $\bs{\hat y}$ channel carries $F_n \to F_0 - F_2$ from the weight $s_\phi^2$. At $\mu = 0$ and for $\bs{q} = q\,\bs{\hat x}$, the static interband response follows from the average of Appendix~\ref{app:vnrxringscreen},
\begin{align}
\epsilon_{b,x} = 1 + \frac{g\,e^2\,k_0}{4} \int_0^{2\pi} \frac{d\phi}{2\,\pi}\, \frac{c_\phi^2}{\tilde{\eta}\,|c_\phi|} = 1 + \frac{g\,e^2\,k_0}{4\,\tilde{\eta}}\,\frac{2}{\pi} = 1 + \frac{g\,e^2\,k_0}{2\,\pi\,\tilde{\eta}}\,.
\end{align}
It is finite because the weight $c_\phi^2$ vanishes at the touching points.

For the $\bs{\hat z}$ channel, the weight is $1$, and the integrand has a logarithmic singularity at the two angles $\pm\,\phi_b$, where $\mu_{\rm eff}$ vanishes. We use
\begin{align}
\varepsilon^{\rm RPA}_{\rm inter} (q\,\bs{\hat z}, \omega) = \frac{g\,e^2\,k_0}{4\,\omega} \int_0^{2\pi} \frac{d\phi}{2\,\pi} \left[ \ln \left| \frac{\omega + 2\,|\mu_{\rm eff}|}{\omega - 2\,|\mu_{\rm eff}|} \right| + i\,\pi\,\Theta (\omega - 2\,|\mu_{\rm eff}|) \right].
\end{align}
Near each crossing, $|\mu_{\rm eff}| \simeq \bar{w}\,|\phi \mp \phi_b|$. With $\tilde{y} = 2\,\bar{w}\,|\phi \mp \phi_b|/\omega$, the integral over both sides of one crossing is $(\omega/\bar{w}) \int_0^{\tilde{Y}} d\tilde{y} \ln |(1 + \tilde{y})/(1 - \tilde{y})|$. The antiderivative gives $(1 + \tilde{Y})\,\ln (1 + \tilde{Y}) + (1 - \tilde{Y})\,\ln |1 - \tilde{Y}| \simeq 2\,\ln \tilde{Y} + 2$ for $\tilde{Y} \gg 1$, and the upper cut-off is $\tilde{Y} \sim \tilde{\eta}/\omega$. The regular part away from the crossings has the same $\omega$-dependence as a constant times $\omega$, and it only adds to the constant. Two crossings and the factor $1/(2\,\pi)$ then give
\begin{align}
\varepsilon^{\rm RPA}_{\rm inter} (q\,\bs{\hat z}, \omega) \simeq \frac{g\,e^2\,k_0}{2\,\pi\,\bar{w}} \left[ \ln \frac{\tilde{\eta}}{\omega} + c \right] + i\,\frac{g\,e^2\,k_0}{4\,\bar{w}}\,.
\end{align}
The imaginary part follows from $F_0 \simeq \omega/(\pi\,\bar{w})$, and it equals $(g\,e^2\,k_0/2\,\pi\,\bar{w})\,(i\,\pi/2)$. This is the form quoted in the main text. The constant $c$ is real, of order unity, and depends on $\mu/\tilde{\eta}$ only. In the $\bs{\hat x}$ and $\bs{\hat y}$ channels, the weights at the crossings are $c_{\phi_b}^2 = \mu^2/\tilde{\eta}^2$ and $s_{\phi_b}^2 = \bar{w}^2/\tilde{\eta}^2$. This gives the extra factors stated in the main text, to leading logarithmic accuracy.

For the damping, we write $\Re \varepsilon^{\rm RPA} \simeq \epsilon_{b,i} (\omega) - \omega_{p,i}^2/\omega^2$ and $\Im \varepsilon^{\rm RPA} = \pi\,g\,e^2\,k_0\,F_n/(4\,\omega)$. The real part vanishes at $\tilde\omega_{p,i}$, where $\omega_{p,i}^2 = \epsilon_{b,i}\,\tilde\omega_{p,i}^2$. Neglecting the logarithmic derivative of $\epsilon_{b,i}$, its slope there is $2\,\omega_{p,i}^2/\tilde\omega_{p,i}^3 = 2\,\epsilon_{b,i}/\tilde\omega_{p,i}$. The zero of the complex $\varepsilon^{\rm RPA}$ is then shifted to $\omega = \tilde\omega_{p,i} - i\,\Gamma_i$, with
\begin{align}
\Gamma_i = \frac{\Im \varepsilon^{\rm RPA}}{\partial_\omega \Re \varepsilon^{\rm RPA}} = \frac{\pi\,g\,e^2\,k_0\,F_n (\tilde\omega_{p,i})}{4\,\tilde\omega_{p,i}}\,\frac{\tilde\omega_{p,i}}{2\,\epsilon_{b,i}} = \frac{\pi\,g\,e^2\,k_0}{8\,\epsilon_{b,i}}\,F_n (\tilde\omega_{p,i})\,.
\end{align}
Here $F_n$ is replaced by $F_0$ for $i = z$, by $F_2$ for $i = x$, and by $F_0 - F_2$ for $i = y$. The last is positive because $F_0 - F_2 = (1/\pi) \int_{a_\omega}^{b_\omega} \sqrt{1 - c^2}\,dc$ has a positive integrand. At $\mu = 0$ and for $g\,e^2\,k_0 \ll \tilde{\eta}$, we have $\omega_{p,z}^2/\tilde{\eta}^2 = 2\,g\,e^2\,k_0/(\pi\,\tilde{\eta}) \ll 1$. The mode then lies well below $2\,\tilde{\eta}$, where $F_0 = (2/\pi)\arcsin [\omega/(2\,\tilde{\eta})] \simeq \omega/(\pi\,\tilde{\eta})$. With $\epsilon_{b,z} \simeq 1$ up to logarithmic factors, we obtain $\Gamma_z = g\,e^2\,k_0\,\omega_{p,z}/(8\,\tilde{\eta})$, and $\Gamma_z/\omega_{p,z} = g\,e^2\,k_0/(8\,\tilde{\eta})$. In the opposite limit $g\,e^2\,k_0 \gg \tilde{\eta}$, the static background of the $\bs{\hat x}$ channel is $\epsilon_{b,x} \simeq g\,e^2\,k_0/(2\,\pi\,\tilde{\eta})$. The static estimate $\tilde\omega_{p,x}^2 = \omega_{p,x}^2/\epsilon_{b,x}$ is then of order $\tilde{\eta}^2$, where the frequency dependence of $\epsilon_{b,x}$ is no longer negligible. The plasmons are strongly screened in this limit, as stated in the main text.

%%%%%%%%%%%%%%%%%%
\subsection{Toroidal-angle dependence of the overlap factor at finite tilt}
\label{app:vnrxvortex}

We give the steps behind Eqs.~\eqref{eq:imdpixvnr} and~\eqref{eq:debxvnr}. The second term of Eq.~\eqref{eq:fsplit} vanishes for $\bs{q} = q\,\bs{\hat z}$, because $\phi' = \phi$. For $\bs{q} = q\,\bs{\hat x}$, Appendix~\ref{app:vnrdirac} gives $\phi' - \phi \simeq -\,q\,s_\phi/k_0$. For $\bs{q} = q\,\bs{\hat y}$, we have $k_x' = k_x$ and $k_y' = k_y + q$, and the relation $k_\perp'\,\sin (\phi' - \phi) = k_y'\,c_\phi - k_x'\,s_\phi = q\,c_\phi$ gives $\phi' - \phi \simeq q\,c_\phi/k_0$. To leading order, $\sin\gamma\,\sin\gamma' \to s_\gamma^2$ and $1 - \cos (\phi' - \phi) \simeq (\phi' - \phi)^2/2$, and
\begin{align}
\delta\mathcal{F}_x = \frac{q^2\,s_\gamma^2\,s_\phi^2}{4\,k_0^2}\,, \quad
\delta\mathcal{F}_y = \frac{q^2\,s_\gamma^2\,c_\phi^2}{4\,k_0^2}\,.
\end{align}
The weights $s_\phi^2$ and $c_\phi^2$ are interchanged relative to $\check w_x$ and $\check w_y$ of Eq.~\eqref{eq:pinterxvnr}. The tilt term is proportional to the identity, and it leaves $\delta\mathcal{F}$ unchanged.

Inserting $\delta\mathcal{F}$ into Eq.~\eqref{eq:polinter} with the bracket $4\,\epsilon_k/[(\Omega + i\,0^+)^2 - 4\,\epsilon_k^2]$ and the measure $k_0\,\kappa\,d\kappa\,d\gamma\,d\phi$, we obtain
\begin{align}
\delta\Pi (q\,\bs{\hat e}, \omega) = \frac{g\,v_0\,q^2}{8\,\pi^3\,k_0} \int_0^{2\pi} d\phi\, \check w^{\rm v}_e (\phi) \int_0^{2\pi} d\gamma\, s_\gamma^2 \int_{\kappa_F (\phi, \gamma)}^{\Lambda} \frac{\kappa^2\,d\kappa}{(\Omega_e + i\,0^+)^2 - 4\,v_0^2\,\kappa^2}\,,
\end{align}
with $\check w^{\rm v}_x = s_\phi^2$, $\check w^{\rm v}_y = c_\phi^2$, $\Omega_x = \Omega$, and $\Omega_y = \omega$. The prefactor is $g\,k_0/(2\,\pi)^3 \times q^2/(4\,k_0^2) \times 4\,v_0$. The lower limit is the Pauli-blocking radius. As in the main text, we replace it by $\kappa_F (\phi) = |\mu_{\rm eff} (\phi)|/v_0$, which neglects $\eta_x$ in $\zeta_\pm$ and is accurate to relative order $\eta_x$. For $\eta_x \to 0$, both angular integrals give $\pi$, and Eq.~\eqref{eq:dpivnr} is recovered.

The imaginary part follows as in Appendix~\ref{app:vnrvortex}, the radial integral has the imaginary part $-\,\pi\,\Omega_e/(16\,v_0^3)$ at the resonant radius $\kappa_0 = \Omega_e/(2\,v_0)$, provided $\kappa_0 > \kappa_F (\phi)$. The latter condition is $\Omega_e > 2\,|\mu_{\rm eff} (\phi)|$, and it is independent of $\gamma$. The $\gamma$-integral gives $\pi$, and the $\phi$-integral gives $2\,\pi\,(F_0 - F_2)$ for $\check w^{\rm v}_x = s_\phi^2 = 1 - c_\phi^2$ and $2\,\pi\,F_2$ for $\check w^{\rm v}_y = c_\phi^2$, with $F_n$ of Eq.~\eqref{eq:fnvnr}. Therefore
\begin{align}
\Im \delta\Pi (q\,\bs{\hat x}, \omega) \simeq -\, \frac{g\,q^2\,\Omega}{64\,v_0^2\,k_0}\, (F_0 - F_2)\,, \quad
\Im \delta\Pi (q\,\bs{\hat y}, \omega) \simeq -\, \frac{g\,q^2\,\omega}{64\,v_0^2\,k_0}\, F_2\,.
\end{align}
We set $\Omega \to \omega$, which changes the result by the relative amount $v_0\,\eta_x\,q/\omega$ on a term that is already small. This is Eq.~\eqref{eq:imdpixvnr}. The ratio to the leading terms $-\,g\,k_0\,q^2\,F_n/(16\,\omega)$ of Eq.~\eqref{eq:imhornvnr} is $(\omega/2\,v_0\,k_0)^2$ times a ratio of the functions $F_n$. For $\omega > 2\,(\tilde{\eta} + \mu)$, $F_0 - F_2 = F_2 = 1/2$, and both channels reduce to Eq.~\eqref{eq:imdpivnr} at $\mu = 0$. In the ring-cyclide regime, all $F_n$ vanish below $\omega_{\rm th}$. In the horn-cyclide regime, they are nonzero at all frequencies.

For the real part, we proceed as before. For $\omega \ll v_0\,\Lambda$, the same steps as in Appendix~\ref{app:vnrvortex} give $\Re \int_{\kappa_F}^{\Lambda} \kappa^2\,d\kappa/(\omega^2 - 4\,v_0^2\,\kappa^2) \simeq -\,(\Lambda - \kappa_F)/(4\,v_0^2)$. The remainder is smaller by a factor of order $\omega/(v_0\,\Lambda)$ up to a logarithm. Using $\int_0^{2\pi} d\gamma\, s_\gamma^2 = \pi$, we obtain
\begin{align}
\Re \delta\Pi (q\,\bs{\hat e}, \omega) \simeq -\, \frac{g\,q^2}{32\,\pi\,v_0\,k_0} \left( \Lambda - \bar\kappa_{F, e} \right),
\quad
\bar\kappa_{F, x} = \frac{\mathcal{J}_0 - \mathcal{J}_2}{\pi\,v_0}\,, \quad
\bar\kappa_{F, y} = \frac{\mathcal{J}_2}{\pi\,v_0}\,,
\end{align}
where we have used $\int d\phi\, s_\phi^2\,|\mu_{\rm eff}| = \mathcal{J}_0 - \mathcal{J}_2$ and $\int d\phi\, c_\phi^2\,|\mu_{\rm eff}| = \mathcal{J}_2$ at leading order in $\eta_x$. The shift of the background constants is $\delta\epsilon_{b,e} = -\,4\,\pi\,e^2\,\Re \delta\Pi/q^2$, which gives
\begin{align}
\delta\epsilon_{b,e} \simeq \frac{g\,e^2 \left( \Lambda - \bar\kappa_{F, e} \right)}{8\,v_0\,k_0}\,, \quad \delta\epsilon_{b,z} = 0\,.
\end{align}
In the ring-cyclide regime, $\mathcal{J}_0 = 2\,\pi\,\mu$ and $\mathcal{J}_2 = \pi\,\mu$ exactly in $\tilde{\eta}$, and both $\bar\kappa_{F,x}$ and $\bar\kappa_{F,y}$ equal $\kappa_F = \mu/v_0$. The result then coincides with Eq.~\eqref{eq:debvnr}. In the horn-cyclide regime, the two differ. For example, at $\mu = 0$, we have $\bar\kappa_{F,x} = 4\,\tilde{\eta}/(3\,\pi\,v_0)$ and $\bar\kappa_{F,y} = 8\,\tilde{\eta}/(3\,\pi\,v_0)$. The tilt-induced splitting between $x$ and $y$ therefore appears at relative order $\bar\kappa_F/\Lambda$, and Eq.~\eqref{eq:debxvnr} holds to leading order in $\kappa_F/\Lambda$.

For the comparison with the leading background constant, we take the ring-cyclide value $\epsilon_{b,x} - 1 = g\,e^2\,k_0\,\mu/[4\,\tilde w\,(\mu + \tilde w)]$ of Eq.~\eqref{eq:ebxvnr}. The ratio is
\begin{align}
\frac{\delta\epsilon_{b,x}}{\epsilon_{b,x} - 1} \simeq  \frac{\Lambda\,\tilde w\,(\mu + \tilde w)}{2\,v_0\,k_0^2\,\mu}\,.
\end{align}
Since $\tilde w \leq \mu$, we have $\tilde w\,(\mu + \tilde w)/\mu \leq 2\,\mu$, and the ratio is at most $\Lambda\,\mu/(v_0\,k_0^2) = \kappa_F\,\Lambda/k_0^2$. It tends to this value for $|\tilde{\eta}| \ll \mu$, where $\tilde w \to \mu$. The shift is thus of relative order $\kappa_F\,\Lambda/k_0^2$, and it depends on $\Lambda$. It is of the same order as the corrections of relative size $\kappa/k_0$ that were dropped from the leading terms, as stated in the main text.

%%%%%%%%%%%%%%%
\section{VNR tilted with respect to the \texorpdfstring{$k_z$}{kz}-axis}
\label{app:ztilt}

This appendix gives the intermediate steps behind the results of Sec.~\ref{secaxial}, where $\bs{\eta} = \eta_z\,\bs{\hat z}$. The subsections follow the order of Secs.~\ref{secazintr} to~\ref{secazvortex}. We write $\zeta (\gamma) = 1 + \eta_z \sin\gamma$ and take $\eta_z > 0$ when we state arcs and thresholds. The joint substitution $\gamma \to -\,\gamma$ and $\eta_z \to -\,\eta_z$ leaves $\zeta$, $\cos^2\gamma$, $\sin^2\gamma$, and $\left( \sin\gamma + \eta_z \right)^2$ unchanged. All angular integrals below therefore depend on $|\eta_z|$ only, and the sign of $\eta_z$ survives only in the shift $v_0\,\eta_z\,q$ of Eq.~\eqref{eq:polinter}.

At fixed $\phi$ we use the local coordinates $u = \kappa\cos\gamma$ and $w = \kappa\sin\gamma = k_z$ of Appendix~\ref{app:vnrdirac}. The tilt term $v_0\,\eta_z\,k_z$ is proportional to the identity in band space. It commutes with the matrix $U_\phi$ of that appendix, and the full local Hamiltonian becomes
\begin{align}
\mathcal{H}_\ell + v_0\,\eta_z\,k_z = U_\phi \left[ h (u, w) + v_0\,\eta_z\,w \right] U_\phi^\dagger \,.
\end{align}
Each copy at fixed $\phi$ is thus a massless two-dimensional Dirac cone that is tilted along $w$, with the same tilt parameter $\eta_z$ for all $\phi$. The eigenvalues are
\begin{align}
\label{eq:azeig}
\varepsilon_s (\bs{k}) = v_0\,\kappa\,\zeta_s (\gamma) \,, \quad
\zeta_s = (-1)^s + \eta_z \sin\gamma \,.
\end{align}
For $|\eta_z| < 1$ we have $\zeta_2 = \zeta \geq 1 - |\eta_z| > 0$ and $\zeta_1 \leq -\left( 1 - |\eta_z| \right) < 0$. Hence $\varepsilon_1 < 0 < \varepsilon_2$ for every $\gamma$, and no pocket of the opposite carrier type forms. The velocity of the band $s = 2$ is $\bs{V}^{(2)} = v_0\,\bs{\hat V}$, with $\bs{\hat V} = \bs{\hat\kappa} + \eta_z\,\bs{\hat z}$ and $\bs{\hat\kappa}$ the radial unit vector of Eq.~\eqref{eq:berry}. In components,
\begin{align}
\bs{\hat V} = \left\lbrace \cos\gamma \cos\phi,\, \cos\gamma \sin\phi,\, \sin\gamma + \eta_z \right\rbrace.
\end{align}
This follows from the gradient of Eq.~\eqref{eq:azeig}, with $\bs{\nabla}_{\bs{k}}\,\kappa = \bs{\hat\kappa}$ and $\bs{\nabla}_{\bs{k}}\,k_z = \bs{\hat z}$.

The exact measure is $d^3 \bs{k} = \kappa \left( k_0 + \kappa \cos\gamma \right) d\kappa\, d\gamma\, d\phi$. The energies and the blocking radius depend on $\gamma$ only through $\sin\gamma$. In Appendices~\ref{app:vnrzdrude} and~\ref{app:vnrzscreen} every integrand multiplied by the term $\kappa\cos\gamma$ of the Jacobian is $\cos\gamma$ times a function of $\sin\gamma$. Such a term integrates to zero over $\gamma$, as in Appendix~\ref{app:vnrdrude}, and we use the measure $k_0\,\kappa\,d\kappa\,d\gamma\,d\phi$ there.

For $\bs{q} = q\,\bs{\hat z}$ we have exactly $\phi' = \phi$, $u' = u$, and $w' = w + q$ by Appendix~\ref{app:vnrdirac}. The tilt therefore adds $v_0\,\eta_z \left( k_z' - k_z \right) = v_0\,\eta_z\,q$ to $\varepsilon_s (\bs{k}') - \varepsilon_s (\bs{k})$ for every $\bs{k}$. This is the shift $v_0\,\bs{\eta} \cdot \bs{q}$ of Eq.~\eqref{eq:polinter}. For $\bs{q} = q\,\bs{\hat x}$ and $\bs{q} = q\,\bs{\hat y}$ we have $k_z' = k_z$, and the shift vanishes. The overlap factor of Eq.~\eqref{eq:formfactor} does not involve the tilt, because the tilt term is proportional to the identity.

The function $\zeta$ is smallest at $\sin\gamma = -\,\mathrm{sgn}\,\eta_z$, where it equals $1 - |\eta_z|$. The largest Fermi momentum is therefore $\kappa_F^{\max} = \mu/\left[ v_0 \left( 1 - |\eta_z| \right) \right]$, as follows from Eq.~\eqref{eq:kappaFzvnr}. The expansion in $\kappa/k_0$ requires $\kappa_F^{\max} \ll k_0$. For $\eta_z = 0.75$ the factor $1/(1 - |\eta_z|)$ equals $4$, and $\kappa_F^{\max}/k_0$ equals $0.4$ and $0.8$ for $\mu = 0.1\,v_0\,k_0$ and $0.2\,v_0\,k_0$, respectively. This is why the tilted case at the larger $\mu$ lies at the edge of validity of the expansions.

%%%%%%%%%%%%%%%%%%
\subsection{Intrinsic point}
\label{app:vnrzintr}

We give the steps behind Eqs.~\eqref{eq:imintrzvnr} and~\eqref{eq:epsintrzvnr} and the statements made after them. At $\mu = 0$ the band $s = 1$ is filled and the band $s = 2$ is empty for every $\gamma$, because $\varepsilon_1 < 0 < \varepsilon_2$ by Eq.~\eqref{eq:azeig}. The occupation factors are therefore those of the untilted ring. The FS is the nodal ring $\kappa = 0$ for every $\eta_z$, because the offset $v_0\,\eta_z\,k_z$ vanishes at $k_z = 0$. No pocket forms, and the Drude weight vanishes, because the ring occupies no phase-space volume.

For $\bs{q} = q\,\bs{\hat z}$, both denominators of Eq.~\eqref{eq:polinter} contain the shift $v_0\,\eta_z\,q$. With $\Omega = \omega - v_0\,\eta_z\,q$, the bracket depends on $\omega$ and $\eta_z$ only through $\Omega$. The overlap factor and the measure do not involve the tilt, and therefore
\begin{align}
\Pi (q\,\bs{\hat z}, \omega) = \Pi_{\eta_z = 0} (q\,\bs{\hat z}, \Omega) = k_0\,\Pi_{\rm 2D} (q, \Omega) \,,
\end{align}
where $\Pi_{\rm 2D}$ is the polarisability of the untilted cone of Appendix~\ref{app:vnrintrinsic}. That appendix gives $\Pi_{\rm 2D} $ to be $ -\, g\,q^2/\big( 16 \sqrt{v_0^2\,q^2 - \Omega^2} \big)$ for $|\Omega| < v_0\,q$. For $\Omega > v_0\,q$ it gives the value $ -\, i\,g\,q^2/\big( 16 \sqrt{\Omega^2 - v_0^2\,q^2} \big)$. The imaginary part is therefore nonzero if and only if $\Omega > v_0\,q$, which is $\omega > v_0\,q \left( 1 + \eta_z \right)$. The condition $\Omega < -\,v_0\,q$ cannot be met for $\omega > 0$ and $|\eta_z| < 1$. Multiplying by $k_0$ gives the first expression of Eq.~\eqref{eq:imintrzvnr}. For $\omega \gg v_0\,q$ we expand
\begin{align}
\frac{1}{\sqrt{\Omega^2 - v_0^2\,q^2}} \simeq \frac{1}{\Omega} \simeq \frac{1}{\omega} \left( 1 + \frac{v_0\,\eta_z\,q}{\omega} \right).
\end{align}
The neglected term is of relative order $(v_0\,q/\omega)^2$, whereas the retained correction is of first order in $v_0\,\eta_z\,q/\omega$.

For $\bs{q} = q\,\bs{\hat x}$ and $\bs{q} = q\,\bs{\hat y}$ the shift vanishes, and the response is that of the untilted ring. By Appendix~\ref{app:vnrintrinsic} the average of $\cos^2\phi$, or of $\sin^2\phi$, equals $1/2$, and this gives $\Im \Pi \simeq -\, g\,k_0\,q^2/(32\,\omega)$ in both channels. This is the second relation of Eq.~\eqref{eq:imintrzvnr}.

Below the onset $\omega < v_0\,q \left( 1 + \eta_z \right)$ the polarisability is real. Inserting $\Pi (q\,\bs{\hat z}, \omega) = -\, g\,k_0\,q^2/\big( 16 \sqrt{v_0^2\,q^2 - \Omega^2} \big)$ into $\varepsilon^{\rm RPA} = 1 - 4\,\pi\,e^2\,\Pi/q^2$ gives Eq.~\eqref{eq:epsintrzvnr}. The argument of the square root factorises as
\begin{align}
v_0^2\,q^2 - \left( \omega - v_0\,\eta_z\,q \right)^2 = \left[ v_0\,q \left( 1 + \eta_z \right) - \omega \right] \left[ v_0\,q \left( 1 - \eta_z \right) + \omega \right].
\end{align}
Both factors are positive for $0 \leq \omega < v_0\,q \left( 1 + \eta_z \right)$ and $|\eta_z| < 1$. The second term of Eq.~\eqref{eq:epsintrzvnr} is therefore real and positive, $\varepsilon^{\rm RPA} > 1$, and no zero exists. For $\bs{q} = q\,\bs{\hat x}$, the same argument as in Appendix~\ref{app:vnrintrinsic} shows that the local threshold $v_0\,q\,|\cos\phi|$ vanishes at $\phi = \pi/2$. The set of angles that contribute has a finite measure for every $\omega > 0$, and $\varepsilon^{\rm RPA}$ is complex at all frequencies.

For the vortex term, no state is blocked at $\mu = 0$, and the tilt does not enter the denominators of the in-plane channels. The steps of Appendix~\ref{app:vnrvortex} therefore apply without change, and they give Eq.~\eqref{eq:imdpivnr} without the step function. This coincides with the limit $\tau \leq -1$ of Eq.~\eqref{eq:imdpizvnr}, which is derived in Appendix~\ref{app:vnrzvortex}.

%%%%%%%%%%%%%%%%%%
\subsection{Doped regime: density-of-states, Drude weights, and plasmons}
\label{app:vnrzdrude}

We give the steps behind Eqs.~\eqref{eq:kappaFzvnr} to~\eqref{eq:wpzvnr}. For $\mu > 0$ only the band $s = 2$ has a FS, because $\varepsilon_1 < 0$. By Eq.~\eqref{eq:azeig}, $\varepsilon_2 = v_0\,\kappa\,\zeta$, and therefore
\begin{align}
\delta (\varepsilon_2 - \mu) = \frac{\delta (\kappa - \kappa_F)}{v_0\,\zeta} \,, \quad
\kappa_F (\gamma) = \frac{\mu}{v_0\,\zeta (\gamma)} \,.
\end{align}
The second relation is the first expression of Eq.~\eqref{eq:kappaFzvnr}. In the poloidal plane, $\kappa_F (\gamma)$ is the polar equation of an ellipse with eccentricity $|\eta_z|$, semi-latus rectum $\ell = \mu/v_0$, and the node $\kappa = 0$ at one focus.

Inserting this into Eq.~\eqref{eqdos}, with the Jacobian given above, we find
\begin{align}
\rho (\mu) = \frac{g}{(2\,\pi)^3} \int_0^{2\pi} d\phi \int_0^{2\pi} d\gamma\, \frac{\kappa_F \left( k_0 + \kappa_F \cos\gamma \right)}{v_0\,\zeta}
= \frac{g\,k_0\,\mu}{4\,\pi^2\,v_0^2} \int_0^{2\pi} \frac{d\gamma}{\zeta^2} \,.
\end{align}
The piece proportional to $\kappa_F \cos\gamma$ is $\mu^2 \cos\gamma/(v_0^3\,\zeta^3)$, and it integrates to zero. We need the integrals $I_n = \int_0^{2\pi} d\gamma\, \zeta^{-n}$. The standard result $\int_0^{2\pi} d\gamma/(\lambda + \eta_z \sin\gamma) = 2\,\pi/\sqrt{\lambda^2 - \eta_z^2}$ for $\lambda > |\eta_z|$ gives $I_1$. Its derivative with respect to $\lambda$ at $\lambda = 1$ gives $I_2$. Together with $I_0 = 2\,\pi$,
\begin{align}
\label{eq:azIn}
I_0 = 2\,\pi \,, \quad
I_1 = \frac{2\,\pi}{\sqrt{1 - \eta_z^2}} \,, \quad
I_2 = \frac{2\,\pi}{\left( 1 - \eta_z^2 \right)^{3/2}} \,.
\end{align}
The last relation gives the second expression of Eq.~\eqref{eq:kappaFzvnr}. As a check, the area of the ellipse is $\tfrac{1}{2} \int_0^{2\pi} \kappa_F^2\, d\gamma = \tfrac{1}{2}\,\ell^2\,I_2 = \pi\,\ell^2/\left( 1 - \eta_z^2 \right)^{3/2}$. This equals $\pi\,\tilde{a}\,\tilde{b}$ with the semi-axes $\tilde{a} = \ell/\left( 1 - \eta_z^2 \right)$ and $\tilde{b} = \ell/\sqrt{1 - \eta_z^2}$. The DOS is even in $\eta_z$, it increases with $|\eta_z|$, and it diverges as $|\eta_z| \to 1$.

With $\bs{V}^{(2)} = v_0\,\bs{\hat V}$ and $v_0\,\kappa_F\,\zeta = \mu$, Eq.~\eqref{eqdrude} becomes
\begin{align}
D_{ij} = \frac{\pi\,g\,e^2}{(2\,\pi)^3}\, \frac{v_0^2}{v_0}\, k_0 \int_0^{2\pi} d\phi \int_0^{2\pi} d\gamma\, \frac{\kappa_F}{\zeta}\, \hat V_i\,\hat V_j
= \frac{g\,e^2\,k_0\,\mu}{8\,\pi^2} \int_0^{2\pi} d\phi \int_0^{2\pi} d\gamma\, \frac{\hat V_i\,\hat V_j}{\zeta^2} \,.
\end{align}
The Jacobian piece proportional to $\kappa_F \cos\gamma$ drops out as before. The off-diagonal entries are proportional to $\int_0^{2\pi} d\phi \cos\phi \sin\phi$, $\int_0^{2\pi} d\phi \cos\phi$, or $\int_0^{2\pi} d\phi \sin\phi$, and they vanish. The $\phi$-integrals $\int_0^{2\pi} d\phi \cos^2\phi = \int_0^{2\pi} d\phi \sin^2\phi = \pi$ give $D_{xx} = D_{yy}$ and
\begin{align}
D_{zz} = \frac{g\,e^2\,k_0\,\mu}{4\,\pi} \int_0^{2\pi} d\gamma\, \frac{\left( \sin\gamma + \eta_z \right)^2}{\zeta^2} \,, \quad
D_{xx} = D_{yy} = \frac{g\,e^2\,k_0\,\mu}{8\,\pi} \int_0^{2\pi} d\gamma\, \frac{\cos^2\gamma}{\zeta^2} \,.
\end{align}
We eliminate $\sin\gamma = (\zeta - 1)/\eta_z$, which gives $\sin\gamma + \eta_z = \left[ \zeta - \left( 1 - \eta_z^2 \right) \right]/\eta_z$ and $\cos^2\gamma = 1 - (\zeta - 1)^2/\eta_z^2$. Then
\begin{align}
\frac{\left( \sin\gamma + \eta_z \right)^2}{\zeta^2} = \frac{1}{\eta_z^2} \left[ 1 - \frac{2 \left( 1 - \eta_z^2 \right)}{\zeta} + \frac{\left( 1 - \eta_z^2 \right)^2}{\zeta^2} \right], \quad
%%%%%%%%%%%%%%%%%%%%
\frac{\cos^2\gamma}{\zeta^2} = \frac{1}{\eta_z^2} 
\left[ -\, \frac{1 - \eta_z^2}{\zeta^2} + \frac{2}{\zeta} - 1 \right].
\end{align}
Integrating with Eq.~\eqref{eq:azIn}, we get
\begin{align}
\int_0^{2\pi} d\gamma\, \frac{\left( \sin\gamma + \eta_z \right)^2}{\zeta^2} 
= \frac{2\,\pi}{\eta_z^2} \left( 1 - \sqrt{1 - \eta_z^2} \right),
\quad
\int_0^{2\pi} d\gamma\, \frac{\cos^2\gamma}{\zeta^2} = \frac{2\,\pi}{\eta_z^2} \left( \frac{1}{\sqrt{1 - \eta_z^2}} - 1 \right),
\end{align}
and Eq.~\eqref{eq:dzvnr} follows. The anisotropy ratio is
\begin{align}
\frac{D_{zz}}{D_{xx}} = \frac{2 \left( 1 - \sqrt{1 - \eta_z^2} \right)}{1/\sqrt{1 - \eta_z^2} - 1} = 2\,\sqrt{1 - \eta_z^2} \,,
\end{align}
because $1/\sqrt{1 - \eta_z^2} - 1 = \left( 1 - \sqrt{1 - \eta_z^2} \right)/\sqrt{1 - \eta_z^2}$.

The derivation of Eq.~\eqref{eqintra} in Appendix~\ref{app:vnrdrude} uses only $\delta (\varepsilon_2 - \mu)$, $\bs{V}^{(2)}$, and the fact that $\mathcal{F}_{22} = 1 - \mathcal{O} (q^2)$. It therefore holds for the tilted bands, and $\omega_{p,i}^2 = 4\,D_{ii}$ gives Eq.~\eqref{eq:wpzvnr}. For the expansions we use $\sqrt{1 - x} = 1 - x/2 - x^2/8 - \dots$ with $x = \eta_z^2$, which gives
\begin{align}
\frac{1 - \sqrt{1 - \eta_z^2}}{\eta_z^2} = \frac{1}{2} + \frac{\eta_z^2}{8} + \mathcal{O} (\eta_z^4) \,, \quad
\frac{1}{\eta_z^2} \left( \frac{1}{\sqrt{1 - \eta_z^2}} - 1 \right) = \frac{1}{2} + \frac{3\,\eta_z^2}{8} + \mathcal{O} (\eta_z^4) \,.
\end{align}
The squared frequencies then become $\omega_{p,z}^2 \simeq g\,e^2\,k_0\,\mu \left( 1 + \eta_z^2/4 \right)$ and $\omega_{p,x}^2 \simeq \tfrac{1}{2}\,g\,e^2\,k_0\,\mu \left( 1 + 3\,\eta_z^2/4 \right)$. Their ratio is $\omega_{p,z}/\omega_{p,x} = \left( 2 \sqrt{1 - \eta_z^2} \right)^{1/2} = \sqrt{2} \left( 1 - \eta_z^2 \right)^{1/4} \simeq \sqrt{2} \left( 1 - \eta_z^2/4 \right)$.

It is convenient to introduce $\lambda = g\,e^2\,k_0/(4\,\mu)$ as in Appendix~\ref{app:vnrdrude} and the functions
\begin{align}
\label{eq:azPz}
\mathcal{P}_z (\eta_z) = \frac{2 \left( 1 - \sqrt{1 - \eta_z^2} \right)}{\eta_z^2} \,, \quad
\mathcal{P}_x (\eta_z) = \frac{\mathcal{P}_z (\eta_z)}{\sqrt{1 - \eta_z^2}} \,.
\end{align}
Both tend to unity for $\eta_z \to 0$, and both are larger than unity otherwise. The bare frequencies are $\omega_{p,z}^2 = 4\,\mu^2\,\lambda\,\mathcal{P}_z$ and $\omega_{p,x}^2 = 2\,\mu^2\,\lambda\,\mathcal{P}_x$, using $g\,e^2\,k_0\,\mu = 4\,\mu^2\,\lambda$. These forms are used in Appendix~\ref{app:vnrzscreen}.

%%%%%%%%%%%%%%%%%%
\subsection{Interband screening and the screened plasmons}
\label{app:vnrzscreen}

We give the steps behind Eqs.~\eqref{eq:pinterzvnr} to~\eqref{eq:imzvnr}, together with the statements on the zero of $\Re \varepsilon^{\rm RPA}$ made after them. We take $\mu > 0$ and $\omega \gg v_0\,q$, and we evaluate the denominators of Eq.~\eqref{eq:polinter} at $q \to 0$. The shift $v_0\,\eta_z\,q$ is then dropped in the $\bs{\hat z}$ channel, with a relative error of order $v_0\,\eta_z\,q/\omega$. It vanishes in the in-plane channels. The bracket becomes $4\,v_0\,\kappa/\left[ (\omega + i\,0^+)^2 - 4\,v_0^2\,\kappa^2 \right]$, as in Appendix~\ref{app:vnrscreen}.

The band $s = 1$ is filled, and the transitions are allowed only into the empty states of the band $s = 2$, which have $\kappa > \kappa_F (\gamma)$. The blocking radius depends on $\gamma$ through $\zeta$, whereas the transition energy $2\,v_0\,\kappa$ does not. We keep the first term of Eq.~\eqref{eq:qmetric}, which gives $\mathcal{F} \simeq q^2 \cos^2\gamma/(4\,\kappa^2)$ for $\bs{q} = q\,\bs{\hat z}$. For $\bs{q} = q\,\bs{\hat x}$ and $\bs{q} = q\,\bs{\hat y}$ it gives $q^2 \sin^2\gamma \cos^2\phi/(4\,\kappa^2)$ and $q^2 \sin^2\gamma \sin^2\phi/(4\,\kappa^2)$, respectively. With the measure $k_0\,\kappa\,d\kappa\,d\gamma\,d\phi$ and $\int_0^{2\pi} d\phi = 2\,\pi$, the $\bs{\hat z}$ channel reads
\begin{align}
\Pi_{\rm inter} (q\,\bs{\hat z}, \omega) = \frac{g\,k_0\,q^2\,v_0}{4\,\pi^2} \int_0^{2\pi} d\gamma\, \cos^2\gamma \int_{\kappa_F (\gamma)}^\infty \frac{d\kappa}{(\omega + i\,0^+)^2 - 4\,v_0^2\,\kappa^2} \,.
\end{align}
The factors of $\kappa$ cancel, as in Appendix~\ref{app:vnrscreen}. The derivative of
$\ln \left| \frac{\omega + 2\,v_0\,\kappa} {\omega - 2\,v_0\,\kappa } \right|$ with respect to $\kappa$ equals $4\,v_0\,\omega/(\omega^2 - 4\,v_0^2\,\kappa^2)$, and this function vanishes as $\kappa \to \infty$. The real part of the radial integral is therefore $-\,\ln \left| (\omega + E_F)/(\omega - E_F) \right|/(4\,v_0\,\omega)$, with $E_F = 2\,v_0\,\kappa_F = 2\,\mu/\zeta$. Its imaginary part is $-\,\pi \int d\kappa\, \delta (\omega^2 - 4\,v_0^2\,\kappa^2) = -\,\pi\,\Theta (\omega - E_F)/(4\,v_0\,\omega)$. Multiplying numerator and denominator inside the logarithm by $\zeta$ gives
\begin{align}
\label{eq:azpiz}
\Pi_{\rm inter} (q\,\bs{\hat z}, \omega) = -\, \frac{g\,k_0\,q^2}{16\,\pi^2\,\omega} \int_0^{2\pi} d\gamma\, \cos^2\gamma \left[ \ln \left| \frac{2\,\mu + \omega\,\zeta}{2\,\mu - \omega\,\zeta} \right| + i\,\pi\,\Theta (\omega\,\zeta - 2\,\mu) \right].
\end{align}
For $\bs{q} = q\,\bs{\hat x}$ the $\phi$-integral is $\int_0^{2\pi} d\phi \cos^2\phi = \pi$, which is half of the value for $\bs{\hat z}$. The weight $\cos^2\gamma$ is replaced by $\sin^2\gamma$, and the prefactor is halved. The $\bs{\hat y}$ channel is identical. For $\eta_z = 0$ we have $\int_0^{2\pi} d\gamma \cos^2\gamma = \pi$, and Eq.~\eqref{eq:azpiz} reduces to the expression of Appendix~\ref{app:vnrscreen}.

We now take the static limit. For $\omega \to 0$ the logarithm tends to $\omega\,\zeta/\mu$, and Eq.~\eqref{eq:azpiz} gives
\begin{align}
\Pi_{\rm inter} (q\,\bs{\hat z}, 0) = -\, \frac{g\,k_0\,q^2}{16\,\pi^2\,\mu} \int_0^{2\pi} d\gamma\, \cos^2\gamma \left( 1 + \eta_z \sin\gamma \right) = -\, \frac{g\,k_0\,q^2}{16\,\pi\,\mu} \,.
\end{align}
The term odd in $\sin\gamma$ integrates to zero, and the result does not depend on $\eta_z$. The same holds for the weight $\sin^2\gamma$, which gives half of this value. Equation~\eqref{eq:ebzvnr} follows from $\epsilon_{b,i} = 1 - 4\,\pi\,e^2\,\Pi_{\rm inter}/q^2$. Equivalently, the static response is proportional to the $\gamma$-average of $1/\kappa_F \propto \zeta$, in which the odd term drops out.

For the real part below the onset, we note that the argument of the second logarithm in Eq.~\eqref{eq:azpiz} stays positive for all $\gamma$ if and only if $\omega \left( 1 + |\eta_z| \right) < 2\,\mu$. This defines the onset $\omega_{\rm th} = 2\,\mu/\left( 1 + |\eta_z| \right)$, which is the smallest value of $E_F (\gamma) = 2\,\mu/\zeta$. Below it, the step function vanishes and no absolute value is needed. With $A_\pm = 2\,\mu \pm \omega$ we have
\begin{align}
\ln \left| \frac{2\,\mu + \omega\,\zeta}{2\,\mu - \omega\,\zeta} \right| = \ln \left( A_+ + \omega\,\eta_z \sin\gamma \right) - \ln \left( A_- - \omega\,\eta_z \sin\gamma \right).
\end{align}
The angular integrals follow from a Fourier expansion. For $A > |B|$ we set $r = \sqrt{A^2 - B^2}$ and $\varrho = |B|/(A + r)$. For $B > 0$, direct multiplication shows that
\begin{align}
A + B \sin\gamma = \frac{A + r}{2} \left( 1 + \varrho^2 + 2\,\varrho \sin\gamma \right),
\end{align}
and the case $B < 0$ follows from $\gamma \to -\,\gamma$. Since $1 + \varrho^2 + 2\,\varrho \sin\gamma = \left| 1 + \varrho\, e^{i (\gamma - \pi/2)} \right|^2$, we have
\begin{align}
\ln \left( 1 + \varrho^2 + 2\,\varrho \sin\gamma \right) = 2 \sum_{n \geq 1} \frac{(-1)^{n-1}\, \varrho^n}{n} \cos \left[ n \left( \gamma - \frac{\pi}{2} \right) \right].
\end{align}
The mean of this series vanishes, and its term with $n = 2$ equals $\varrho^2 \cos 2\gamma$. Only this term and the constant $\ln [(A + r)/2]$ survive in the integrals with $\cos^2\gamma = (1 + \cos 2\gamma)/2$ and $\sin^2\gamma = (1 - \cos 2\gamma)/2$, and
\begin{align}
\int_0^{2\pi} d\gamma \cos^2\gamma \ln (A + B \sin\gamma) 
= \pi \left[ \ln \!\left( \frac{A + r}{2} \right)  + \frac{\varrho^2}{2} \right], \quad
\int_0^{2\pi} d\gamma \sin^2\gamma \ln (A + B \sin\gamma) 
= \pi \left[ \ln \!\left( \frac{A + r}{2} \right)  - \frac{\varrho^2}{2} \right].
\end{align}
Applying this to both logarithms, with $(A, B) = (A_+, \omega\,\eta_z)$ and $(A_-, -\,\omega\,\eta_z)$, we obtain Eq.~\eqref{eq:pinterzvnr} with $r_\pm = \sqrt{A_\pm^2 - \eta_z^2\,\omega^2}$ and $\varrho_\pm = |\eta_z|\,\omega/(A_\pm + r_\pm)$. The prefactors are $-\,g\,k_0\,q^2/(16\,\pi\,\omega)$ and $-\,g\,k_0\,q^2/(32\,\pi\,\omega)$ for the $\bs{\hat z}$ and the in-plane channels, respectively. The condition $r_- > 0$ is again $\omega < \omega_{\rm th}$. At $\omega \to \omega_{\rm th}$ we have $r_- \to 0$ and $\varrho_- \to 1$, and the expressions stay finite. For $\eta_z = 0$ the denominator $A_- + r_- = 2\,(2\,\mu - \omega)$ vanishes at $\omega = 2\,\mu$ instead, and $\Pi_{\rm inter}$ diverges there.

Above the onset, for $\omega > \omega_{\rm th}$, the step function in Eq.~\eqref{eq:azpiz} selects the arc where $\omega\,\zeta > 2\,\mu$. For $\eta_z > 0$ this is $\sin\gamma > \tau$ with $\tau = \left( 2\,\mu/\omega - 1 \right)/\eta_z$. For $|\tau| < 1$ the arc is $\gamma \in (\check{a}, \pi - \check{a})$ with $\check{a} = \arcsin \tau$. The integrals over the arc are
\begin{align}
\int_{\check{a}}^{\pi - \check{a}} d\gamma \left\lbrace \cos^2\gamma, \sin^2\gamma \right\rbrace = \frac{\pi}{2} - \check{a} \mp \tau \sqrt{1 - \tau^2} = \pi\,\mathcal{G}_\pm (\tau) \,,
\end{align}
where the upper sign refers to $\cos^2\gamma$, and $\sin 2\,\check{a} = 2\,\tau \sqrt{1 - \tau^2}$ has been used. This is the function $\mathcal{G}_\pm (\tau) = 1/2 - \left[ \arcsin \tau \pm \tau \sqrt{1 - \tau^2} \right]/\pi$ of Eq.~\eqref{eq:imzvnr}. For $\tau \geq 1$ the arc is empty and $\mathcal{G}_\pm = 0$. For $\tau \leq -1$ it is the full circle and $\mathcal{G}_\pm = 1$. Both limits agree with the formula at $\tau = \pm 1$. The condition $\tau = 1$ is $\omega = \omega_{\rm th}$. Inserting the imaginary part of Eq.~\eqref{eq:azpiz} gives
\begin{align}
\Im \Pi (q\,\bs{\hat z}, \omega) = -\, \frac{g\,k_0\,q^2}{16\,\omega}\, \mathcal{G}_+ (\tau) \,, \quad
\Im \Pi (q\,\bs{\hat x}, \omega) = -\, \frac{g\,k_0\,q^2}{32\,\omega}\, \mathcal{G}_- (\tau) \,.
\end{align}
The relation $\Im \varepsilon^{\rm RPA} = -\,4\,\pi\,e^2\,\Im \Pi/q^2$ then yields Eq.~\eqref{eq:imzvnr}. For $\eta_z < 0$, the substitution $\gamma \to -\,\gamma$ shows that the same expressions hold with $\eta_z \to |\eta_z|$.

We turn to the zero of the real part. Below the onset, $\Re \varepsilon^{\rm RPA} = \epsilon_{b,i} (\omega) - \omega_{p,i}^2/\omega^2$, with $\epsilon_{b,i} (\omega) = 1 - 4\,\pi\,e^2\,\Pi_{\rm inter}/q^2$. Since $\ln \left[ (1 + x)/(1 - x) \right] = 2 \operatorname{arctanh} x$, the $\bs{\hat z}$ channel can be written as
\begin{align}
\epsilon_{b,z} (\omega) = 1 + \frac{g\,e^2\,k_0}{4\,\pi\,\mu} \sum_{n \geq 0} \frac{1}{2\,n + 1} \left( \frac{\omega}{2\,\mu} \right)^{2\,n} \int_0^{2\pi} d\gamma\, \cos^2\gamma\, \zeta^{2\,n + 1} \,.
\end{align}
The series converges for $\omega < \omega_{\rm th}$, because $\omega\,\zeta/(2\,\mu) < 1$ for all $\gamma$ there. Every coefficient is positive, since $\zeta > 0$. The same holds in the in-plane channels with the weight $\sin^2\gamma$ and the prefactor halved. Hence $\epsilon_{b,i} (\omega)$ increases with $\omega$ on $(0, \omega_{\rm th})$, and $\epsilon_{b,i} (\omega) \geq \epsilon_{b,i} (0)$. The Drude term $-\,\omega_{p,i}^2/\omega^2$ increases as well and tends to $-\,\infty$ for $\omega \to 0$. Therefore $\Re \varepsilon^{\rm RPA}$ is strictly increasing and continuous on $(0, \omega_{\rm th})$, and its value at $\omega_{\rm th}$ is finite by the remark above. A zero below the onset exists if and only if $\Re \varepsilon^{\rm RPA} (\omega_{\rm th}) > 0$, and it is then unique. Since $\epsilon_{b,i} (\omega) \geq \epsilon_{b,i} (0)$, the zero cannot exceed $\tilde\omega_{p,i}$, as stated in the main text. If $\tilde\omega_{p,i} < \omega_{\rm th}$, then $\Re \varepsilon^{\rm RPA} (\omega_{\rm th}) \geq \epsilon_{b,i} (0) - \omega_{p,i}^2/\omega_{\rm th}^2 > 0$, and a zero exists. This is the sufficient condition quoted in the main text.

The condition at the onset can be made explicit. At $\omega = \omega_{\rm th}$ we have $A_- = |\eta_z|\,\omega_{\rm th}$ and $A_+ = \left( 2 + |\eta_z| \right) \omega_{\rm th}$. With $t = \sqrt{1 + |\eta_z|}$ this gives $A_+ + r_+ = \left( 1 + t \right)^2 \omega_{\rm th}$ and $\varrho_+ = (t - 1)/(t + 1)$. We also use $g\,e^2\,k_0/(4\,\omega_{\rm th}) = \lambda \left( 1 + |\eta_z| \right)/2$. Then
\begin{align}
\epsilon_{b,z} (\omega_{\rm th}) 
= 1 + \frac{\lambda \left( 1 + |\eta_z| \right)}{2} \left[ 2 \ln (1 + t) - \ln |\eta_z| - \frac{1 - \varrho_+^2}{2} \right], \quad
\epsilon_{b,x} (\omega_{\rm th}) 
= 1 + \frac{\lambda \left( 1 + |\eta_z| \right)}{4} \left[ 2 \ln (1 + t) - \ln |\eta_z| + \frac{1 - \varrho_+^2}{2} \right].
\end{align}
The Drude terms at the onset are $\omega_{p,z}^2/\omega_{\rm th}^2 = \lambda \left( 1 + |\eta_z| \right)^2 \mathcal{P}_z$ and $\omega_{p,x}^2/\omega_{\rm th}^2 = \lambda \left( 1 + |\eta_z| \right)^2 \mathcal{P}_x/2$, with $\mathcal{P}_{z,x}$ of Eq.~\eqref{eq:azPz}. The criterion $\Re \varepsilon^{\rm RPA} (\omega_{\rm th}) > 0$ is thus a condition on $\lambda$ and $\eta_z$ only. For $\lambda \gg 1$ the unity on the left-hand side is negligible, and the criterion becomes a condition on $|\eta_z|$ alone, which does not depend on $\mu$. We have evaluated this criterion numerically. It gives a critical tilt that is smaller along $\bs{\hat z}$ than along $\bs{\hat x}$ and $\bs{\hat y}$, and it is nearly the same for the two values of $\mu$ because $\lambda \gg 1$ there. This agrees with the statement of Sec.~\ref{secazscreen}.

The sufficient condition has a simple form. With $\tilde\omega_{p,z}^2 = 4\,\mu^2\,\lambda\,\mathcal{P}_z/(1 + \lambda)$, $\tilde\omega_{p,x}^2 = 4\,\mu^2\,\lambda\,\mathcal{P}_x/(2 + \lambda)$, and $\omega_{\rm th}^2 = 4\,\mu^2/\left( 1 + |\eta_z| \right)^2$, the condition $\tilde\omega_{p,i} < \omega_{\rm th}$ reads
\begin{align}
\frac{\lambda\,\mathcal{P}_z}{1 + \lambda} < \frac{1}{\left( 1 + |\eta_z| \right)^2} \,, \quad
\frac{\lambda\,\mathcal{P}_x}{2 + \lambda} < \frac{1}{\left( 1 + |\eta_z| \right)^2} \,.
\end{align}
For $\eta_z = 0$ we have $\mathcal{P}_z = \mathcal{P}_x = 1$, and the left-hand sides are below unity for every $\lambda > 0$. A zero below the onset then exists for every coupling, as in Appendix~\ref{app:vnrscreen}. For $\lambda \gg 1$ and $\eta_z \neq 0$ the left-hand sides tend to $\mathcal{P}_z$ and $\mathcal{P}_x$, which exceed unity, while the right-hand sides are below unity. The sufficient condition then fails, and the screened frequencies lie above the onset. This is the situation at large tilt described in the main text. The tilt lowers $\omega_{\rm th}$ linearly in $|\eta_z|$, whereas $\mathcal{P}_z \simeq 1 + \eta_z^2/4$ and $\mathcal{P}_x \simeq 1 + 3\,\eta_z^2/4$ raise $\tilde\omega_{p,i}$ only at order $\eta_z^2$. The plasmon is therefore pushed towards the interband continuum as $|\eta_z|$ grows.

%%%%%%%%%%%%%%%%%%
\subsection{Toroidal-angle dependence of the overlap factor at finite tilt}
\label{app:vnrzvortex}

We give the steps behind Eqs.~\eqref{eq:dpizvnr} to~\eqref{eq:debzvnr}. By Appendix~\ref{app:vnrvortex}, the second term of Eq.~\eqref{eq:fsplit} vanishes for $\bs{q} = q\,\bs{\hat z}$, because $\phi' = \phi$. For $\bs{q} = q\,\bs{\hat x}$ it equals $\delta\mathcal{F}_x = q^2 \sin^2\gamma \sin^2\phi/(4\,k_0^2)$. For $\bs{q} = q\,\bs{\hat y}$ we have $\phi' - \phi \simeq q \cos\phi/k_0$, and the weight $\sin^2\phi$ is replaced by $\cos^2\phi$. The tilt does not enter $\delta\mathcal{F}$, because the tilt term is proportional to the identity. It does not enter the energy denominators of the in-plane channels either, because $v_0\,\bs{\eta} \cdot \bs{q} = 0$ there. It enters only through the blocking radius $\kappa_F (\gamma)$, which depends on $\gamma$ and not on $\phi$.

Inserting $\delta\mathcal{F}$ into Eq.~\eqref{eq:polinter}, with the bracket $4\,v_0\,\kappa/\left[ (\omega + i\,0^+)^2 - 4\,v_0^2\,\kappa^2 \right]$ and the measure $k_0\,\kappa\,d\kappa\,d\gamma\,d\phi$, we obtain
\begin{align}
\delta\Pi (q\,\bs{\hat x}, \omega) = \frac{g\,k_0}{(2\,\pi)^3}\, \frac{q^2}{4\,k_0^2} \int_0^{2\pi} d\phi \sin^2\phi \int_0^{2\pi} d\gamma \sin^2\gamma \int_{\kappa_F (\gamma)}^{\Lambda} \kappa\, d\kappa\, \frac{4\,v_0\,\kappa}{(\omega + i\,0^+)^2 - 4\,v_0^2\,\kappa^2} \,.
\end{align}
The $\phi$-integral equals $\pi$ for $\sin^2\phi$ and for $\cos^2\phi$, and the prefactor becomes $g\,v_0\,q^2/(8\,\pi^2\,k_0)$. This is Eq.~\eqref{eq:dpizvnr}, and it shows that $\delta\Pi (q\,\bs{\hat x}, \omega) = \delta\Pi (q\,\bs{\hat y}, \omega)$. For $\eta_z \to 0$ the $\gamma$-integral gives $\pi$, and the prefactor becomes $g\,v_0\,q^2/(8\,\pi\,k_0)$ of Eq.~\eqref{eq:dpivnr}.

The imaginary part follows as in Appendix~\ref{app:vnrvortex}. The radial integral has the imaginary part $-\,\pi\,\omega/(16\,v_0^3)$ at the resonant radius $\kappa_0 = \omega/(2\,v_0)$, provided that $\kappa_F (\gamma) < \kappa_0 < \Lambda$. The first condition is $\omega\,\zeta > 2\,\mu$, which is $\sin\gamma > \tau$ with $\tau$ of Appendix~\ref{app:vnrzscreen}. The second is $\omega < 2\,v_0\,\Lambda$. The integral of $\sin^2\gamma$ over the arc equals $\pi\,\mathcal{G}_- (\tau)$, and we obtain
\begin{align}
\Im \delta\Pi (q\,\bs{\hat x}, \omega) = \frac{g\,v_0\,q^2}{8\,\pi^2\,k_0} \left( -\, \frac{\pi\,\omega}{16\,v_0^3} \right) \pi\,\mathcal{G}_- (\tau) = -\, \frac{g\,q^2\,\omega}{128\,v_0^2\,k_0}\, \mathcal{G}_- (\tau) \,.
\end{align}
This is Eq.~\eqref{eq:imdpizvnr}, and it reduces to Eq.~\eqref{eq:imdpivnr} for $\tau \leq -1$. The leading term in the same channel is $-\,g\,k_0\,q^2\,\mathcal{G}_- (\tau)/(32\,\omega)$ by Appendix~\ref{app:vnrzscreen}. The ratio of the two is $\omega^2/(4\,v_0^2\,k_0^2) = (\omega/2\,v_0\,k_0)^2$. It does not depend on the tilt, because both terms carry the same function $\mathcal{G}_-$. Both vanish below $\omega_{\rm th}$.

For the real part, we take $\omega \ll v_0\,\Lambda$ and use the identity and the antiderivative of Appendix~\ref{app:vnrvortex}. The radial integral has the real part $-\left( \Lambda - \kappa_F \right)/(4\,v_0^2)$, and the remainder is smaller by a factor of order $\omega/(v_0\,\Lambda)$ up to a logarithm. Here $\kappa_F$ depends on $\gamma$, and the $\gamma$-integral with the weight $\sin^2\gamma$ gives
\begin{align}
\Re \delta\Pi (q\,\bs{\hat x}, \omega) = \Re \delta\Pi (q\,\bs{\hat y}, \omega) \simeq -\, \frac{g\,q^2}{32\,\pi\,v_0\,k_0} \left( \Lambda - \bar\kappa_F \right),
\quad
\bar\kappa_F = \frac{1}{\pi} \int_0^{2\pi} d\gamma \sin^2\gamma\, \kappa_F (\gamma) \,.
\end{align}
The same value follows from the static limit of Eq.~\eqref{eq:polinterstatic}, with the second term of Eq.~\eqref{eq:qmetric} and the integrand $-\,g\,\delta\mathcal{F}/\epsilon_k$ over $\kappa_F (\gamma) < \kappa < \Lambda$. To evaluate $\bar\kappa_F$ we write $\sin^2\gamma = (\zeta - 1)^2/\eta_z^2$. This gives $\sin^2\gamma/\zeta = \left( \zeta - 2 + 1/\zeta \right)/\eta_z^2$. With $\int_0^{2\pi} d\gamma\, \zeta = 2\,\pi$ and Eq.~\eqref{eq:azIn},
\begin{align}
\int_0^{2\pi} d\gamma\, \frac{\sin^2\gamma}{\zeta} = \frac{1}{\eta_z^2} \left( 2\,\pi - 4\,\pi + \frac{2\,\pi}{\sqrt{1 - \eta_z^2}} \right) = \frac{2\,\pi}{\eta_z^2} \left( \frac{1}{\sqrt{1 - \eta_z^2}} - 1 \right).
\end{align}
Multiplying by $\mu/(\pi\,v_0)$ gives Eq.~\eqref{eq:kbarzvnr}, which can be written as $\bar\kappa_F = (\mu/v_0)\,\mathcal{P}_x (\eta_z)$. Hence $\bar\kappa_F \simeq (\mu/v_0) \left( 1 + 3\,\eta_z^2/4 \right)$ at small $\eta_z$, and $\bar\kappa_F \geq \mu/v_0$. The quantity $\bar\kappa_F$ is an average of $\kappa_F (\gamma)$ with the normalised weight $\sin^2\gamma/\pi$. It is therefore bounded by $\kappa_F^{\max}$, and $\Lambda > \kappa_F^{\max}$ implies $\Lambda > \bar\kappa_F$.

The shift of the background constants follows from $\delta\epsilon_{b,x} = -\,4\,\pi\,e^2\,\Re \delta\Pi/q^2$, which gives Eq.~\eqref{eq:debzvnr}. It is positive for $\Lambda > \bar\kappa_F$, and $\delta\epsilon_{b,z} = 0$ because $\delta\mathcal{F}_z = 0$. The axial tilt lowers the shift relative to the untilted one by $g\,e^2 \left( \bar\kappa_F - \mu/v_0 \right)/(8\,v_0\,k_0)$. Relative to $\epsilon_{b,x} - 1 = g\,e^2\,k_0/(8\,\mu)$ of Eq.~\eqref{eq:ebzvnr}, the ratio is
\begin{align}
\frac{\delta\epsilon_{b,x}}{\epsilon_{b,x} - 1} = \frac{g\,e^2 \left( \Lambda - \bar\kappa_F \right)/(8\,v_0\,k_0)}{g\,e^2\,k_0/(8\,\mu)} = \frac{\left( \Lambda - \bar\kappa_F \right) \kappa_F}{k_0^2} \,,
\end{align}
where $\kappa_F = \mu/v_0$ has been used. For $\eta_z = 0$ and $\Lambda = k_0$ it equals $\left( 1 - \kappa_F/k_0 \right) \kappa_F/k_0$. The requirement $\kappa_F^{\max} < \Lambda \leq k_0$ is equivalent to $|\eta_z| < 1 - \mu/(v_0\,k_0)$ for $\Lambda = k_0$, which is the range of validity of Eq.~\eqref{eq:debzvnr} stated in the main text. The shift depends on $\Lambda$, and it is of the same order as the corrections of relative size $\kappa/k_0$ that were dropped from the leading terms.

%%%%%%%%%%%%%%%%%%%%%%%%%
\bibliography{ref_pl}

\end{document}